\documentclass[aps,amsmath,amssymb,superscriptaddress,prb,reprint,twocolumn,longbibliography]{revtex4-2}
\usepackage{braket,bm}
\usepackage{float}
\usepackage{mathptmx}
\DeclareMathAlphabet{\mathcal}{OMS}{cmsy}{m}{n}
\usepackage[dvipdfmx]{graphicx}
\usepackage{xcolor}
\usepackage{mathtools}
\usepackage{setspace} 

\usepackage{booktabs}
\usepackage{multirow}

\usepackage{datetime2}

\begin{document}
	\title{Shaping Maximally Localized Wannier Functions via Discrete Adiabatic Transport}
	\author{Yuji Hamai}
	\email{hamai.sg@gmail.com}
	\affiliation{%
		Department of Nanotechnology for Sustainable Energy,
		School of Science and Technology, Kwansei Gakuin University,
		Gakuen-Uegahara 1, Sanda, 669-1330, Japan
	} %
	\author{Katsunori Wakabayashi}%
	\email{ka.wakaba@gmail.com}
	\affiliation{%
		Department of Nanotechnology for Sustainable Energy,
		School of Science and Technology, Kwansei Gakuin University,
		Gakuen-Uegahara 1, Sanda, 669-1330, Japan
	} %
	\affiliation{
		National Institute for Materials Science (NIMS), 
		Namiki 1-1, Tsukuba 305-0044, Japan
	}%
\date{\today}
	\begin{abstract}
		Maximally localized Wannier functions (MLWFs) are conventionally constructed by iteratively minimizing a spread functional over a high-dimensional gauge landscape. In this work, we present a non-variational constructive algorithm that unifies gauge smoothing and the eigenvalue problem of the projected position operator into a single deterministic framework. 
		We demonstrate that discrete adiabatic transport across band degeneracies emerges naturally as an integral part of the solution procedure for the position eigenvectors. In this transport-aligned gauge, the Bloch overlaps exhibit an approximately linear phase dependence, allowing the Wannier centers to be extracted via deterministic fixed-point iterations and self-consistent updates rather than spread-functional minimization. Benchmark calculations for one- and two-dimensional systems yield spreads and orbital shapes in good agreement with standard minimization schemes. Furthermore, this analytical approach transparently isolates the physical origin of the $\mathcal{O}(L)$ mesh-dependent spread scaling ($L$ being the boundary seam resolution) observed in graphene, demonstrating that it is an intrinsic geometric manifestation of non-commuting projected position operators forcing finite gauge defects to accumulate along a one-dimensional boundary seam.
	\end{abstract}
	\maketitle
	\section{INTRODUCTION}
	Wannier functions (WFs) provide an orthogonal, localized, and translationally
	invariant real-space framework for electronic states and have been
	systematically analyzed since the early days of solid-state
	physics~\cite{PhysRev.52.191,adams_chem,adams_phys,Blount,kivelson1982}. A
	central theme in the historical development of Wannier theory has been the
	search for exponentially localized WFs
	(ELWFs)~\cite{PhysRev.115.809,PhysRev.135.A685,PhysRevB.8.2485,PhysRevB.47.10112}. Motivated both by theoretical interest and practical needs for localized basis sets, these studies laid the foundation for the framework of maximally localized WFs (MLWFs), which has since become a standard tool across condensed matter, materials science, and photonics~\cite{MarzariVanderbilt1997,SouzaMarzariVanderbilt2001,MarzariRMP2012,Wannier90_2008,Wannier90_2020}.
	
	MLWFs enable the construction of tight-binding-like Hamiltonians and facilitate
	quantitative analyses of
	polarization~\cite{VanderbiltKingSmith1993,Resta1994,PhysRevLett.97.107602},
	orbital magnetization~\cite{PhysRevB.85.014435}, and related properties,
	chemical bonding~\cite{PhysRev.129.554,PhysRevB.64.245108}, and photonic
	confinement~\cite{Kurt2003,Busch2011,PhysRevB.88.075201}. They also provide the
	foundation for developments in topological electronic and photonic
	devices~\cite{Liu2022,Pham2016,PhysRevB.74.195116}. In parallel, the Berry phase
	and Wilson-loop frameworks have clarified the gauge structure behind WFs and
	their connection to band
	topology~\cite{Simon1983,Berry1984,Yu2011,SoluyanovVanderbilt2011,PhysRevB.89.155114}.
	
	The canonical approach to constructing MLWFs, established by Marzari and
	Vanderbilt (MV)~\cite{MarzariVanderbilt1997,SouzaMarzariVanderbilt2001},
	involves the numerical minimization of a spread functional with respect to
	unitary gauge transformations. While MV analytically showed that the minimum
	spread condition corresponds to a maximally smooth Berry connection, perfectly
	flattening this gauge is only possible in one-dimensional systems. 
	In two or higher dimensions, finding such a globally smooth gauge is fundamentally obstructed by the geometric frustration arising from the non-commutativity of the projected position operators, as explicitly formulated in the foundational works of MV \cite{MarzariVanderbilt1997,Brouder2007}. Standard algorithms resolve this non-commutativity via an iterative spread minimization, which inherently acts as a variational global optimization that distributes this unavoidable geometric gauge mismatch across the entire Brillouin zone \cite{MarzariVanderbilt1997}.

	To make the gauge search more robust against the choice of initial conditions, significant advancements have been integrated into widely used open-source codes~\cite{Wannier90_2008,Wannier90_2020}.
	These powerful refinements include optimized projections~\cite{PhysRevB.92.165134}, density-matrix based localized orbitals (SCDM-$k$)~\cite{ct500985f,damle2016scdmk}, and direct constructions via projected position operators in three dimensions~\cite{PhysRevB.103.075125}.
	
	In this paper, we propose a non-variational constructive algorithm
	by unifying gauge smoothing and the eigenvalue problem of the projected position operator (equivalently, the projected translation operator $e^{i\delta\boldsymbol k\cdot\hat{\boldsymbol x}}$) into a single deterministic framework. We demonstrate that discrete adiabatic transport, justified by Kato's adiabatic theorem~\cite{Kato1950, AvronElgart1998, AvronElgart1999,Teufel2003}, naturally emerges as an integral part of the solution procedure for the position eigenvectors. Rather than performing a variational global optimization of a spread functional over the high-dimensional gauge manifold, this transport directly aligns the periodic parts of the Bloch functions across internal degeneracies to exhibit a linear phase dependence.
	
	In this transport-aligned gauge, we derive a trigonometric equation for the Wannier centers, which is solved by a deterministic fixed-point iteration termed the ``sinc-loop''. In the multiband case, the remaining numerical loops are fixed-point iterations and self-consistent updates for explicit center equations and projected-position matrices; they are not variational gauge searches over the spread functional. 
	Crucially, because our sequential extraction approximately flattens the interior gauge along chosen one-dimensional strings instead of globally distributing the geometric mismatch, it transparently isolates the gauge frustration inherent in 2D systems. By applying our framework to graphene, we rigorously demonstrate that the anomalous $\mathcal{O}(L)$ mesh-dependent spread scaling ($L$ being the boundary seam resolution) is not a numerical artifact, but an intrinsic geometric manifestation of non-commuting projected position operators forcing finite $\mathcal{O}(L^0)$ gauge defects to accumulate along a one-dimensional boundary seam.
	
	The remainder of the paper is organized as follows: Section~\ref{sec:description}
	defines the system and notations. Section~\ref{sec:peeling} details the band peeling process via discrete adiabatic
	transport and its formal equivalence to the projected position operator framework.
	Sections~\ref{sec:xeigen} and \ref{sec:single-band} introduce the constructive phase-alignment
	procedure and the sinc-loop iteration for multi-band systems. Section~\ref{sec:results} presents numerical
	benchmarks in 1D and 2D systems, culminating in the explicit reciprocal-space analysis of the boundary seam and the macroscopic spread divergence in graphene. Finally, Section~\ref{sec:summary} summarizes our conclusions.
	\section{Description of the System}\label{sec:description} 
	This section introduces the notation and definitions 
	used for 1D and 2D isolated composite-band systems.
	\subsection{Bloch Functions}\label{sec:bloch}
	The Bravais lattice of the system consists of $L^d (L\in \mathbb Z)$ unit cells,
	$\Omega=[0,1)^d$, which are
	labeled as follows:
	\begin{equation}
		\mathbb M \;=\; (\mathbb Z_L)^d, 
	\end{equation}
	where,
	\begin{equation}
		\mathbb{Z}_L=\left\{ M\in\mathbb{Z}:-\frac{L}{2}\le M<\frac{L}{2}\right\},
	\end{equation}
	and it is assumed $L$ is even.
	The set of lattice momenta of the system is given as follows: 
	\begin{equation}\label{eq:K-def}
		\mathbb K \;=\; \Bigl\{\, \boldsymbol k = \delta k\,\boldsymbol n \;\Big|\; 
		\boldsymbol n\in(\mathbb Z_L)^d \Bigr\}, 
		\qquad \delta k = \frac{2\pi}{L}.
	\end{equation}
	The interior of the lattice momentum (open) space is also defined:
	\begin{equation}\label{eq:K_I-def}
		\mathbb K_I =\Bigl\{\, \boldsymbol k = \delta k \boldsymbol n \;\Big|\; 
		\boldsymbol n\in(\mathbb Z_I)^d \Bigr\}, 
	\end{equation}
	where,
	\begin{equation}
		\mathbb{Z}_I=\left\{ M\in\mathbb{Z}:-\frac{L}{2} < M<\frac{L}{2}\right\}.
	\end{equation}
	The Schrödinger equation of the system is: 
	\begin{equation}
		\hat{H}|\psi_{\boldsymbol{k}}^n\rangle=\epsilon_{\boldsymbol{k}}^n|\psi_{\boldsymbol{k}}^n\rangle,\boldsymbol{k}\in\mathbb{K},n\in\mathbb{Z}_{>0},
	\end{equation}
	where $\hat{H}$ is the Hamiltonian operator of the system and
	$\{\vert\psi_{\boldsymbol{k}}^n\rangle\}$ and
	$\{\epsilon_{\boldsymbol{k}}^n\}$ are the eigenvectors and eigenvalues of the
	operator, respectively. The superscript $n$ denotes the energy level or band
	index of the eigenvector pertaining to the wave number $\boldsymbol{k}$. The
	eigenvectors, namely BFs, are further decomposed into the
	following form: \begin{equation}\label{eq:bloch}
		\begin{aligned}\vert\psi_{\boldsymbol{k}}^n\rangle=L^{-d/2}e^{i\boldsymbol{k}\cdot\hat{\boldsymbol{x}}}\vert v_{\boldsymbol{k}}^n\rangle,\end{aligned}
	\end{equation}
	where $\{\vert v_{\boldsymbol{k}}^n\rangle\}$ is the set of cell-periodic
	solutions of the following Bloch Hamiltonian eigenvalue equation: 
	\begin{equation}
		\hat{H}(\boldsymbol{k})\vert v_{\boldsymbol{k}}^n\rangle=\epsilon_{\boldsymbol{k}}^n\vert v_{\boldsymbol{k}}^n\rangle,
	\end{equation}
	where, 
	\begin{equation}
		\hat{H}(\boldsymbol{k})=e^{-i\boldsymbol{k}\cdot\hat{\boldsymbol{x}}}\hat{H}e^{i\boldsymbol{k}\cdot\hat{\boldsymbol{x}}},
	\end{equation}
	and $\{\vert v_{\boldsymbol{k}}^n\rangle\}$ are normalized as follows:
	\begin{equation}
		\langle v_{\boldsymbol{k}}^n\vert v_{\boldsymbol{k}}^n\rangle=\int_{\mathrm{cell}}\overline{v}_{\boldsymbol{k}}^n(\boldsymbol{x})v_{\boldsymbol{k}}^n(\boldsymbol{x})d^d\boldsymbol{x}=1.
	\end{equation}
	The overline, e.g., $\overline{v}_{\boldsymbol{k}}^n$, denotes the complex conjugate of the respective variable.
	Since
	\begin{equation}
		\begin{aligned}
			&\forall\,\boldsymbol{k}_{1}, \boldsymbol{k}_{2}\in\mathbb{K},\\
			&\mathrm{Span}\{|v_{\boldsymbol{k}_{1}}^m\rangle\mid m\in\mathbb{Z}_{\ge 0}\}
			=\mathrm{Span}\{|v_{\boldsymbol{k}_{2}}^m\rangle\mid m\in\mathbb{Z}_{\ge 0}\},
		\end{aligned}
	\end{equation}
	for all $\vert f\rangle$ satisfying: 
	\begin{equation}\label{eq:v_periodic}
		\begin{aligned}
			\forall\boldsymbol{M}\in\mathbb{M},
			\langle\boldsymbol{x}+\boldsymbol{M}\vert f\rangle=\langle \boldsymbol{x} \vert f\rangle,\end{aligned}
	\end{equation}
	the following holds: 
	\begin{equation}\label{eq:v_span}
		\begin{aligned}\forall\boldsymbol{k}\in\mathbb{K},\ \vert f\rangle & =\sum_{m}\vert v_{\boldsymbol{k}}^m\rangle\langle v_{\boldsymbol{k}}^m\vert f\rangle.\end{aligned}
	\end{equation}
	\subsection{ State Space of Composite Energy Band System}\label{app:op-def} 
	Let the set of the energy bands $\{n_{a},n_{b},\cdots,n_{z}\}$
	composing the composite energy band under consideration be: 
	\begin{equation}
		\mathbb{N}_{C}=\{n_{a},n_{b},\cdots,n_{z}\},
	\end{equation}
	then the state space is spanned by the BFs described in Eq.~(\ref{eq:bloch}):
	\begin{equation}
		\mathcal{H}_{C}={\mathrm{Span}}\{\vert\psi_{\boldsymbol{k}}^n\rangle:\boldsymbol{k}\in\mathbb{K},n\in\mathbb{N}_{C}\},
	\end{equation}
	To close the operation within $\mathcal{H}_{C}$, the following projection
	operator is introduced,
	\begin{equation}
		\begin{aligned}\hat{P}_{\mathcal{H}_{C}} & =\sum_{n\in\mathbb{N}_{C}}\sum_{\boldsymbol{k}\in\mathbb{K}}\vert\psi_{\boldsymbol{k}}^n\rangle\langle\psi_{\boldsymbol{k}}^n\vert.\end{aligned}
	\end{equation}
	By using this, the following band projected operator is defined and
	it suppresses components protruding from $\mathcal{H}_{C}$, as well
	as filters out components outside $\mathcal{H}_{C}$ as the input:
	\begin{equation}
		\begin{aligned}
			\tilde{A} & =\hat{P}_{\mathcal{H}_{C}}\;\hat{A}\;\hat{P}_{\mathcal{H}_{C}}.
		\end{aligned}
		\label{eqap:projection-operator}
	\end{equation}
	Another projection operator pertaining to the space spanned by the local frame at $\boldsymbol k$ is defined
	as follows: 
	\begin{equation}\label{eq:pp-def}
		\hat P(\boldsymbol{k})=\sum_{n\in\mathbb{N}_{C}}\vert v_{\boldsymbol{k}}^n\rangle\langle v_{\boldsymbol{k}}^n\vert.
	\end{equation}
	\subsection{Wannier Functions }
	The WFs are defined as follows: 
	
	\begin{equation}\label{eq:wannier-def}
		\begin{aligned}
			\vert W_{\boldsymbol{M}}^s\rangle 
			&=L^{-d/2}\sum_{\boldsymbol{k},n}e^{-i\boldsymbol{k}\cdot\boldsymbol{M}}f_{\boldsymbol{k}}^{s,n}\vert\psi_{\boldsymbol{k}}^n\rangle\\
			&=L^{-d/2}\sum_{\boldsymbol{k},n}e^{-i\boldsymbol{k}\cdot\boldsymbol{M}}e^{i\boldsymbol{k}\cdot\hat{\boldsymbol{x}}}f_{\boldsymbol{k}}^{s,n}\vert v_{\boldsymbol{k}}^n\rangle,
		\end{aligned}
	\end{equation}
	with the condition: 
	\begin{equation}\label{eq:wannier_orthonormal}
		\langle W_{\boldsymbol{M}_{1}}^{s_{1}}\vert W_{\boldsymbol{M}_{2}}^{s_{2}}\rangle
		=\delta_{\boldsymbol{M}_{1},\boldsymbol{M}_{2}}\delta_{s_{1},s_{2}}, \boldsymbol M_1, \boldsymbol M_2 \in \mathbb M,
	\end{equation}
	where $\boldsymbol{M},\boldsymbol{M}_{1}$ and $\boldsymbol{M}_{2}$ are
	the cell indices,
	and $s$ indicates the series index in the composite band system and
	$n$ denotes the band index. $f_{\boldsymbol{k}}^{s,n}$ is the $(s,n)$ element of a unitary
	matrix transforming the periodic parts pertaining to band $n$ to
	those pertaining to series, $\{\vert u_{k}^s\rangle\}$: 
	\begin{equation}
		\vert u_{\boldsymbol{k}}^s\rangle=\sum_{n}f_{\boldsymbol{k}}^{s,n}\vert v_{\boldsymbol{k}}^n\rangle,\label{eq:composite_periodic}
	\end{equation}
	and hence, the WF belonging to the series $s$ is expressed: 
	\begin{equation}
		\begin{aligned}
			\vert W_{\boldsymbol M}^s\rangle & =L^{-d/2}\sum_{\boldsymbol k}e^{-i\boldsymbol{k}\cdot\boldsymbol{M}}e^{i\boldsymbol{k}\cdot\hat{\boldsymbol{x}}}\vert u_{\boldsymbol{k}}^s\rangle.
		\end{aligned}
		\label{eq:wannier_series_def}
	\end{equation}
	From Eq.~(\ref{eq:wannier_orthonormal}), 
	\begin{equation}
		\begin{aligned}
			\mathcal{H}_{C}
			&={\mathrm{Span}}\{\vert\psi_{\boldsymbol{k}}^n\rangle:\boldsymbol{k}\in\mathbb{K},n\in\mathbb{N}_{C}\}\\
			&={\mathrm{Span}}\{\vert W^s_{\boldsymbol{M}}\rangle:\boldsymbol{M}\in\mathbb{M},s\in\mathbb{N}_{C}\}.
		\end{aligned}
	\end{equation}
	\section{Band peeling for a composite band with internal degeneracies}\label{sec:peeling}
	We address an isolated composite-band subspace that may have internal
	degeneracies, while being separated from the rest of the spectrum by finite gaps
	throughout the Brillouin zone. We will also show that two ostensibly different
	viewpoints---(i) eigenvalue equation for $e^{i\delta \boldsymbol k \cdot
		\hat{\boldsymbol x}}$ and (ii) the multi-band adiabatic transport--- coincide to
	first order in $|\delta\boldsymbol{k}|$, and provide a first-order accurate
	smoothing of gauge discontinuities in $\mathbb K_I$. We call this procedure band
	peeling to distinguish it from the variational disentanglement of
	Souza–Marzari–Vanderbilt~\cite{SouzaMarzariVanderbilt2001}, which selects an
	optimal subspace from energetically entangled bands.
	
	Furthermore, it is shown that the adiabatic transport/projection in the current
	study is equivalent to Kato’s parallel transport consistent with the gapless
	adiabatic framework of Avron and
	Elgart~\cite{Kato1950,AvronElgart1998,AvronElgart1999}, which eventually
	provides a justification of the current procedures.

	Throughout this section and the appendices, $\vert v^n_{\boldsymbol k}\rangle$ denotes the local energy-ordered periodic Bloch frame at each $\boldsymbol k$, indexed by the band label $n$, whereas $\vert u^s_{\boldsymbol k}\rangle$ denotes the peeled/transported frame indexed by the series label $s$.
	\begin{figure}[htbp]
		\centering
		\includegraphics[width=1.0\linewidth]{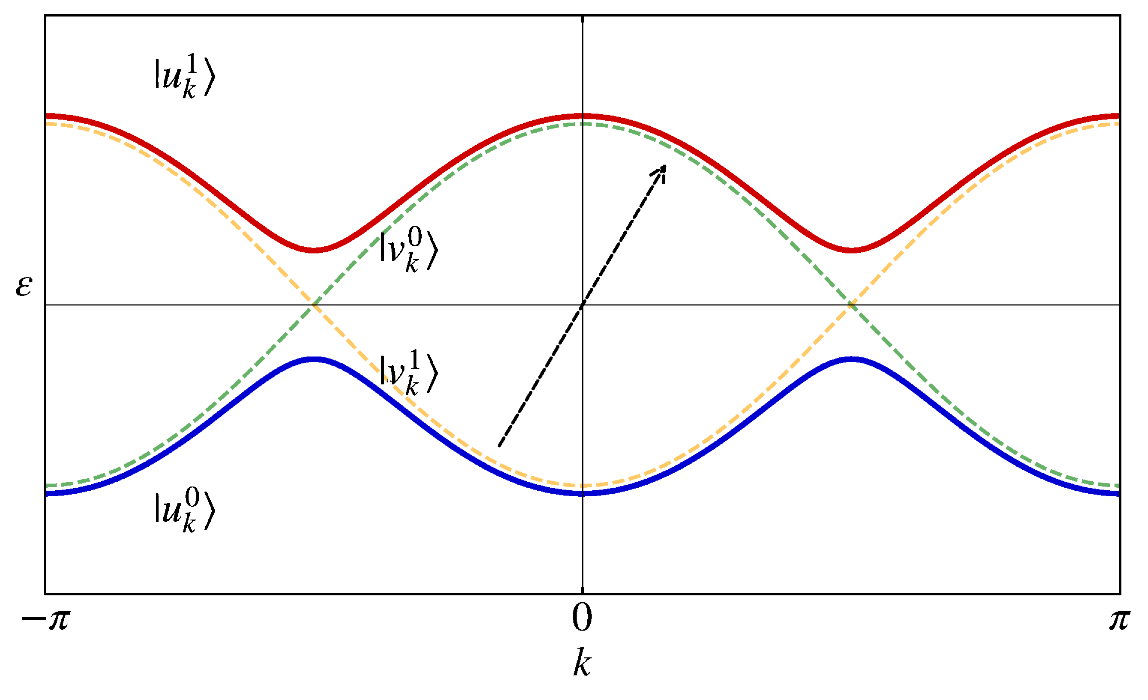}
		\caption{(Color online)
			Schematic diagram of energy band peeling across degeneracy points. In the traditional
			band labeling, as shown with $\vert v^n_{\boldsymbol k}\rangle$, the bands are indexed locally in k-space in ascending order from the bottom. On the other hand, the numbering indicated with $\vert u^s_{\boldsymbol k}\rangle$ provides a \textit{smoother} connection. Mathematically, $\vert u^s_{\boldsymbol k}\rangle$ is projected to Span$\{\vert v^m_{\boldsymbol k+\delta \boldsymbol k} \rangle \vert m=0,1\}$ and becomes $\vert u^s_{\boldsymbol k+\delta \boldsymbol k}\rangle$. 
		}
		\label{fig:bandpeeling}
	\end{figure}
	\subsection{Adiabatic change induced by $e^{i\delta \boldsymbol k \cdot \hat{\boldsymbol x } }$}\label{sec:adiabatic}
	As shown in Fig.~\ref{fig:bandpeeling}, 
	it is reasonable to assume that
	the \textit{natural} connection of energy bands might be achieved
	by extending the energy bands with the adiabatic approximation~\cite{Asboth2016ShortCourse,CayssolFuchs2021TopologicalGeometrical}. 
	And hence, this section describes how the periodic parts of BFs are
	transformed by the adiabatic approximation.
	
	Let $\hat H(\boldsymbol k)$ be the \textit{Hamiltonian} and $\epsilon^n(t)$ be the instantaneous eigenvalue of the system,
	\begin{equation}\label{eq:adiabatic-hamiltonian}
		\hat H(\boldsymbol{k})\vert v_{\boldsymbol{k}}^n\rangle=\epsilon^n(t)\vert v_{\boldsymbol{k}}^n\rangle,
	\end{equation}
	where $\boldsymbol k$ is a function of time $t$:
	\begin{equation}
		\boldsymbol k = \boldsymbol k (t).
	\end{equation}
	By expanding the transient solution $\vert\Psi(t)\rangle$ in the following way:
	\begin{equation}
		\begin{array}{c}
			\begin{aligned}\vert\Psi(t)\rangle & =\sum_{n}c_{\boldsymbol{k}}^ne^{-i\theta^n(t)}\vert v_{\boldsymbol{k}}^n\rangle,
			\end{aligned}
		\end{array}
	\end{equation} 
	and imposing the following equation:
	\begin{equation}\label{eq:dpsidt}
		i\hbar\frac{d}{dt}\vert\Psi(t)\rangle=\hat{H}(\boldsymbol k)\vert\Psi(t)\rangle,
	\end{equation}
	we have (see Appendix~\ref{ap:adiabatic-supplement} for details), 
	\begin{equation}
		\begin{aligned}
			c_{\boldsymbol{k}+\delta\boldsymbol{k}}^m & =\sum_{n}c_{\boldsymbol{k}}^n\langle v_{\boldsymbol{k}+\delta\boldsymbol{k}}^m\vert v_{\boldsymbol{k}}^n\rangle e^{i(\theta^m(t)-\theta^n(t))}+\mathcal O(\delta\boldsymbol{k}^2),
		\end{aligned}
	\end{equation}
	which leads to the following transport picture,
	\begin{equation}\label{eq:transient-adiabatic-solution}
		\vert\Psi(t+\delta t )\rangle=\left\{ \sum_{m}\vert v_{\boldsymbol{k}+\delta\boldsymbol{k}}^m\rangle\langle v_{\boldsymbol{k}+\delta\boldsymbol{k}}^m\vert\right\} \sum_{n}c_{\boldsymbol{k}}^n \vert v_{\boldsymbol{k}}^n\rangle e^{-i\theta^n(t)},
	\end{equation}
	with 
	\begin{equation}
		\begin{array}{c}
			\begin{aligned}
				\theta^n(t) & =\frac{1}{\hbar}\int^t\epsilon^n(\tau)d\tau.
			\end{aligned}
		\end{array}
	\end{equation} 
	When the initial condition is:
	\begin{equation}
		c_{\boldsymbol{k}}^n = \delta_{n,s},
	\end{equation}
	after setting the dynamical factor to unity,
	the state at $\boldsymbol k $ develops to:
	\begin{equation}\label{eq:composite-adiabatic-peeling}
		\begin{aligned}	
			\vert u_{\boldsymbol{k}+\delta\boldsymbol{k}}^s\rangle&=\vert\Psi(\boldsymbol{k}+\delta\boldsymbol{k})\rangle\\
			&=\left\{ \sum_{m}\vert v_{\boldsymbol{k}+\delta\boldsymbol{k}}^m\rangle\langle v_{\boldsymbol{k}+\delta\boldsymbol{k}}^m\vert\right\} \vert u_{\boldsymbol{k}}^s\rangle+\mathcal O(\delta\boldsymbol{k}^2).
		\end{aligned}
	\end{equation}
	Since the projection is summation over all bands in the composite-band system
	and $\vert u^s_{\boldsymbol k}\rangle$ is projected to the closest vector at
	$\boldsymbol k+ \delta \boldsymbol k$, the new indexing at
	$\boldsymbol{k}+\delta\boldsymbol{k}$ no longer depends on a particular local
	labeling of the energy bands denoted by $m$ in
	Eq.~(\ref{eq:composite-adiabatic-peeling}), as shown in
	Fig.~\ref{fig:bandpeeling}. Therefore, it eliminates at least the phase jump
	coming from band indexing.
	%
	\subsection{Eigenvalue Equation for $e^{i\delta \boldsymbol k \cdot \hat{\boldsymbol x } }$}
	This subsection derives the solution procedure of the eigenvalue problem of the translation operator, $e^{i\delta \boldsymbol k \cdot \hat{\boldsymbol x } }$. For this operator the following equations hold:
	\begin{equation}\label{eq:composite-eigenvalue-equation}
		\begin{aligned}
			\langle\psi_{\boldsymbol{k}}^m\vert e^{i\delta\boldsymbol{k}\cdot\hat {\boldsymbol x} }\vert W_{0}^s\rangle & =\sum_{n}f_{\boldsymbol{k}-\delta\boldsymbol{k}}^{s,n}\langle v_{\boldsymbol{k}}^m\vert v_{\boldsymbol{k}-\delta\boldsymbol{k}}^n\rangle +\mathcal O(\delta\boldsymbol{k}^2)\\
			\langle\psi_{\boldsymbol{k}}^m\vert e^{i\delta\boldsymbol{k}\cdot\hat {\boldsymbol x}}\vert W_{0}^s\rangle & =e^{i\delta \boldsymbol k \cdot \boldsymbol x_{0}}f_{\boldsymbol{k}}^{s,m},
		\end{aligned}
	\end{equation}
	and hence, it leads to the following equation composed of the periodic parts (see Appendix~\ref{ap:xeigen-supplement} for details):
	\begin{equation}\label{eq:composite-final-eigenvalue-equation}
		f_{\boldsymbol{k}}^{s,m}=e^{-i\delta \boldsymbol k \cdot \boldsymbol x_{0}}\sum_{n}f_{\boldsymbol{k}-\delta\boldsymbol{k}}^{s,n}\langle v_{\boldsymbol{k}}^m\vert v_{\boldsymbol{k}-\delta\boldsymbol{k}}^n\rangle+\mathcal O(\delta\boldsymbol{k}^2),
	\end{equation}
	By multiplying $\vert v_{\boldsymbol{k}}^n\rangle$ to both sides and 
	summing over $n$, the energy bands are reconstructed and the following projection equation yields:
	\begin{equation}\label{eq:composite-peeling-projection}
		\begin{aligned}
			\vert u_{\boldsymbol{k}}^s\rangle & =e^{-i\delta \boldsymbol k \cdot \boldsymbol x_{0}}\left\{ \sum_{m}\vert v_{\boldsymbol{k}}^m\rangle\langle v_{\boldsymbol{k}}^m\vert\right\} \vert u_{\boldsymbol{k}-\delta\boldsymbol{k}}^s\rangle+\mathcal O(\delta\boldsymbol{k}^2),
		\end{aligned}
	\end{equation}
	The \textit{twist} $e^{-i\delta \boldsymbol k \cdot \boldsymbol x_{0}}$~\cite{Vanderbilt2018Book} has to be
	determined separately, so that it reconciles the phase jumps across the border
	of the k-space after summation over the k-space as seen in
	Eq.~(\ref{eq:composite-eigenvalue-equation}). Before proceeding, we note explicitly that Eq.~(\ref{eq:composite-peeling-projection}) reduces to Eq.~(\ref{eq:composite-adiabatic-peeling}) once the twist is removed, $\boldsymbol x_0=\boldsymbol 0$, and the wave number is shifted as $\boldsymbol k\rightarrow\boldsymbol k+\delta\boldsymbol k$:
	\begin{equation*}
		\text{Eq.~(\ref{eq:composite-peeling-projection})}\ \xrightarrow[\boldsymbol x_0=\boldsymbol 0]{\boldsymbol k\to\boldsymbol k+\delta\boldsymbol k}\ 
		|u_{\boldsymbol k+\delta \boldsymbol k}^{s}\rangle = \left\{ \sum_m |v_{\boldsymbol k+\delta \boldsymbol k}^m\rangle \langle v_{\boldsymbol k+\delta \boldsymbol k}^m| \right\} |u_{\boldsymbol k}^s\rangle + \mathcal{O}(|\delta \boldsymbol k|^2),
	\end{equation*}
	which is precisely Eq.~(\ref{eq:composite-adiabatic-peeling}).

	\subsection{ Underlying Symmetry
		of the Adiabatic Expansion}\label{sec:adiabatic-translation}
	The primary objective of this subsection is to explicitly demonstrate that the discrete adiabatic transport (Eq.~(\ref{eq:composite-adiabatic-peeling})) and the eigenvalue equation of the projected translation operator (Eq.~(\ref{eq:composite-peeling-projection})) are formally equivalent and originate from the same first-order geometric transport.
	
	Dropping the explicit time dependence of
	Eq.~(\ref{eq:transient-adiabatic-solution}) is equivalent to setting the
	left-hand side of Eq.~(\ref{eq:dpsidt}) to zero. Equivalently, for any
	orthonormal frame $\{|v_{\boldsymbol{k}}^m\rangle\}$, at $\boldsymbol k$,
	\begin{equation}\label{eq:Psi=0}
		\left\{ \sum_{m }\vert v_{\boldsymbol{k}}^{m}\rangle\langle v_{\boldsymbol{k}}^{m}\vert\right\} \frac{d}{dt}\vert\Psi(\boldsymbol{k}(t))\rangle=0,
	\end{equation} 
	where the summation is performed over all relevant band indices.
	Since $\vert\Psi(\boldsymbol{k})\rangle$ depends on $t$ only through $\boldsymbol{k}(t)$, the
	following equation then holds:
	\begin{equation}
		\delta t \left\{ \sum_{m }\vert v_{\boldsymbol{k}}^{m}\rangle\langle v_{\boldsymbol{k}}^{m}\vert\right\} \frac{d\boldsymbol{k}}{dt}\nabla_{\boldsymbol{k}}\vert\Psi(\boldsymbol{k}(t))\rangle=0,
	\end{equation}
	where $\delta t$ is an infinitesimally short time duration. 
	If $\{\vert u^s_{\boldsymbol k} \rangle \}$ is a specific 
	realization of $\vert\Psi(\boldsymbol{k})\rangle$, 
	by replacing $\delta t\,(d\boldsymbol{k}/dt)$ with $\delta \boldsymbol{k}$,
	the following equation 
	has to hold:
	\begin{equation}\label{eq:berry-connection-zero}
		\begin{split}
			\left\{ \sum_{m}\vert v_{\boldsymbol{k}}^{m}\rangle\langle v_{\boldsymbol{k}}^{m}\vert\right\} \delta\boldsymbol{k}\nabla_{\boldsymbol{k}}\vert u_{\boldsymbol{k}}^{s}\rangle&=
			\left\{ \sum_{m}\vert v_{\boldsymbol{k}}^{m}\rangle\langle v_{\boldsymbol{k}}^{m}\vert\right\} \left\{ \vert u_{\boldsymbol{k}}^{s}\rangle-\vert u_{\boldsymbol{k}-\delta\boldsymbol{k}}^{s}\rangle\right\} \\&=0.
		\end{split}
	\end{equation}
	Thus we arrive at Eq.~(\ref{eq:composite-peeling-projection}) without the twist.
	\begin{equation}
		\vert u_{\boldsymbol{k}}^{s}\rangle=\left\{ \sum_{m}\vert v_{\boldsymbol{k}}^{m}\rangle\langle v_{\boldsymbol{k}}^{m}\vert\right\} \vert u_{\boldsymbol{k}-\delta\boldsymbol{k}}^{s}\rangle.
	\end{equation} 
	Therefore, the role of the adiabatic expansion/transport in the present construction is to make the transported local subspace spanned by $\{|u_{\boldsymbol{k}}^{s}\rangle\}$ as uniform as possible along the transport direction. In this sense, the transport virtually removes one degree of gauge discontinuity from the Bloch frame by enforcing first-order consistency between neighboring local subspaces.
	
	Although the adiabatic expansion is introduced from the Schr\"odinger equation with the Bloch Hamiltonian $\hat H(\boldsymbol{k})$, the Hamiltonian plays no direct role in the construction from Eq.~(36) onward. Its role in the peeling and in identifying the eigenvectors of the projected position/translation operator is only to specify the family of composite-band subspaces at each $\boldsymbol{k}$. The essential content of the adiabatic transport, as expressed in Eq.~(36), is that the periodic part should remain unchanged to first order along the transport direction $d\boldsymbol{k}/dt$, and hence along $\delta\boldsymbol{k}$.
	
	\subsection{Kato's transport and justification of the adiabatic transport in the presence of internal degeneracies}
	In standard treatments of the adiabatic theorem, one typically assumes that
	the instantaneous eigenstates are nondegenerate. Here we justify the use of
	Eq.~(\ref{eq:composite-adiabatic-peeling}) for isolated composite‑band systems
	in the presence of internal degeneracy.
	
	In the works of Kato, Avron and Elgart~ \cite{AvronElgart1998, AvronElgart1999, Kato1950}, the
	validity of the adiabatic theorem is investigated, and they have shown that the
	approximation is valid even when gaps close, as long as,
	\begin{equation}\label{eq:pp-continuity}
		\hat P(\boldsymbol{k})=\sum_{m\in \mathbb N_C}\vert v_{\boldsymbol{k}}^m\rangle\langle v_{\boldsymbol{k}}^m\vert \in C^1.
	\end{equation}
	In their formula, the state at $\boldsymbol k$ evolves in the following way,
	\begin{equation}
		\vert\Psi_{\boldsymbol k+\delta \boldsymbol k}\rangle = \hat U(\boldsymbol{k} )\vert\Psi_{\boldsymbol k}\rangle,
	\end{equation}
	where,
	\begin{equation}
		\begin{aligned}
			\dot{ \hat{U}}(\boldsymbol{k}) & =-i\hat{E}(\boldsymbol{k}) \hat{U}(\boldsymbol{k})\\
			\hat E(\boldsymbol{k}) & =i\left[\dot{\hat P}(\boldsymbol{k}),\hat P(\boldsymbol{k})\right],
		\end{aligned}
	\end{equation}
	and $\boldsymbol k$ is a function of $t$ and the dot, e.g., $\dot{P}(\boldsymbol{k})$, denotes the time derivative of the respective variable. 
	The first order solution is thus, given as follows:
	\begin{equation}
		\begin{aligned}
			\hat U(\boldsymbol{k}) 
			& =I-i\delta t \hat E(\boldsymbol k)+\mathcal O(\delta k^2),\\
		\end{aligned}
	\end{equation}
	By limiting the initial states at $\boldsymbol k$ and final states at $\boldsymbol k+ \delta \boldsymbol k$
	in the composite energy bands
	(see Appendix~\ref{ap:kato} for details),
	\begin{equation}
		\begin{aligned}
			\hat{P}(\boldsymbol{k}+\delta\boldsymbol{k})\hat{U}(\boldsymbol{k})\hat{P}(\boldsymbol{k}) 
			& =\hat{P}(\boldsymbol{k}+\delta\boldsymbol{k})\hat{P}(\boldsymbol{k})+\left(\delta\boldsymbol{k}\nabla \hat{P}(\boldsymbol{k})\right)^2\hat{P}(\boldsymbol{k})\\
			& =\hat{P}(\boldsymbol{k}+\delta\boldsymbol{k})\hat{P}(\boldsymbol{k})+\mathcal O(\vert\delta\boldsymbol{k}\vert^2).
		\end{aligned}
	\end{equation}
	And hence, the following propagation/translation equation is obtained, 
	\begin{equation}\label{eq:kato-transport}
		\begin{aligned}
			\vert u_{\boldsymbol{k}+\delta\boldsymbol{k}}^s\rangle & =\hat{P}(\boldsymbol{k}+\delta\boldsymbol{k})\hat{U}(\boldsymbol{k})\hat{P}(\boldsymbol{k})\vert u_{\boldsymbol{k}}^s\rangle\\
			& =\hat{P}(\boldsymbol{k}+\delta\boldsymbol{k})\hat{P}(\boldsymbol{k})\vert u_{\boldsymbol{k}}^s\rangle\\
			& =\left\{ \sum_{m_{1}}\vert v_{\boldsymbol{k}+\delta\boldsymbol{k}}^{m_{1}}\rangle\langle v_{\boldsymbol{k}+\delta\boldsymbol{k}}^{m_{1}}\vert\right\} \left\{ \sum_{m_{2}}\vert v_{\boldsymbol{k}}^{m_{2}}\rangle\langle v_{\boldsymbol k}^{m_{2}}\vert\right\} \vert u_{\boldsymbol{k}}^s\rangle\\
			& =\sum_{m}\vert v_{\boldsymbol{k}+\delta\boldsymbol{k}}^m\rangle\langle v_{\boldsymbol{k}+\delta\boldsymbol{k}}^m\vert u_{\boldsymbol{k}}^s\rangle,
		\end{aligned}
	\end{equation}
	which coincides with the geometrical parts of
	Eq.~(\ref{eq:composite-adiabatic-peeling}). Therefore,
	Eq.~(\ref{eq:composite-adiabatic-peeling}) is justified at first order for
	isolated composite-band systems even in the presence of internal degeneracies,
	without assuming a uniform spectral gap, provided $\hat P(\boldsymbol
	k)$ remains $C^1$ along the path.
	\subsection{Correction to
		Eqs.~(\ref{eq:composite-adiabatic-peeling}) and
		(\ref{eq:composite-peeling-projection}) based on Kato's formula } There are two
	practical differences between Kato's transport and the first-order projection
	described by Eqs.~(\ref{eq:composite-adiabatic-peeling}) and
	(\ref{eq:composite-peeling-projection}). First, the latter employ a one-sided
	finite-difference step; this is not essential and can be improved by
	higher-order schemes. Second, the overlap matrix between the frames at
	$\boldsymbol{k}$ and $\boldsymbol{k}+\delta\boldsymbol{k}$ is not exactly
	unitary at finite mesh spacing. We enforce unitarity by replacing it with the
	unitary factor of the polar decomposition via singular value
	decomposition~(SVD)~\cite{Vanderbilt2018Book}.
	
	Writing the projected vectors as,
	\begin{equation}
		\begin{aligned}
			\vert u_{\boldsymbol{k}+\delta\boldsymbol{k}}^s\rangle
			& =\sum_{m\in\mathbb{N}_C}\vert v_{\boldsymbol{k}+\delta\boldsymbol{k}}^m\rangle\langle v_{\boldsymbol{k}+\delta\boldsymbol{k}}^m\vert u_{\boldsymbol{k}}^s\rangle
			=\sum_{m}M_{m,s} \vert v_{\boldsymbol{k}+\delta\boldsymbol{k}}^m\rangle,
		\end{aligned}
	\end{equation}
	where,
	\begin{equation}
		M_{m,s} =
		\langle v^m_{\boldsymbol k +\delta \boldsymbol k}\vert u^s_{\boldsymbol k}\rangle,
	\end{equation}
	we compute the SVD $M = V\,\Sigma\,W^\dagger$ with $\Sigma=\mathrm{diag}(\sigma_1,\dots,\sigma_J)$ and define the unitary polar factor,
	\begin{equation}
		Q =V W^{\dagger}.
	\end{equation}
	The orthonormalized transport step is then,
	\begin{equation}
		\vert u_{\boldsymbol{k}+\delta\boldsymbol{k}}^s\rangle
		= \sum_{m} \vert v_{\boldsymbol k+\delta \boldsymbol k }^m\rangle Q_{m,s}.
	\end{equation}
	\section{Fixing Wannier Center and Initial Condition of $\vert u_{\boldsymbol{k}}^s\rangle$ }\label{sec:xeigen}
	Since the parallel transport Eqs.~(\ref{eq:composite-adiabatic-peeling}) and
	(\ref{eq:composite-peeling-projection}) alone do not provide the initial condition of $\{f_{\boldsymbol{k}}^{s,n} \}$ 
	nor fix the Wannier center $\boldsymbol x_0$, this section is devoted to solving these
	problems. It should be noted, however, that when the following condition,
	\begin{equation}
		\langle u_{\boldsymbol{k}}^{s_{0}}\vert u_{\boldsymbol{k}}^{s_{1}}\rangle=\delta_{s_{0},s_{1}},
	\end{equation}
	is satisfied, the periodic parts remain approximately orthogonal, as
	detailed in Appendix~\ref{ap:k+dk-orthogonal},
	\begin{equation}\label{eq:k+dk-orthogonal}
		\langle u_{\boldsymbol{k}+\delta\boldsymbol{k}}^{s_{0}}\vert u_{\boldsymbol{k}+\delta\boldsymbol{k}}^{s_{1}}\rangle=\delta_{s_{0},s_{1}}+\mathcal{O}(\delta k^2).
	\end{equation}
	And hence, by provisionally setting, 
	\begin{equation}\label{eq:provisional-initial-condition}
		\vert u^s_{\boldsymbol k_0}\rangle = \vert v^s_{\boldsymbol k_0} \rangle,
	\end{equation}
	somewhere in the k-space, where the energy bands are reasonably separated, the
	obtained WFs are approximately orthogonal up to the first order of $\delta k$
	and resemble the right eigenvectors of the projected translation operator
	except the Wannier center and the initial value of $\{f^{s,n}_{\boldsymbol
		k_0}\}$. \subsection{Fixing the Initial Condition of $\{f^{s,n}_{\boldsymbol
			k}\}$} By using the provisional WFs $\{ \vert w^s_{\boldsymbol 0}\rangle\}$, the
	right MLWFs, $\{ \vert W^s_{\boldsymbol 0}\rangle \}$, the eigenvectors of a
	position operator, have to be obtained by applying a unitary transformation,
	\begin{equation}
		\vert\boldsymbol{W}_{\boldsymbol{0}}\rangle=\vert\boldsymbol{w}_{\boldsymbol{0}}\rangle\boldsymbol{F},
	\end{equation}
	if expressed element-by-element:
	\begin{equation}
		\begin{array}{c}
			\begin{aligned}
				\left[\begin{array}{cccc}
					\vert W^0{}_{\boldsymbol{0}}\rangle & \vert W^1{}_{\boldsymbol{0}}\rangle & \ldots & \vert W^{N_C - 1}{}_{\boldsymbol{0}}\rangle\end{array}\right] & =\\
				\left[\begin{array}{cccc}
					\vert w^0_{\boldsymbol{0}}\rangle & \vert w^1_{\boldsymbol{0}}\rangle & \ldots & \vert w^{N_C - 1}_{\boldsymbol{0}}\rangle\end{array}\right] & \left[\begin{array}{cccc}
					F_{0}^0 & F_{0}^1 & \ldots & F_{0}^{N_C-1}\\
					F_{1}^0 & F_{1}^1 & \ldots & F_{1}^{N_C-1}\\
					\vdots & \vdots & \ddots & \vdots\\
					F_{N_C-1}^0 & F_{N_C-1}^1 & \ldots & F_{N_C-1}^{N_C-1}
				\end{array}\right].
			\end{aligned}
		\end{array}
	\end{equation}
	By requiring $\vert \boldsymbol{ W}_{\boldsymbol 0} \rangle$ to be the eigenvectors of the position operator
	$\hat x $, we have the following eigenvalue equation.
	\begin{equation}\label{eq:xeigen-F}
		\begin{split}
			\langle\boldsymbol{W}_{0}\vert\hat{x}\vert\boldsymbol{W}_{0}\rangle&=\boldsymbol{F}^{\dagger}\langle\boldsymbol{w}_{0}\vert\hat{x}\vert\boldsymbol{w}_{0}\rangle\boldsymbol{F}\\&=\boldsymbol{F}^{\dagger}X\boldsymbol{F}\\&=\boldsymbol{\Lambda},
		\end{split}
	\end{equation}
	where, $\boldsymbol X$ is the matrix representation of $\hat{ \boldsymbol x}$
	calculated with the provisional WFs, $\{ \vert w^s_{\boldsymbol 0} \rangle \}$,
	and $\boldsymbol \Lambda$ is the diagonal eigenvalue matrix. $\boldsymbol F$,
	the eigenvector matrix of the eigenvalue equation, is easily obtained with a
	numerical library, such as the GNU Scientific Library~(GSL)~\cite{gsl_mnl}. And
	hence, the arbitrariness of the initial condition of $\{f^{s,n}_{\boldsymbol
		k}\}$ is fixed and the Wannier centers are obtained from $\boldsymbol \Lambda$.
	
	In the current study, $\{f^{s,n}_{\boldsymbol k}\}$ is provisionally fixed at
	$\boldsymbol k$ where all energy bands in the composite system are separated
	with reasonable sizes of energy gaps, e.g., $\boldsymbol k_0 = (\pi, \pi) $ in
	the case of lower energy bands of graphene,
	\begin{equation}
		f^{s,n}_{\boldsymbol k_0}=\delta_{s,n}. 
	\end{equation}
	Accordingly, the initial condition is revised as follows:
	\begin{equation}\label{eq:initial-condition}
		f^{s,n}_{\boldsymbol k_0}= \sum_p \delta_{p,n} F_{s,p}. 
	\end{equation}
	\subsection{Fixing Wannier Centers}\label{sec:fixing-wanneir-centers}
	By solving Eq.~(\ref{eq:xeigen-F}), the Wannier centers are, theoretically,
	obtained as the elements of the diagonal eigenvalue matrix $\boldsymbol
	\Lambda$. In a numerical context, the values of $\boldsymbol X =
	\langle\boldsymbol{w}_{0}\vert\hat{x}\vert\boldsymbol{w}_{0}\rangle$ are affected by the
	concentration of $\boldsymbol w (\boldsymbol x)=\langle \boldsymbol x
	\vert\boldsymbol{w}_{0}\rangle$, which recursively depends on the Wannier centers
	given in $\boldsymbol \Lambda$. Thus, an iterative method to obtain the Wannier
	centers with theoretically guaranteed stability is introduced. In practice, the coupled determination of $\boldsymbol X$ and $\boldsymbol \Lambda$ implied by Eq.~(\ref{eq:xeigen-F}) is handled iteratively; this is a numerical solver for the explicit equations, not a search or minimization over a spread functional.
	
	In this subsection, the WFs are treated series by series
	through Eq.~(\ref{eq:xeigen-F}), and hence the superscript indicating its
	series is dropped.
	\subsubsection{ Wannier Group and Transformation}\label{sec:grouping}
	From the periodic parts $\{ \vert u_{\boldsymbol k} \rangle \}$,
	obtained from Eqs.~(\ref{eq:composite-eigenvalue-equation}), (\ref{eq:xeigen-F})
	and (\ref{eq:initial-condition}), we have a WF with a candidate Wannier center
	$\boldsymbol r$,
	\begin{equation}	\label{eq:translation_wannier}
		\vert W_{\boldsymbol M}(\boldsymbol r)\rangle
		= L^{-d/2}\!\sum_{\boldsymbol k}
		e^{-i\boldsymbol r \cdot \boldsymbol k }
		e^{ i\boldsymbol k\cdot(\hat {\boldsymbol x} - \boldsymbol M)}
		\vert u_{\boldsymbol k} \rangle,\;	\boldsymbol r\in[0,1)^d.
	\end{equation}
	We further define sets labeled by $\boldsymbol r$,
	\begin{equation}
		\mathcal G_{\boldsymbol r}^s = \bigl\{\,
		\ket{W_{\boldsymbol M}(\boldsymbol r)}:\ \boldsymbol M\in\mathbb M
		\,\bigr\}.
	\end{equation}
	Since the space spanned by the elements of each group has to be identical,
	\begin{equation}\label{eq:wannier-group}
		\mathrm{Span}\{\mathcal G_{\boldsymbol p}\}=\mathrm{Span}\{\mathcal G_{\boldsymbol r}\}\quad\forall \boldsymbol r, \boldsymbol p\in
		[0,1)^d.
	\end{equation} 
	A transformation between them must exist, and it is given by:
	\begin{equation}\label{eq:wannier-conversion}
		\begin{split}
			\vert W_{\boldsymbol M}(\boldsymbol r)\rangle 
			=\sum_{\boldsymbol N} 
			\prod_{\mu=1}^d&\operatorname{sinc}\! \bigl( M_\mu -N_\mu + r_\mu-p_\mu \bigr)
			\vert W_{\boldsymbol N }(\boldsymbol p)\rangle, 
		\end{split}
	\end{equation}
	where,
	\begin{equation}\label{eq:def-sinc}
		\operatorname{sinc}x=\frac{\sin\pi x}{\pi x},
	\end{equation}
	and the following approximation is utilized ( see Appendix~\ref{ap:sinc} for details and
	simplified calculations),
	\begin{equation}
		\begin{aligned}
			\frac{1}{L}	\lim_{\delta k\rightarrow0}\sum_{k}e^{-ik(M-N+r-p)} 
			& =\mathrm{sinc}(M-N+r-p).
		\end{aligned}
	\end{equation}
	Matrix elements calculated by basis belonging to different groups are 
	also transformed in the following way,
	\begin{equation}	\label{eq:transformation_rule}
		\begin{aligned}
			X_{\boldsymbol M_1\boldsymbol M_2} (r)
			&= \langle W_{\boldsymbol M_1}(\boldsymbol r)\vert \hat x
			\vert W_{\boldsymbol M_2}(\boldsymbol r)\rangle\\
			&=\!\!\!\!\!\! \sum_{\boldsymbol N_1,\boldsymbol N_2}
			\prod_{\mu=1}^d\operatorname{sinc}\! \bigl( M_{1,\mu}\! -N_{1,\mu} + r_\mu \!-p_\mu \bigr) \cdot \\
			&\quad\quad\; \prod_{\mu=1}^d\operatorname{sinc}\! \bigl( M_{2,\mu} \!-N_{2,\mu} + r_\mu \!-p_\mu \bigr)
			X_{\boldsymbol N_1\boldsymbol N_2}(\boldsymbol p).
		\end{aligned}
	\end{equation}
	\subsubsection{Finding Wannier center of MLWFs}\label{sec:finding_eigenvalue_s} 
	For the sake of conciseness and clarity, the derivation of the method to obtain
	Wannier centers is done in a 1D model. The multidimensional formulation is given
	in Appendix~\ref{ap:multi-sinc}.
	
	A WF in 1D with a fractional shift $r$ is given as follows:
	\begin{equation}\label{eq:wannier-1D}
		\vert W_M(r)\rangle =\frac{1}{\sqrt L}\sum_{k}e^{-ik r}e^{ik (\hat x-M)}\vert u_k \rangle. \\
	\end{equation}
	Let us assume that the right value $r=x_0$ is found, then the following holds:
	\begin{equation}\label{eq:flat}
		\begin{split}
			\langle W_{M}(x_0)\vert\hat{x}\vert W_{N}(x_0)\rangle=\delta_{M,N}(M+x_0).
		\end{split}
	\end{equation}
	Equation~(\ref{eq:transformation_rule}) relates the above to a known, calculable value,
	\begin{equation}\label{eq:x00r}
		\begin{split}
			X_{0,0}(r)=\langle W_{0}(r)\vert\hat{x}\vert W_{0}(r)\rangle,
		\end{split}
	\end{equation}
	and using Eq.~(\ref{eqap:sinc^2_sum}), we have:
	\begin{equation}\label{eq:sinc-00-1d}
		\begin{split}
			X_{0,0}(r)\!\!\!
			&=\!\!\!\sum_{M,N}\!\!\!\mathrm{sinc}(M+x_0-r)\mathrm{sinc}(N + x_0-r)
			\delta_{M,N}(M+x_0)\\
			&=\frac{\sin2\pi( x_0-r)}{2\pi}+r.\\
		\end{split}
	\end{equation}
	$X_{0,0}^r$ is implicitly related to $x_0$
	via the right-hand side of Eq.~(\ref{eq:x00r}).
	And hence, the following two iterative dynamics are identical,
	\begin{equation}\label{eq:sinc-loop-w}
		r^{n+1}
		=\langle W_{0}({\boldsymbol{r}^n})\vert\hat{x}\vert W_{0}({\boldsymbol{r}^n})\rangle,
	\end{equation} 
	\begin{equation}\label{eq:sinc-loop-s}
		r^{n+1} = \frac{\sin\!\bigl(2\pi(x_0-r^n)\bigr)}{2\pi} + r^n,
	\end{equation}
	where $n$ appearing on the shoulder of $r$ is the iteration number.
	The former (Eq.~(\ref{eq:sinc-loop-w})) is used for the actual numerical calculation, while the latter (Eq.~(\ref{eq:sinc-loop-s})) reveals the underlying mathematical structure and is used to prove convergence.
	Through the iteration described in Eqs.~(\ref{eq:sinc-loop-w}) and (\ref{eq:sinc-loop-s}),
	the following holds, as detailed in Appendix~\ref{ap:r-converges-to-x_0}.
	\begin{equation}\label{eq:x_0-convergence}
		\lim_{n\rightarrow\infty} r^n = x_0.
	\end{equation}
	\begin{figure}[htbp]
		\centering
		\includegraphics[width=\linewidth]{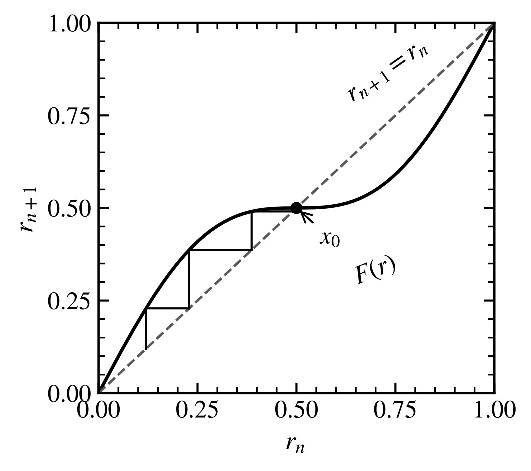}
		\caption{Cobweb diagram for the fixed-point iteration
			$r^{n+1}=r^n+\sin[2\pi(x_0-r^n)]/(2\pi)$\ $(\bmod\,1)$.
			The dashed line denotes $r^{n+1}=r^n$.
			For any start $r_0\not\equiv x_0\pm\tfrac12\ (\bmod\,1)$, the iteration converges to $x_0$
			with a cubic (third-order) local rate.}
		\label{fig:devaneycobwebsinepeeling}
	\end{figure}
	This fixed-point dynamics is illustrated by the cobweb diagram in
	Fig.~\ref{fig:devaneycobwebsinepeeling}.
	
	Note that Eq.~(\ref{eq:sinc-00}) itself is a solvable equation with given $	X_{0,0}(r)$ with no
	iteration. The numerically obtained $X_{0,0}(r)$, in reality, contains errors coming from 
	the specific settings. Applying the iteration, therefore, appears to be a better choice.
	Even in multi-dimensional cases, by letting, 
	\begin{equation}
		\begin{aligned}
			X_{\boldsymbol{M}_{1},\boldsymbol{M}_{2}}(\boldsymbol{x}_{0})\!=\delta_{M_{1,x},M_{2,x}}\delta_{M_{1,y},M_{2,y}}\delta_{M_{1,z},M_{2,z}}(M_{1,x}+x_{0})\end{aligned},
	\end{equation} 
	the following similarly holds, as described in Appendix~{\ref{ap:multi-sinc}},
	\begin{equation}\label{eq:sinc-00}
		\begin{aligned}
			X_{\boldsymbol 0\boldsymbol 0}(\boldsymbol r ) = \frac{\sin 2\pi(x_{0}-r_{x})}{2\pi}+r_{x},
		\end{aligned}
	\end{equation}
	and hence the above argument also applies.
	
	Alternatively, rather classical derivation and the numerical verification
	are found in Appendix~\ref{ap:direct-derivation}.

\begin{figure}[htbp]
		\centering
		\includegraphics[width=0.7\linewidth]{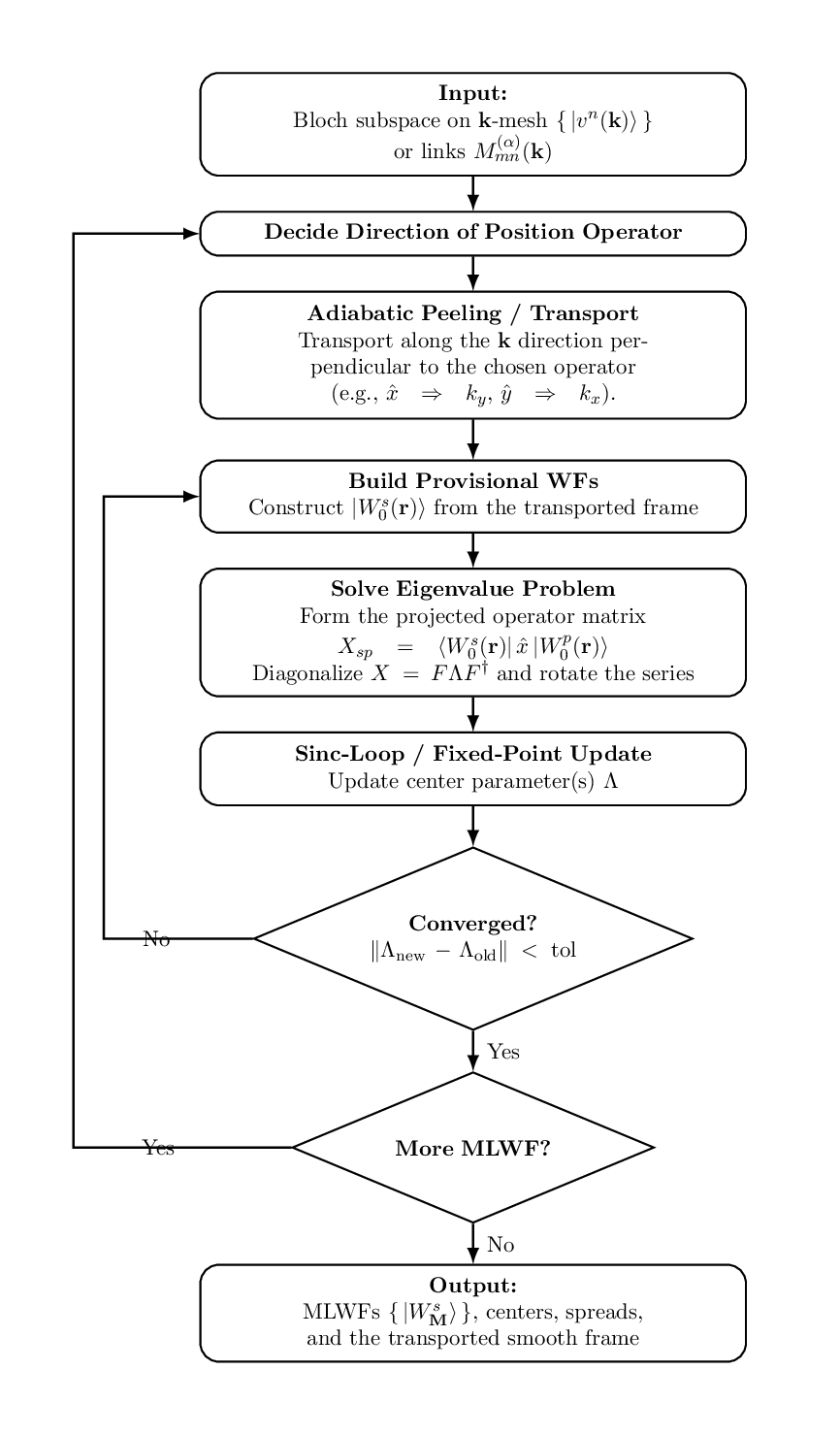}
		\caption{Flowchart of the constructive numerical workflow used in the later results sections. The algorithm combines adiabatic peeling with fixed-point iterations and self-consistent updates of the Wannier centers and projected-position matrices, followed by sequential extraction of localized orbitals.}
		\label{fig:numerical_procedure}
	\end{figure}

	For later numerical sections, the overall constructive numerical workflow is summarized in Fig.~\ref{fig:numerical_procedure}. The iterative part of this workflow should be understood as a solver comprising fixed-point iterations and self-consistent updates for center parameters and projected-position matrices, not as a variational spread minimization.

	\section{Single-Band System}\label{sec:single-band}
	Before moving to benchmarks, we specialize the general framework to an isolated
	single band, where the gauge group reduces to $U(1)$. In this limit, it is easier
	to see how the phase flattening done by the transport actually contributes to
	minimizing the spread, since the spread functional is reduced to an analytic
	Euler–Lagrange equation, a k-space Poisson problem.
	
	In a single band system, $f_{k}^{s,n}$ is reduced to: 
	\begin{equation}
		f_{\boldsymbol{k}}^{s,n}\rightarrow e^{i\phi_{\boldsymbol{k}}},\phi_{\boldsymbol{k}}\in\mathbb{R},
	\end{equation}
	and hence, the BFs and WFs are expressed as follows: 
	\begin{equation}\label{eq:single_band_Wannier}
		\begin{cases}
			&\vert\psi_{\boldsymbol{k}}\rangle=L^{-d/2}e^{i\boldsymbol{k}\cdot\hat{\boldsymbol{x}}}\vert v_{\boldsymbol{k}}\rangle\\
			& \vert W_{\boldsymbol{M}}\rangle=\frac{1}{\sqrt{L}}\sum_{\boldsymbol k}e^{-i\boldsymbol{k}\cdot\boldsymbol{M}}e^{i\boldsymbol{k}\cdot\hat{\boldsymbol{x}}}\vert u_{\boldsymbol{k}}\rangle\\
			& \vert u_{\boldsymbol k}\rangle=e^{i\phi_{\boldsymbol{k}}}\vert v_{\boldsymbol{k}}\rangle.
		\end{cases}
	\end{equation}
	\subsection{Berry Connection}
	\label{sec:berry} 
	By introducing $\theta_{\boldsymbol{k}}(\boldsymbol{x})$,
	the phase of $v_{\boldsymbol k}(\boldsymbol x)$, as follows:
	\begin{equation}
		v_{\boldsymbol{k}}(\boldsymbol{x})=\vert v_{\boldsymbol{k}}(\boldsymbol{x})\vert e^{i\theta_{\boldsymbol k}(\boldsymbol{x})}\ (\theta_{\boldsymbol k}(\boldsymbol{x})\in\mathbb{R}),
	\end{equation}
	the Berry connection $\langle v_{\boldsymbol k}\vert v_{\boldsymbol k-\delta \boldsymbol k}\rangle$
	for $-\pi<k<\pi$ is calculated as follows: 
	\begin{equation}\label{eq:v_k_connection}
		\begin{aligned}
			\langle v_{\boldsymbol{k}}\vert v_{\boldsymbol{k}-\delta\boldsymbol{k}}\rangle 
			&=\langle v_{\boldsymbol{k}}\vert v_{\boldsymbol{k}}\rangle-\delta\boldsymbol{k}\langle v_{\boldsymbol k}\vert\boldsymbol{\nabla_{\boldsymbol{k}}}\vert v_{\boldsymbol k}\rangle+\mathcal O(\delta\boldsymbol{k}^2)\\
			&=1- i\delta\boldsymbol{k} \cdot	\boldsymbol A_{\boldsymbol k}+\mathcal O(\delta\boldsymbol{k}^2),\\
		\end{aligned}
	\end{equation}
	where,
	\begin{equation}\label{eq:i<u|d|u>=A}
		\begin{aligned}
			\boldsymbol A_{\boldsymbol k} 
			&=-i\langle v_{\boldsymbol k}\vert\boldsymbol{\nabla_{\boldsymbol{k}}}\vert v_{\boldsymbol k}\rangle.\\
		\end{aligned}
	\end{equation}
	By utilizing,
	\begin{equation}
		\boldsymbol{\nabla_{\boldsymbol{k}}}\int d^d\boldsymbol{x}\vert v_{\boldsymbol{k}}(\boldsymbol x)\vert^2=0,
	\end{equation}
	$\boldsymbol A_{\boldsymbol k}$ is also expressed as follows:
	\begin{equation}\label{eq:definition-A}
		\begin{aligned}
			\boldsymbol A_{\boldsymbol k} 
			&=\int d^d\boldsymbol{x}\vert v_{\boldsymbol{k}}(\boldsymbol{x})\vert^2\boldsymbol{\nabla_{\boldsymbol{k}}}\theta_{\boldsymbol k}(\boldsymbol{x}).\\
		\end{aligned}
	\end{equation}
	Note that $\boldsymbol A_{\boldsymbol k}$ is decomposed into the transverse and longitudinal part:
	\begin{equation}
		\boldsymbol A_{\boldsymbol k}=\boldsymbol A^{\mathrm T}_{\boldsymbol k}
		+\nabla_{\boldsymbol{k}}\varphi_{\boldsymbol k},\qquad
		\nabla_{\boldsymbol{k}}\cdot \boldsymbol A^{\mathrm T}_{\boldsymbol k}=0.
	\end{equation}
	\subsection{Minimum Spread Condition }
	To find out a constraint on $\phi^n_{\boldsymbol k}$ in $\mathbb K_I$, the functional equation
	for the minimum spread is reviewed.
	The spread of a WF with its center at $\boldsymbol{x}_{0}$ is calculated as follows~\cite{MarzariVanderbilt1997}:
	\begin{equation}
		\begin{aligned}
			\sigma(\boldsymbol{x}^2) 
			& =\langle W_{\boldsymbol 0}\vert(\hat{\boldsymbol{x}}-\boldsymbol{x}_{0})^2\vert W_{\boldsymbol 0}\rangle\\
			& =\langle W_{\boldsymbol 0}\vert\hat{\boldsymbol{x}}^2\vert W_{\boldsymbol 0}\rangle-\boldsymbol{x}_{0}^2,
		\end{aligned}
	\end{equation}
	where $\boldsymbol{x}_0$ is the Wannier center. 
	The minimum spread is therefore achieved, when the following is minimized:
	\begin{equation}
		\begin{aligned}\langle\hat{\boldsymbol{x}}^2\rangle & =\langle W_{0}\vert\hat{\boldsymbol{x}}^2\vert W_{0}\rangle\\
			& =\frac{1}{L^d}\sum_{\boldsymbol k}\langle\nabla_{\boldsymbol{k}}u_{\boldsymbol{k}}\vert\nabla_{\boldsymbol{k}}u_{\boldsymbol{k}}\rangle\\
			& =\frac{1}{L^d}\sum_{\boldsymbol k}\int d^d\boldsymbol{x}\left(\nabla_{\boldsymbol{k}}\vert v_{\boldsymbol{k}}(\boldsymbol{x})\vert\right)^2+
			\vert v_{\boldsymbol{k}}(\boldsymbol{x})\vert^2
			\left\{\nabla_{\boldsymbol{k}}
			\left(
			\theta_{\boldsymbol{k}} + \phi_{\boldsymbol k } \right)\right\}^2.
		\end{aligned}
		\label{eq:x2}
	\end{equation}
	By utilizing Eq.~(\ref{eq:definition-A}) and approximating the summation with integration, the functional derivative of Eq.~(\ref{eq:x2}) is given as follows
	~\cite{EngelDreizler2011_AppA, GelfandFomin1963, Olver_CalcVar, EoM_FunctionalDerivative},
	\begin{equation}
		\begin{aligned}
			\left\langle \frac{\delta\hat{\boldsymbol{x}}^2\left[\phi\right]}{\delta\phi},\lambda \right\rangle 
			&=\frac{1}{\pi^d}\frac{d}{d\epsilon} \int d^d\boldsymbol{k} \int d^d\boldsymbol{x} \,
			\left| v_{\boldsymbol{k}}(\boldsymbol{x}) \right|^2\cdot\\
			&\quad\quad\quad\left|
			\nabla_{\boldsymbol{k}}\left(\theta_{\boldsymbol{k}}(\boldsymbol{x})
			+ \phi_{\boldsymbol{k}} +\epsilon\lambda_{\boldsymbol{k}}\right)
			\right|^2 \Bigg|_{\epsilon=0}\\
			&= \frac{2}{\pi^d} \int d^d\boldsymbol{k} \left( \boldsymbol{A}_{\boldsymbol{k}} + \nabla_{\boldsymbol{k}}\phi_{\boldsymbol{k}} \right) \cdot \nabla_{\boldsymbol{k}}\lambda_{\boldsymbol{k}} \\
			&=-\frac{2}{\pi^d}\int d^d\boldsymbol{k} \, \lambda_{\boldsymbol{k}}
			\left(
			\nabla_{\boldsymbol{k}} \cdot \boldsymbol{A}_{\boldsymbol{k}}
			+\Delta_{\boldsymbol{k}}\phi_{\boldsymbol{k}}
			\right),
		\end{aligned}
	\end{equation}
	where $\Delta_{\boldsymbol k}$ is the Laplacian operator in the k-space, and 
	$\lambda_{\boldsymbol k}$ is an arbitrary trial function satisfying the
	following(see Eq.~(\ref{eq:K-def})):
	\begin{equation}
		\lambda_{\boldsymbol k} \big|_{\partial\mathbb K} = 0.
	\end{equation}
	Thus, the minimum spread condition is expressed in the following partial
	differential equation (see Eq.~(\ref{eq:K_I-def})),
	\begin{equation}\label{eq:spread_requirement}
		\Delta_{\boldsymbol k } \phi_{\boldsymbol k} +	\nabla_{\boldsymbol k} \cdot \boldsymbol A_{\boldsymbol k}
		=0, \;\boldsymbol k \in \mathbb K_I,
	\end{equation}
	Denoting a particular solution by $\phi^0_{\boldsymbol k}$, we have a family of solutions,
	\begin{equation}
		\phi_{\boldsymbol k} =\phi^0_{\boldsymbol k} + \boldsymbol k \cdot \boldsymbol x_0,
	\end{equation}
	where $\boldsymbol k \cdot \boldsymbol x_0$ represents a homogeneous solution.
	Hence, the Berry connection made of $\{\vert u_{\boldsymbol k}\rangle \}$ is,
	\begin{equation}
		\tilde{\boldsymbol A}_{\boldsymbol k} 
		= \boldsymbol A_{\boldsymbol k} + \nabla_{\boldsymbol k} \phi_{\boldsymbol k}.
	\end{equation}
	Thus,
	\begin{equation}\label{eq:nabla-a=0}
		\nabla_{\boldsymbol k} \cdot \tilde{\boldsymbol A}_{\boldsymbol k}=0. 
	\end{equation}
	If the band admits an ELWF and an MLWF, and if, in addition,
	\begin{equation}
		\boldsymbol A^T_{\boldsymbol k}=0,
	\end{equation}
	then Eqs.~(\ref{eq:spread_requirement}) and (\ref{eq:nabla-a=0}) have a solution of the form,
	\begin{equation}\label{eq:solution-ak}
		\tilde{ \boldsymbol A}_{\boldsymbol k} = -\boldsymbol{x}_{0}.
	\end{equation}
	From the above, 
	\begin{equation}\label{eq:connection-flat}
		\begin{split}
			\langle u_{\boldsymbol k}\vert u_{\boldsymbol k+\delta \boldsymbol k}\rangle
			&=\langle u_{\boldsymbol k}\vert \{ \vert u_{\boldsymbol k}\rangle
			+ \delta \boldsymbol k \cdot \nabla_{\boldsymbol k} \vert u_{\boldsymbol k}\rangle\} + \mathcal O(\delta \boldsymbol k^2)\\
			&=\langle u_{\boldsymbol k}\vert u_{\boldsymbol k}\rangle
			+ i\delta \boldsymbol k \cdot \tilde{\boldsymbol A}_{\boldsymbol k}+ \mathcal O(\delta \boldsymbol k^2)\\
			&=1-i\delta \boldsymbol k \cdot \boldsymbol x_0 + \mathcal O(\delta \boldsymbol k^2).
		\end{split}
	\end{equation}
	The last line is the infinitesimal form of a phase-plane relation.
	Indeed, to first order in the mesh spacing,
	\begin{equation}
		1-i\delta\boldsymbol k\cdot\boldsymbol x_0
		=
		e^{-i\delta\boldsymbol k\cdot\boldsymbol x_0}
		+\mathcal O(\delta\boldsymbol k^2).
	\end{equation}
	Equivalently, if $\boldsymbol k(t)$ is a path in $\mathbb K_I$ and
	\[
	\theta(t)=
	\arg\langle u_{\boldsymbol k(0)}\vert u_{\boldsymbol k(t)}\rangle ,
	\]
	then Eq.~(\ref{eq:connection-flat}) gives the first-order phase equation
	\begin{equation}
		\frac{d\theta}{dt}
		=
		-\dot{\boldsymbol k}(t)\cdot\boldsymbol x_0 .
	\end{equation}
	Integrating this relation along a link from $\boldsymbol k_1$ to
	$\boldsymbol k_2$ gives the finite-link phase factor.  For a finite
	link, the numerical overlap may also have a positive modulus.  Since the
	present discussion concerns the phase-plane relation, this positive
	modulus is understood to be separated off, and only the unit-modulus
	phase factor is written explicitly below.  In vector form and in the
	two-dimensional component form used below, this phase factor is
	\begin{equation}\label{eq:arg}
		\begin{split}
			e^{i\arg\langle u_{\boldsymbol k_1}\vert u_{\boldsymbol k_2}\rangle}
			&=
			e^{-i(\boldsymbol k_2-\boldsymbol k_1)\cdot\boldsymbol x_0}\\
			&=
			e^{i(k_{1_x}-k_{2_x})x_0}
			e^{i(k_{1_y}-k_{2_y})y_0},
		\end{split}
	\end{equation}
	where $\boldsymbol k_1,\boldsymbol k_2\in\mathbb K_I$. 
	The first line follows from the constant-connection condition
	Eq.~(\ref{eq:solution-ak}), while the second line is the
	two-dimensional component form of the same phase factor.
	
	Equivalently, Eq.~(\ref{eq:arg}) gives the following phase-plane
	equation:
	\begin{equation}\label{eq:arg-plane}
		\begin{split}
			\arg\langle u_{\boldsymbol k_1}\vert u_{\boldsymbol k_2}\rangle
			&=
			(\boldsymbol k_1-\boldsymbol k_2)\cdot\boldsymbol x_0\\
			&=
			(k_{1_x}-k_{2_x})x_0+(k_{1_y}-k_{2_y})y_0.
		\end{split}
	\end{equation}
	Thus Eq.~(\ref{eq:arg-plane}) is the equation of a tilted plane for
	the overlap phase. The slopes of this phase plane are the components of
	the Wannier center. This relation is the link between the
	transport-aligned gauge and the later center-fixing iteration.
	The first line of Eq.~(\ref{eq:arg}) with
	$\boldsymbol x_0=\boldsymbol 0$ is also obtained from
	Eq.~(\ref{eq:composite-peeling-projection}) and
	Eq.~(\ref{eq:k+dk-orthogonal}). Hence, $\{\vert u^s_{\boldsymbol k} \rangle \}$
	generated by the adiabatic transport satisfy the minimum spread condition,
	in this curvature-free case.
	\subsection{Wannier Center}\label{sec:group_transform_s}
	The above argument fixes the gauge only up to a global constant in $\mathbb K_I$
	and the specific value of $\boldsymbol x_0$ is not determined.
	We determine the $\boldsymbol x_0$ using Eqs.~(\ref{eq:sinc-loop-w}) and (\ref{eq:sinc-loop-s})
	by reconciling the slope in the interior and the jump across the boundary of the k-space.
	\section{ Results and Discussion }\label{sec:results}
	The purpose of the section is to validate the formulations in the previous 
	sections. Therefore, not only the comparison of spreads with published
	results but also the key equations are numerically verified with
	graphical representation of the results.
	
	The calculation of the matrix elements of the projected position operators is
	the most important part of the procedure and there are two ways to
	perform it. In Secs.~\ref{sec:1d} and \ref{sec:square}, the relevant projected-position matrices are first evaluated directly in real space from the coordinate representation of the WFs, and the spreads are then computed from the resulting MLWFs via Eq.~(\ref{eq:spread-def}). In Sec.~\ref{sec:graphene-matrix-element}, the same projected-position information is extracted in k-space from the overlap links.
	
	These two approaches are theoretically equivalent. The former is more
	straightforward, whereas the latter is more computationally efficient.
	Thus, this section discusses not only the numerical results and the actual
	implementation, but also the procedural differences.
	\subsection{1D Systems }\label{sec:1d} 
	This section focuses on verifying the formulations in 1D systems, for
	it allows us to see the essential parts of the formulation 
	in simple settings.
	
	As the solution procedure of the BFs, a scaling function method~\cite{hamawaka}
	with Symlet-4~\cite{Daub10-6,Daubechies+2006+564+652} is employed. The number of divisions
	in one unit cell, $N$, is indicated in figures and tables. As a reference for
	the spatial accuracy, the position eigenvectors obtained by spatially solving the
	eigenvalue equations of $N_C L $-by-$N_C L$ position operator matrices are also
	shown~\cite{hamawaka} and denoted "Matrix" in the figures.
	
	Since all energy bands appearing in this section do not have degeneracy,
	the numerical procedure begins from the eigenvalue problem described in
	Sec.~\ref{sec:xeigen}.
	\subsubsection{Matrix elements of $\hat x$} 
	\begin{figure*}[htbp]
		\centering
		\includegraphics[width=0.8\linewidth]	{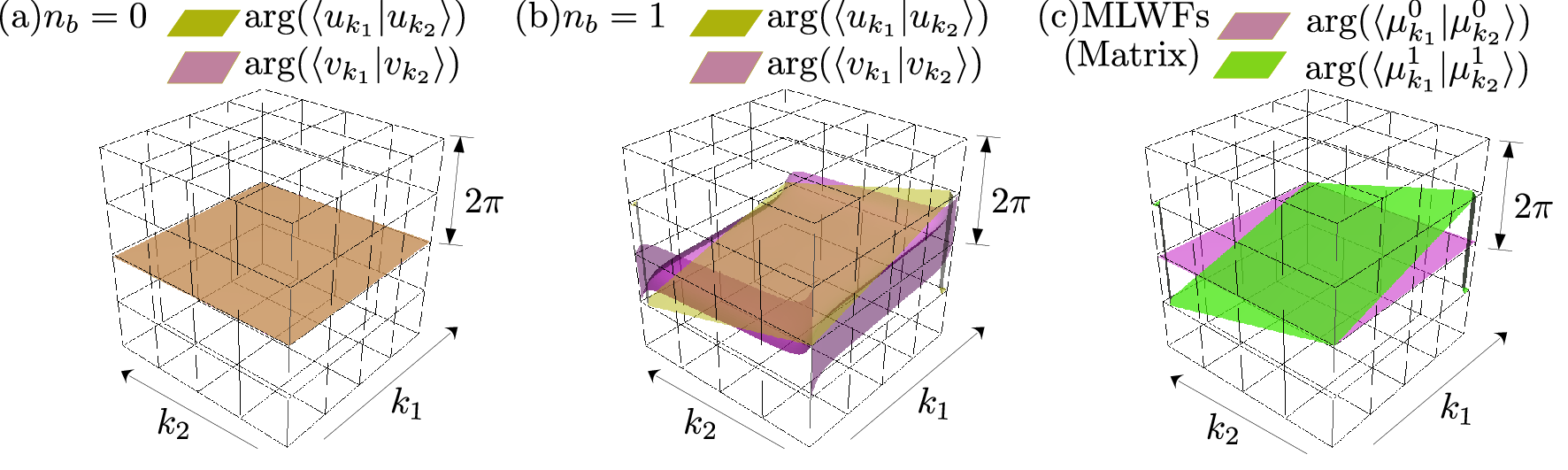}
		\caption{(Color online)
			Phases of the inner products of the periodic parts of BFs. $n_b$ 
			denotes the band index. The axes are scaled by
			$\pi$. In panels~(a) and (b), 
			$\textrm{arg}(\langle v_{k_{1}}\vert v_{k_{2}}\rangle)$ is evaluated from the raw unprocessed periodic parts and is shown as the corrugated surface (purple in the online version),
			whereas $\textrm{arg}(\langle u_{k_{1}}\vert u_{k_{2}}\rangle)$ is evaluated after the transport/phase-alignment procedure and is shown as the flatter surface (yellow in the online version).
			Accordingly, the raw overlap phase need not be planar, while the processed overlap phase becomes approximately planar; this distinction is especially visible in panel~(b), where the raw surface remains corrugated whereas the processed surface is flattened. Regions where the two semi-transparent surfaces overlap appear darker (brown in the online version).
			In panel~(c), $\vert \mu_{k}\rangle$ denotes	
			the periodic parts of the position eigenvectors obtained by 
			solving the spatial position operator matrix.
			The band indices 
			are denoted as the superscripts of $\mu$, showing that the overlap phases are also approximately planar, with slopes corresponding to the Wannier centers.
		}\label{fig:fractional_phase}
	\end{figure*} 
	\begin{figure}[htbp] \centering
		\includegraphics[width=1.0\linewidth]{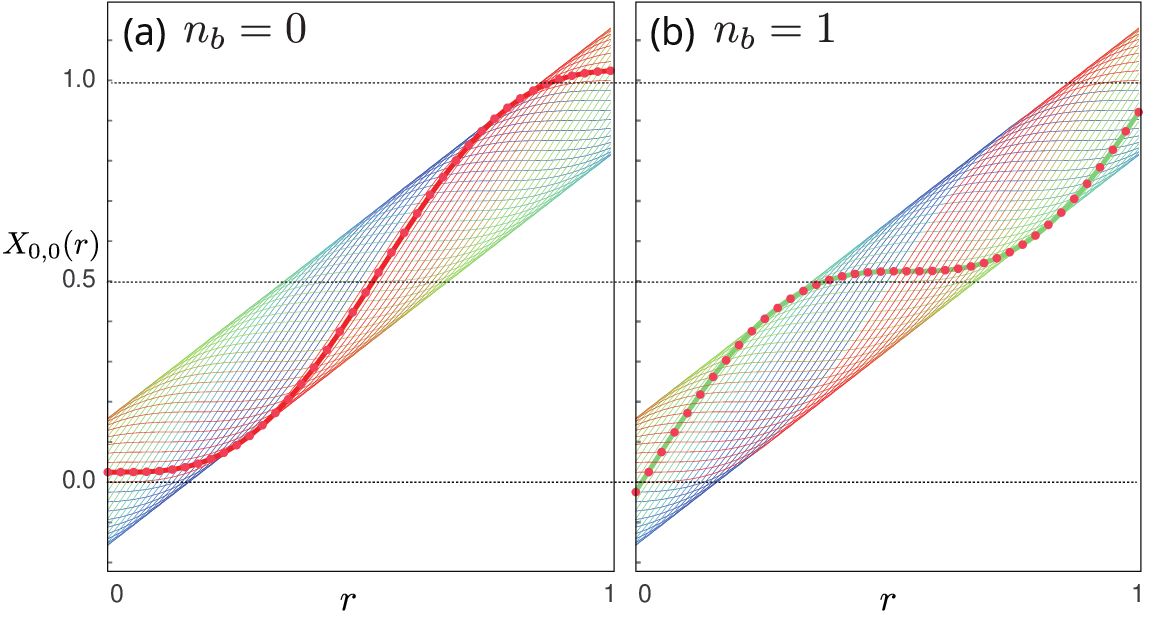}
		\caption{(Color online)
			A graphical representation of Eq.~(\ref{eq:sinc-00-1d}). Thin curves,
			$X_{0,0}(r)=\sin 2\pi( x_0-r)/2\pi+r$, 
			are drawn with varying $x_0$, and the thick ones are drawn with
			$x_0$ obtained through the iteration of Eqs.~(\ref{eq:sinc-loop-w}) and (\ref{eq:sinc-loop-s}).
			Each dot is plotted by calculating 
			$X_{0,0}(r)=\langle W_{0}(r)\vert\hat{x}\vert W_{0}(r)\rangle$ with
			the actual WFs with varying $s$. 
		}\label{fig:fraciona_sinc}
	\end{figure}
	\begin{figure}[htbp]
		\centering
		\includegraphics[width=0.9\linewidth]{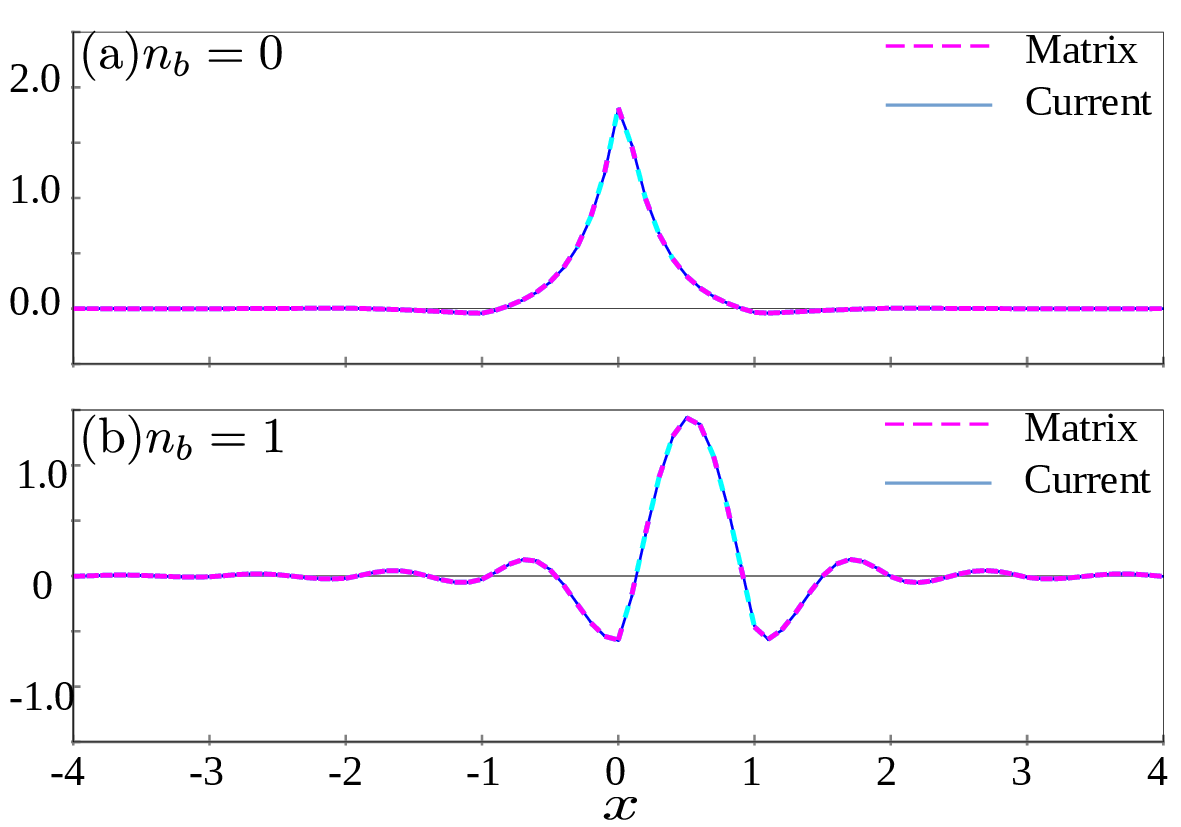}
		\caption{(Color online)
			Profiles of the MLWFs. Those obtained by the current method are 
			indicated "Current". The results indicated "Matrix" are obtained by 
			solving the spatial position operator matrix.
			$n_b$ at the left upper corner of each panel denotes
			the band index. In all panels, $N=40$, $L=200$ and $V(x)=\sum V_0 \delta(x-n)$ with $V_0=-0.2(2\pi^2)$~\cite{VELLASCO2020}.
		}\label{fig:fraciona_wannier}
	\end{figure}
	\begin{table}[htbp] %
		\caption{Comparison of spreads, $\sigma_x^2 =\langle W\vert \{ \hat{x} - \langle W\vert \hat{x} \vert W\rangle \}^2\vert W \rangle $, obtained from different methods.}
		\label{tab:frac} 
		\centering	
		\begin{tabular}{cccccc}
			\hline 
			\multicolumn{2}{c}{Calculation condition} & & \multicolumn{3}{c}{$\sigma_{x}^2$}\tabularnewline
			\hline 
			$V_{0}$ & Band & & Ref. ~\cite{VELLASCO2020} & Matrix & Current \tabularnewline
			\hline 
			$-0.2\times2\pi^2$ & 0 & & 0.03 & 0.03 & 0.03 \tabularnewline
			$-0.2\times2\pi^2$ & 1 & & 0.12 & 0.12 & 0.12 \tabularnewline
			\hline 
		\end{tabular}
	\end{table}
	In 1D cases, the real-space evaluation of the projected-position matrix elements $X_{s_1s_2}$ is explicitly calculated as follows:
	\begin{equation}
		\begin{split}
			\langle W_0^{s_1}\vert \hat x \vert W_0^{s_2} \rangle
			&= \langle W_0^{s_1}\vert \lbrace \int dx\vert x\rangle \langle x \vert \rbrace \hat x \vert W_0^{s_2} \rangle\\ 
			&= \int \langle W_0^{s_1}\vert x\rangle dx x \langle x \vert W_0^{s_2} \rangle\\ 
			&= \int \overline{W}_0^{s_1}(x)x W_0^{s_2}(x)dx. 
		\end{split}\label{eq:1d-matrix}
	\end{equation} 
	In the overall algorithmic workflow, this real-space matrix is subsequently diagonalized via Eq.~(\ref{eq:xeigen-F}) to determine the Wannier centers. The final spreads are then evaluated from the resulting MLWFs via Eq.~(\ref{eq:spread-def}).

	\subsubsection{Single Band Systems} 
	The potential energy employed is of
	Kr\"onig-Penney type, $V(x)=\sum V_0 \delta(x-n)$, with $V_0=-0.2(2\pi^2)$~\cite{VELLASCO2020,hamawaka}.
	The spatial and k-space resolutions are $N=40$ and $L=200$, respectively.
	
	As seen in Fig.~\ref{fig:fractional_phase}, the overlap phase of $\langle u^s_{k_{1}}\vert u^s_{k_{2}}\rangle$ exhibits an approximately linear dependence on $k_1-k_2$, indicating that the phase surface is approximately planar, consistent with Eq.~(\ref{eq:arg-plane}). The visually apparent $180^{\circ}$ folds are simply branch cuts arising from the principal-value phase wrapping into $[-\pi,\pi)$ and do not imply any non-planarity of the underlying geometric phase surface.
	
	Fig.~\ref{fig:fraciona_sinc} graphically shows the validity of Eq.~(\ref{eq:sinc-00-1d}).
	Although $x_0$ is obtained through the loop formed by Eqs.~(\ref{eq:sinc-loop-w}) and
	(\ref{eq:sinc-loop-s}), the right Wannier center $x_0$ can also be obtained by fitting
	a few plots of $X^s_{0,0}$ with the sinc function.
	
	Figure~\ref{fig:fraciona_wannier} shows the profiles of the MLWFs, and 
	the curves obtained by the current method agree well with those
	obtained by the spatial matrix solver.
	The calculated minimum spreads are summarized in Table \ref{tab:frac}, and the values obtained with the current method agree well with previous studies.
	\subsubsection{Composite Band Systems}\label{sec:composite-band}
\begin{figure}[htbp] \centering
		\includegraphics[width=1.0\linewidth]{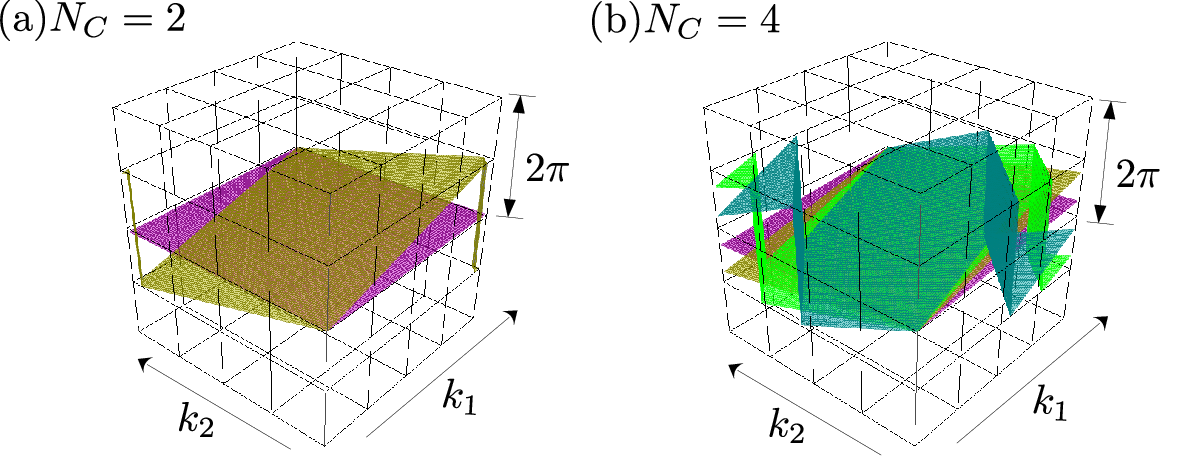}
		\caption{(Color online)
			Phases of the inner products of the periodic parts of BFs. The composite band
			consists of \#0 and \#1 in panel~(a), and \#0 to \#3 in panel~(b)~\cite{Wang2014}. 
			In panel~(a),
			$\textrm{arg}(\langle u_{k_{1}}^s\vert u_{k_{2}}^s\rangle)$ of series
			\#0 and \#1 of the two-band system with $V_0=-0.661$ are plotted on
			$k_{1}$-$k_{2}$-plane.
			In panel~(b), those of \#0 to \#3 with $V_0=+0.661$ are plotted. The axes
			are scaled by $\pi$ in the panels, and the visible folds occur when the principal value crosses $\pm\pi$ because the plotted range is restricted to $[-\pi,+\pi)$.
		}\label{fig:sel-phase}
	\end{figure}

\begin{figure}[htbp]
		\centering
		\includegraphics[width=1.0\linewidth]{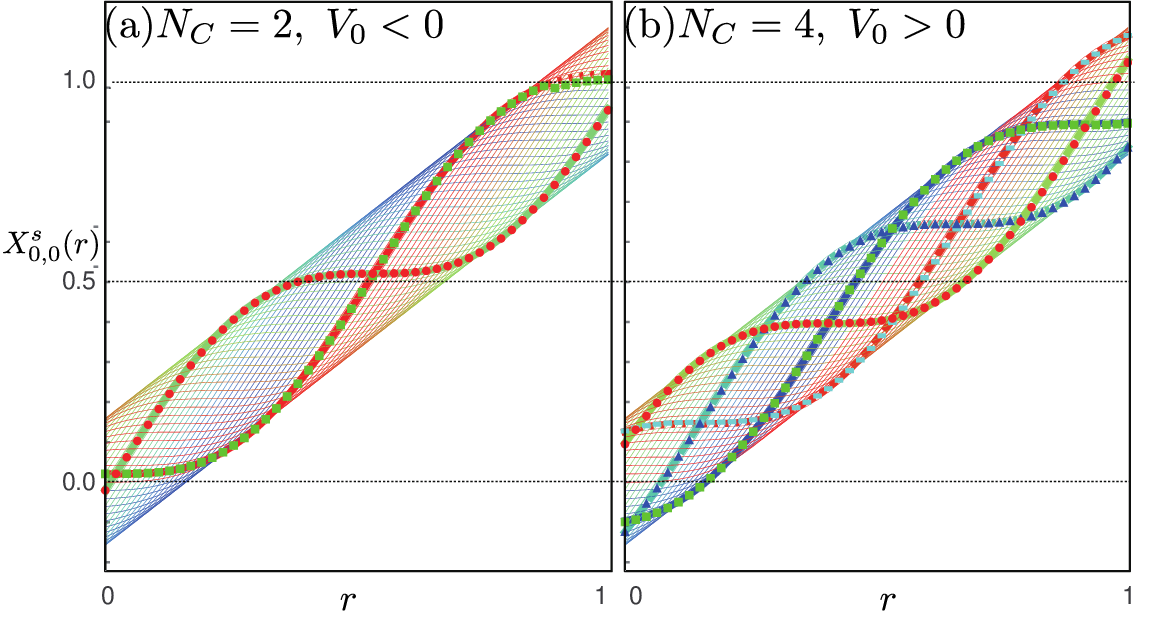}
		\caption{(Color online)
			A graphical representation of Eq.~(\ref{eq:sinc-00-1d}) in 
			composite band systems. 
			The composite bands
			consist of \#0 and \#1 with $V_0=0.661$ in panel~(a) and \#0 to \#3 with $V_0=+0.661$ in panel~(b)~\cite{Wang2014}.
			Thin curves, $X^s_{0,0}(r)=\sin 2\pi( x^s_0-r)/2\pi+r$, are drawn with
			varying $x_0$. The thick ones are drawn with $x_0$ obtained through the
			iteration of Eqs.~(\ref{eq:sinc-loop-w}) and (\ref{eq:sinc-loop-s}). Each
			dot is plotted by calculating $X^s_{0,0}(r)=\langle
			W^s_{0}(r)\vert\hat{x}\vert W^s_{0}(r)\rangle$ with the actual WFs with varying $s$.
		}\label{fig:sel-sinc}
	\end{figure}

\begin{figure}[htbp]
		\centering
		\includegraphics[width=0.9\linewidth]{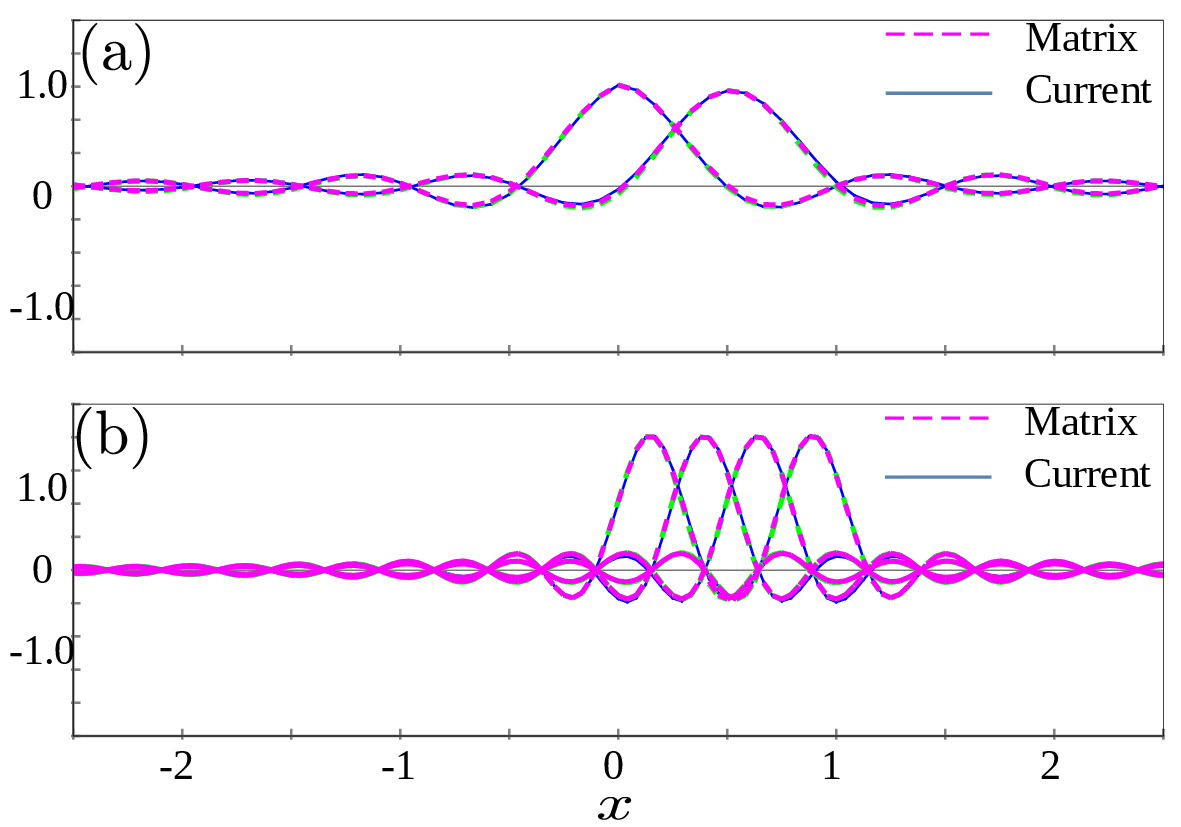}
		\caption{(Color online)
			The profiles of the MLWFs in composite band systems. Those obtained by the current method are 
			indicated "Current". The results indicated "Matrix" are obtained by 
			solving the spatial position operator matrix. The composite band
			consists of \#0 and \#1 in panel~(a), and \#0 to \#3 in panel~(b).
			In both panels, the spatial and k-space resolutions are $N=48$ and $L=200$,
			respectively, and the potential energy is $V(x)=\sum V_0 \delta( x -n)$.
			In panel~(a) $V_0=0.661$, and in panel~(b) $V_0=+0.661$~\cite{Wang2014}.
		}\label{fig:sel-wannier}
	\end{figure}

	The calculation results are compared with published results
	~\cite{Wang2014,hamawaka}. The potential energy employed is of
	Kr\"onig-Penney type, $V(x)=\sum V_0 \delta(x-n)$.
	The spatial and k-space resolutions are $N=48$ and $L=200$, respectively.
	The specific value of $V_0$ is indicated in the captions of Figs.~\ref{fig:sel-phase}
	to \ref{fig:sel-wannier}.
	
	As seen in Fig.~\ref{fig:sel-phase}, the overlap phase surfaces appear to be approximately planar, consistent with Eq.~(\ref{eq:arg-plane}). Similar to the single-band case, the visible sharp steps are merely branch cuts due to phase wrapping modulo $2\pi$.

As shown in Fig.~\ref{fig:sel-sinc}, as in the
single-band cases, the actually calculated $\{X^s_{0,0}\}$
represented by dots lie on the minimum-spread lines represented
by thick curves.
	
	Figure~\ref{fig:sel-wannier} shows the profiles of the MLWFs and 
	the curves obtained by the current method agree well with those
	obtained by the spatial matrix solver, series by series.
	The calculated minimum spreads are listed in the
	Table~\ref{tab:sel} and those obtained by the current method
	are as small as the published results.

	
	\begin{table}[htbp]
		\caption{Comparison of spreads, $\sigma_x^2 =\langle W\vert \{ \hat{x} - \langle W\vert \hat{x} \vert W\rangle \}^2\vert W \rangle $, obtained from different methods.}
		\label{tab:sel}
		\centering
		\begin{tabular}{c c c c c c}
			\hline
			\hline
			\multicolumn{2}{c}{Calculation condition} && \multicolumn{3}{c}{$\sigma_x^2$}\\
			\cline {1-2} \cline{4-6}
			$V_0$ & Bands && Ref. ~\cite{Wang2014} & Matrix & Current \\ \hline
			0.661 & 0 to 3 && 0.29 & 0.29 & 0.31 \\
			-0.661 & 0 to 1 && 0.27 & 0.27 & 0.29 \\
			\hline \hline
		\end{tabular}
	\end{table}
	\subsection{2D Square Potential }\label{sec:square}
	This section presents the calculation procedures, implementation and results of
	a numerical model. The system consists of a unit square with a
	step-function-like potential well,
	\begin{equation}\label{eq:s-potential}
		V(x,y) =
		\begin{cases}
			&1 \;( x, y \in [1/2, 3/4] )\\
			&0\; ( \textrm{otherwise} ).
		\end{cases}
	\end{equation}
	The periodic boundary condition is employed on the four sides and the calculations are carried out
	in the unit cell and k-space both having $32\times 32$ resolution.
	\subsubsection { Numerical procedure }\label{sec:s-procedure}
	The energy bands are first peeled from each other with the following equation derived from Eq.~(\ref{eq:composite-adiabatic-peeling}):
	\begin{equation}
		\label{eq:simple-peeling}
		\begin{aligned}	
			\vert u_{\boldsymbol{k}+\delta\boldsymbol{k}}^s\rangle&
			&=\left\{ \sum_{m}\vert v_{\boldsymbol{k}+\delta\boldsymbol{k}}^m\rangle\langle v_{\boldsymbol{k}+\delta\boldsymbol{k}}^m\vert\right\} \vert u_{\boldsymbol{k}}^s\rangle,
		\end{aligned}
	\end{equation}
	with the initial condition:
	\begin{equation}\tag{\ref{eq:provisional-initial-condition}}
		\vert u^s_{\boldsymbol k_0}\rangle = \vert v^s_{\boldsymbol k_0} \rangle,
	\end{equation}
	where, $\boldsymbol k_0=(0,0)$ and $s$ indicates the peeled \textit{energy band} (series) number.

\begin{figure}[htbp]
		\centering
		\includegraphics[width=0.6\linewidth]{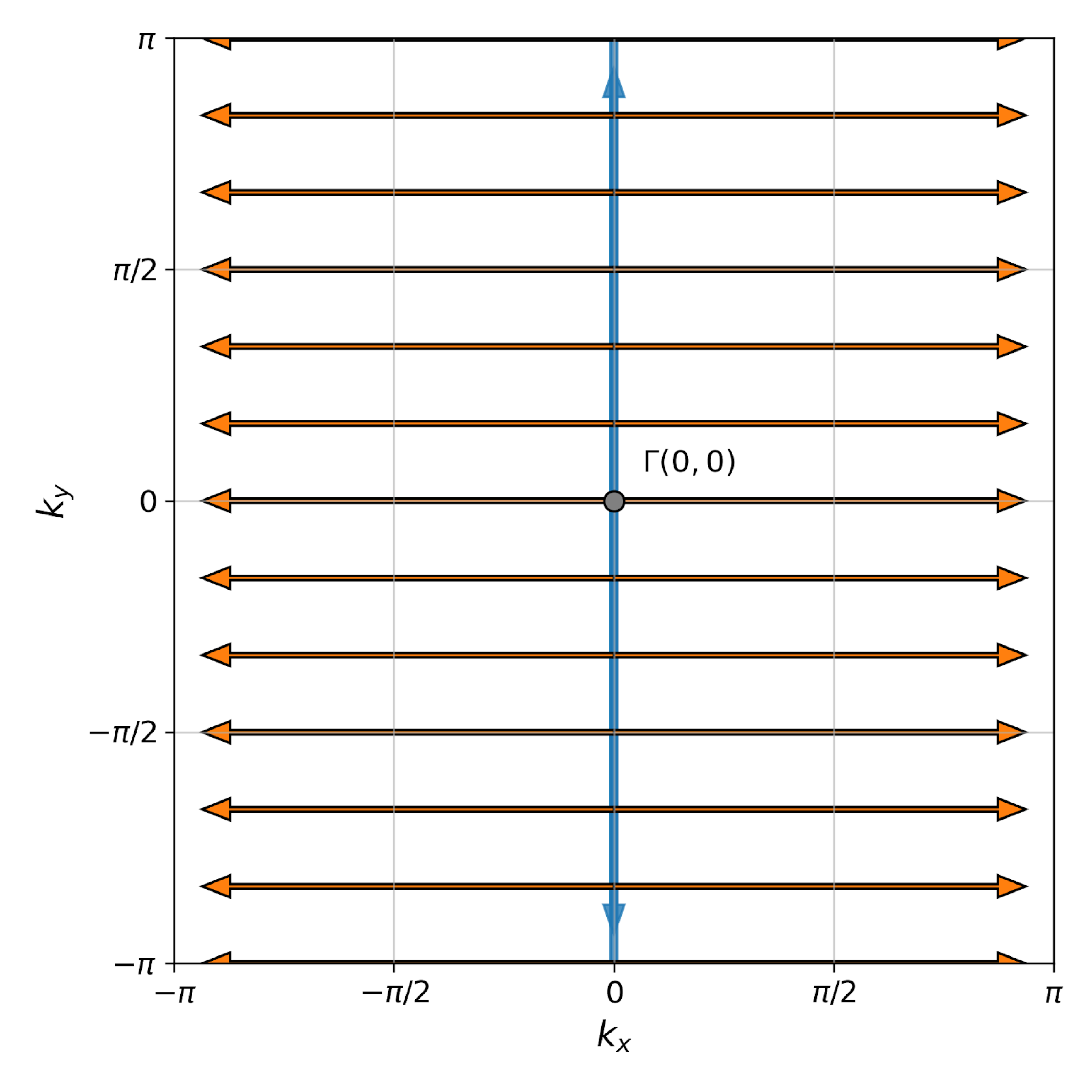}
		\caption{(Color online)
			A schematic diagram of transporting the bases at the center all over the
			k-space.}
		\label{fig:ptg}
	\end{figure}

	The projection (transport) is applied outward from the center to all k-points, as schematically depicted in Fig.~\ref{fig:ptg}. Specifically, the bases defined on the line $k_{x}=0$ are transported along the $k_{y}$ direction toward the boundaries at $k_{x}=\pm\pi$.
	
	The directions of the transport and that of the position operator
	are, theoretically, desired to be consistent. In this calculation,
	the direction of the transport does not show any discernible effect
	on the results. 
	The square-potential calculation below follows the general workflow already summarized in Fig.~\ref{fig:numerical_procedure}. After the peeling, the following two equations are repeated alternately 
	until a predetermined convergence criterion is met.
	The matrix elements are first calculated with a set of provisional WFs with $\{ \vert u^s_{\boldsymbol k} \rangle \}$
	calculated by Eq.~(\ref{eq:simple-peeling}),
	\begin{equation}\label{eq:s-XeqWxW}
		\boldsymbol X^n=\langle\boldsymbol{W}_{0}(\boldsymbol \Lambda^{n-1})\vert\hat{x}\vert\boldsymbol{W}_{0}(\boldsymbol\Lambda^{n-1})\rangle.
	\end{equation}
	With the estimated matrix, $\boldsymbol X^n$, 
	the following eigenvalue problem is solved, 
	\begin{equation}\label{eq:s-lambda-eigenvalue}
		\begin{split}
			\boldsymbol{\Lambda}^{n+1}=\boldsymbol{F}^{\dagger}\boldsymbol X^n\boldsymbol{F},
		\end{split}
	\end{equation}
	and hence the unitary matrix $\boldsymbol F$ and the Wannier centers are obtained.
	As the iteration proceeds, the matrix $\boldsymbol{X}$ approaches a diagonal form, namely, 
	\begin{equation}
		\lim_{n \rightarrow \infty} \Vert \boldsymbol \Lambda^n - \boldsymbol X^n \Vert \rightarrow \boldsymbol 0,
	\end{equation}
	Because of the convergent nature of the sinc-loop described in
	Sec.~IV B 2, we use $\boldsymbol\Lambda=\boldsymbol 0$ as the initial
	value in the current study.
	For the same reason, Eq.~(\ref{eq:s-XeqWxW}) can be repeated
	with a convergence criterion before proceeding to Eq.~(\ref{eq:s-lambda-eigenvalue}).
	
	Since $y_0$ cannot be obtained from Eq.~(\ref{eq:s-XeqWxW}) alone, 
	\begin{equation}
		y^{s, m+1}_0= \langle W_0^{s,n}(x_0^{s,n}, y_0^{s,m} )\vert \hat y \vert W_0^{s,n}(x_0^{s,n}, y_0^{s,m} )\rangle,
	\end{equation}
	is inserted to have convergence of $y_0$, prior to Eq.~(\ref{eq:s-lambda-eigenvalue}).
	
	When the convergence is reached, the series with the smallest spread is removed
	from the bases. From the symmetry of the system, $\hat x$ is replaced with
	$\hat y$ and the same iteration is repeated. For the final WF, Wannier center is
	obtained by simply repeating the single-element version of Eq.~(\ref{eq:s-XeqWxW}).
	\subsubsection{Peeling}\label{sec:s-peeling}
\begin{figure}[htbp]
		\centering
		\includegraphics[width=1.0\linewidth]{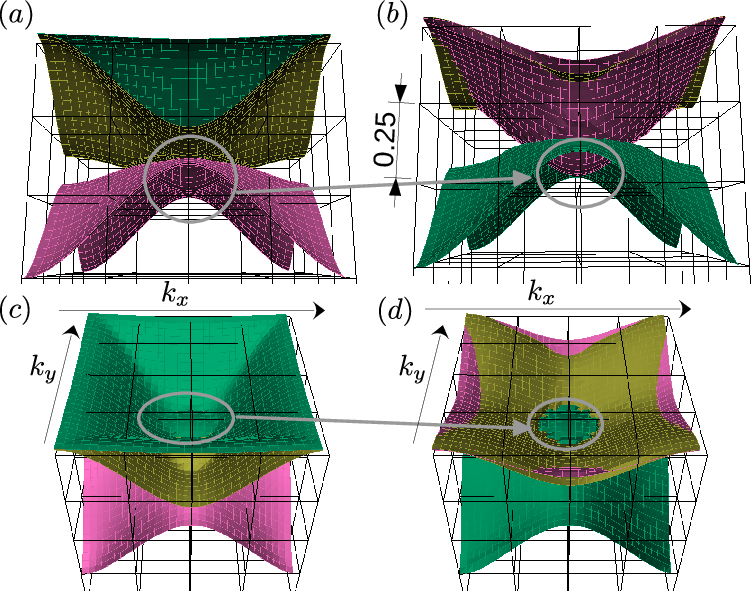}
		\caption{(Color online)
			Energy bands \#3, \#4 and \#5 of the 2D square potential system ($L$=36). The left half
			shows the original energy bands and the right half shows adiabatically
			peeled \textit{energy bands}. Strong bends are observed in the circled parts of
			the original energy band curvatures on the left side, while the bends are
			relaxed by the adiabatic transport via connecting multiple numbers of the
			original curvatures. }
		\label{fig:s-peeling}
	\end{figure}

	The left half of Fig.~\ref{fig:s-peeling} shows the energy bands obtained by solving
	the Schr\"odinger equation of the system. 
	The right half shows the peeled \textit{energy bands} defined as follows:
	\begin{equation}\label{eq:s-eband}
		e^s_{\boldsymbol k} = \sum_n \vert \langle u^s_{\boldsymbol k}\vert v^n_{\boldsymbol k}\rangle \vert^2 \epsilon^n_{\boldsymbol k},
	\end{equation}
	where $\epsilon^n_{\boldsymbol k}$ is the original energy as a function of the energy band index $n$ and 
	the wavenumber $\boldsymbol k$.
	As circled in Fig.~\ref{fig:s-peeling}, the original energy band
	curvatures have rather abrupt folds, for the band numbering is simply done
	according to the magnitude of the local energy in the k-space. After performing
	the procedure described by Eq.~(\ref{eq:simple-peeling}), the folds disappear by
	\textit{naturally} connecting and bending the original energy band curvatures.
	This is only for visualization of the effect of the adiabatic transport, 
	and the reconstruction of the band curvature per se
	does not have any effect on the calculation of the spread.
	
\begin{figure}[htbp]
		\centering
		\includegraphics[width=1.0\linewidth]{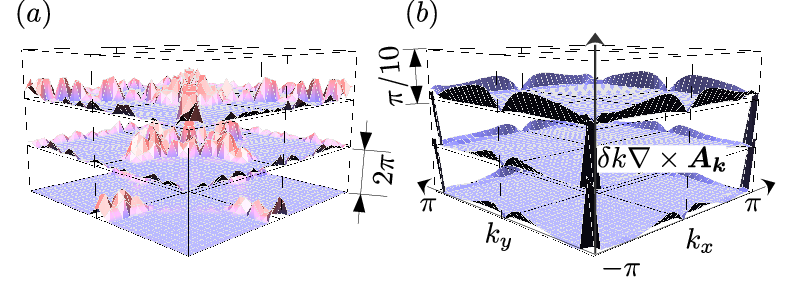}
		\caption{(Color online)
			Scaled curls of the diagonal connection (used here as frame-smoothness diagnostics) of the square potential model ($L$=36), defined as $\delta k \vert \nabla \times \boldsymbol A_{\boldsymbol k} \vert$, where $\boldsymbol A_{\boldsymbol k}$ denotes the corresponding diagonal connection of the frame being plotted, before and after the adiabatic transport. A scale factor of $40$ is applied to the smoother low-amplitude diagnostic surface obtained after transport so that it can be shown alongside the rough high-amplitude diagnostic surface before transport in the same figure; these two surfaces are distinguished by color in the online version. After realigning the original Bloch frames, the curl of the diagonal connection becomes approximately zero in the interior, while finite values remain on the boundary seam. Most of the rough pre-transport variations in this case are caused by the intertwining of the locally energy-ordered bands.}
		\label{fig:srota}
	\end{figure}

	The significance of Eq.~(\ref{eq:simple-peeling}) is also seen in the discrete curl of the diagonal connection $\boldsymbol A^s_{\boldsymbol k} = -i\langle u^s_{\boldsymbol k}|\nabla_{\boldsymbol k}|u^s_{\boldsymbol k}\rangle$ evaluated on the original and processed periodic parts of the BFs, namely $\{ \vert v^n_{\boldsymbol k} \rangle \}$ and $\{ \vert u^s_{\boldsymbol k} \rangle \}$, as shown in Fig.~\ref{fig:srota}. Here, this quantity is used solely as a frame-smoothness diagnostic.

	For a smoothly tracked nondegenerate Hamiltonian eigenband in a spinless $\mathcal{PT}$-symmetric system, the gauge-invariant Abelian Berry curvature vanishes pointwise wherever it is defined~\cite{Ahn2017,Fang2015}.
	The quantities shown in Figs.~\ref{fig:srota} and \ref{fig:grota} are not intended to represent the physical Berry curvature. Before transport (Fig.~\ref{fig:srota}(a)), the raw energy-ordered frame changes discontinuously near internal band intertwinings, and the discrete curl of the diagonal connection is contaminated by band-index switching artifacts. After transport (Fig.~\ref{fig:srota}(b)), the curl of the diagonal connection becomes approximately zero in the interior, confirming a smooth transported frame. The residual finite values strictly localized near the boundary are the geometric seam defects of the transported frame, which is consistent with the exact diagnostics ($\omega_\beta^1$ and $\sigma_{\min}$) analyzed later in Section \ref{sec:graphene}.
	\subsubsection{ Solving the Position Eigenvalue Problem and Identifying the Wannier Center}\label{sec:solve-wannier-center}
\begin{figure}[htbp]
		\centering
		\includegraphics[width=1\linewidth]{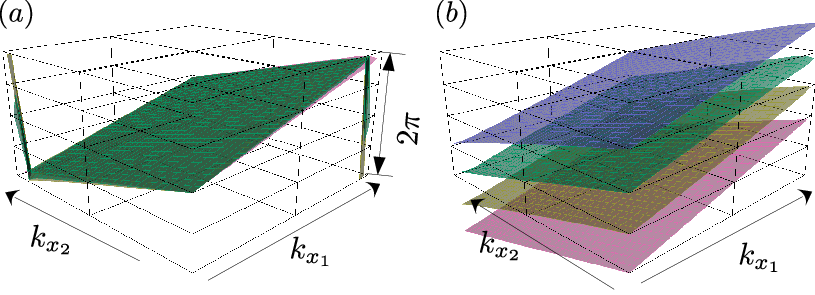}
		\caption{(Color online)
			Phase of the overlap $\langle u^s_{\boldsymbol k_1} \vert
			u^s_{\boldsymbol k_2} \rangle$ with fixed $k_{y}$ for the square potential 
			model ($L$=36). The figure on the left side
			shows the three overlap-phase surfaces of the three MLWFs with $k_y=0$. They
			are approximately the same plane, since all three MLWFs share the same Wannier
			center $\boldsymbol x_0 = ( 1/2, 1/2 )$ as the slope indicates. The visible folds occur when the principal value crosses $\pm\pi$, because the plotted range is restricted to
			$[-\pi,+\pi)$. The four surfaces on the right side correspond to
			one of the MLWFs at $k_y=-\pi, -\pi/2, 0, +\pi/2$, respectively. To show them
			distinctively, the planes are elevated and separated by $\pi/4$, even though they are 
			essentially the same plane.
		}	\label{fig:sphase}
	\end{figure}

	Following the adiabatic transport, the iteration based on
	Eqs.~(\ref{eq:s-XeqWxW}) and (\ref{eq:s-lambda-eigenvalue}) is applied.
	The purpose of the following phase plot is to verify the phase-plane
	relation in Eq.~(\ref{eq:arg-plane}) in two dimensions. The second
	lines of Eqs.~(\ref{eq:arg}) and (\ref{eq:arg-plane}) are precisely the
	component form used in this numerical plot. As seen in
	Fig.~\ref{fig:sphase}, the processed overlap phase is well approximated
	by this planar surface. The visible folds are artifacts of the
	principal-value wrapping of the phase into $[-\pi,\pi)$ and do not
	indicate a breakdown of the underlying phase-plane relation. The same
	behavior is observed when the roles of $k_x$ and $k_y$ are interchanged.
	
	\begin{figure}[htbp]
		\centering
		\includegraphics[width=1.0\linewidth]{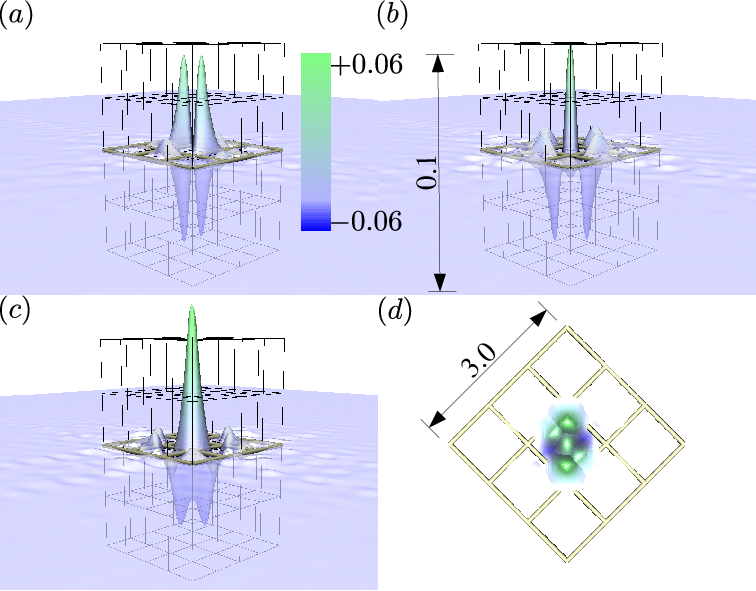}
		\caption{(Color online)
			The three MLWFs are shown in panels~(a), (b) and (c). To show the negative
			parts of the MLWFs, they are drawn with transparent surfaces. Panel~(d)
			shows all three MLWFs. The white rectangle objects are added to indicate
			$3\times3$ unit cells.
		}
		\label{fig:wanniers}
	\end{figure}
	\begin{table}[htbp] %
		\caption{Wannier Centers and Spreads of Square Potential Model.}
		\label{tab:sig-s}
		\centering
		\begin{tabular}{cc|c|c|cccc}
			\hline
			& Series\# & $x_0$ & $y_0$ & $\sigma^2_s$ & & & \tabularnewline
			\hline
			\hline
			& 1 & 0.50 & 0.50 & 0.24 & & & \tabularnewline
			& 2 & 0.50 & 0.50 & 0.24 & & & \tabularnewline
			& 3 & 0.50 & 0.50 & 0.28 & & & \tabularnewline
			\hline
		\end{tabular}
	\end{table}
	Figure~\ref{fig:wanniers} shows the MLWFs obtained from the periodic parts and the Wannier centers.
	The MLWFs in Fig.~\ref{fig:wanniers}(a),(b) and (c) are solutions of the following equations, 
	\begin{equation}\label{eq:sq-MLWF-equation}
		\begin{split}
			\hat x \vert W^a_{\boldsymbol 0} \rangle &= x_0\vert W^a_{\boldsymbol 0} \rangle\\
			\hat y \vert W^b_{\boldsymbol 0} \rangle &= y_0\vert W^b_{\boldsymbol 0} \rangle\\
			(\hat x + \hat y) \vert W^c_{\boldsymbol 0} \rangle &= c_0\vert W^c_{\boldsymbol 0} \rangle.\\
		\end{split}
	\end{equation}
	Since they have the same Wannier centers, it is possible to achieve stronger localization by linearly combining
	some of them, although it is not pursued in the current study. 
	
	The spatial calculation of Wannier centers and the spreads are defined by the following equations, 
	\begin{equation}\label{eq:spread-def}
		\begin{split}
			\boldsymbol x_0 &= \langle W^s_{\boldsymbol 0}\vert \hat {\boldsymbol x} \vert W^s_{\boldsymbol 0}\rangle\\
			\sigma_s^2 &=\langle W^s_{\boldsymbol 0}\vert ( \hat{\boldsymbol x} - \boldsymbol x_0 )^2\vert W^s_{\boldsymbol 0} \rangle,
		\end{split}
	\end{equation}
	and the results are listed in Table~\ref{tab:sig-s}.
	
	\subsection{Graphene}\label{sec:graphene}
	This section focuses on MLWFs of graphene.
	The entire calculation is performed in the oblique coordinate system and 
	hence the coordinate $(\xi, \eta)$ is transformed to $(x, y)$,
	\begin{equation}\label{eq:oblique-transform}
		\begin{pmatrix} x \\ y \end{pmatrix} 
		= \begin{pmatrix} 1 & -1/2 \\ 0 & \sqrt{3}/2 \end{pmatrix} \begin{pmatrix} \xi \\ \eta \end{pmatrix}.
	\end{equation}
	The wavenumbers are also in the oblique coordinate system and denoted $(k_{\xi},
	k_{\eta})$. The final spreads and Wannier centers are presented in the Cartesian
	coordinate system.
	
	First-principles calculations were carried out using \textsc{Quantum ESPRESSO}
	(QE)~\cite{QE_2009,QE_2017,QE_Website}. The spatial resolutions were set
	identical to the k-space resolutions. The calculations are carried out by
	changing the resolution from $12\times12$ to $36\times36$ in the k-space;
	the corresponding centers and spreads are summarized later in
	Table~\ref{tab:centers_match_no_blank}. 
	The link variables were taken as the raw nearest-neighbor overlaps
	$\langle v^m_{\boldsymbol{k}}\vert v^n_{\boldsymbol{k}+\delta\boldsymbol{k}_{\alpha}}\rangle$
	provided by the \textsc{Wannier90} (W90)~\cite{Wannier90_2008,Wannier90_2020,Wannier90_Guide}
	interface, where $\delta\boldsymbol{k}_{\alpha}$ connects a mesh point to its near neighbors
	(see Appendix~\ref{ap:kspace-def} for notation).
	These links are used to evaluate the matrix elements of the position operators,
	as shown later in this section.
	
	As a setting of the QE calculation, we choose five Wannier functions consisting
	of two $C-p_z$ orbitals ($\pi$ manifold) and three bond-centered orbitals ($\sigma$ bonds),
	providing a minimal chemically-motivated set for graphene within the five-band
	subspace used to build the overlap matrices ( see
	Appendix~\ref{ap:qe-conditions} for details).
	
	While five energy bands, from \#0 to \#4 are provided to the W90, lower four
	energy bands, \#0 to \#3, are fed into the current computational program.
	Eventually, three MLWFs are obtained both from W90 and the current program.
	
	In the case of graphene, the chosen sequential extraction from the composite subspace isolates a component that inevitably accumulates a topologically obstructed intermediate state. Therefore, the specific 4-band pruning and deflation sequence detailed in Fig.~\ref{fig:flowchartdeflation4band} is devised to execute the ``More MLWF? '' loop of the general flowchart (Fig.~\ref{fig:numerical_procedure}).
	
	The calculation results are presented in this section and the spreads are
	compared with those obtained from W90 under the very same conditions to check the
	validity and utility of the current method. \subsubsection{Matrix elements of
		$\hat x$}\label{sec:graphene-matrix-element} Instead of using BFs of graphene,
	the links,$\{\langle v^{n_1}_{{\boldsymbol k}_1} \vert v^{n_2}_{{\boldsymbol
			k}_2} \rangle\}$, generated by W90 are utilized. By using the links, the
	calculation of the matrix elements of the position operator, for example, are
	performed as follows:
	\begin{equation}\label{eq:link-x}
		\begin{aligned}\langle W_{\boldsymbol{0}}^{s_{1}}\vert\hat{x}\vert W_{\boldsymbol{0}}^{s_{2}}\rangle & =\frac{1}{L^2}\sum_{\boldsymbol{k}_{1},\boldsymbol{k}_{2}}\langle u_{\boldsymbol{k}_{1}}^{s_{1}}\vert e^{-i\boldsymbol{k}_{1}\cdot\hat{{\boldsymbol{x}}}}\hat{x}e^{i\boldsymbol{k}_{2}\cdot\hat{{\boldsymbol{x}}}}\vert u_{\boldsymbol{k}_{2}}^{s_{2}}\rangle\\
			& =i\sum_{\boldsymbol{k}_{1},\boldsymbol{k}_{2}}\langle u_{\boldsymbol{k}_{1}}^{s_{1}}\vert\partial_{k_{2x}}\vert u_{\boldsymbol{k}_{2}}^{s_{2}}\rangle\delta_{\boldsymbol{k}_{1},\boldsymbol{k}_{2}}\\
			& =i\sum_{\boldsymbol{k}}\langle u_{\boldsymbol{k}}^{s_{1}}\vert\partial_{k_{x}}\vert u_{\boldsymbol{k}}^{s_{2}}\rangle\\
			& =\frac{i}{2\delta k}\!\sum_{\boldsymbol{k}}\!\!\!\!\left\{ \langle u_{(k_{x},k_{y})}^{s_{1}}\vert u_{(k_{x}+\delta k,k_{y})}^{s_{2}}\rangle-\langle u_{(k_{x},k_{y})}^{s_{1}}\vert u_{(k_{x}-\delta k,k_{y})}^{s_{2}}\rangle\right\} \\
			& =\frac{i}{2\delta k}\sum_{\boldsymbol{k},n_{1},n_{2}}\bar{f}_{(k_{x},k_{y})}^{s_{1},n_{1}}\left\{ \langle v_{(k_{x},k_{y})}^{n_{1}}\vert v_{(k_{x}+\delta k,k_{y})}^{n_{2}}\rangle f_{(k_{x}+\delta k,k_{y})}^{s_{2},n_{2}}\right.\\
			& \;\;\;\;\;\;\;\;\;\;\;\;\;\;\;\;\;\;\;\;\;\;\;\;\;\;\;-\left.\langle v_{(k_{x},k_{y})}^{n_{1}}\vert v_{(k_{x}-\delta k,k_{y})}^{n_{2}}\rangle f_{(k_{x}-\delta k,k_{y})}^{s_{2},n_{2}}\right\}\\
			& \;\;\;\;\;\;\;\;\;\;\;\;\;\;\;\;\;\;\;\;\;\;\;\;\;\;\;
			\;\;\;\;\;\;\;\;\;\;\;\;\;\;\;\;\;\;\;\;+\mathcal{O}(\delta k^2)
		\end{aligned}
	\end{equation}
	At the very beginning of the calculation, the basis vectors are expressed as,
	\begin{equation}
		\begin{split}
			f^0_{\boldsymbol k}&=\{1, 0, 0 \ldots, 0 \}\\
			f^1_{\boldsymbol k}&=\{0, 1, 0 \ldots, 0 \}\\
			\vdots \\
			f^{N_C -1}_{\boldsymbol k}&=\{0, 0, 0 \ldots, 1 \},
		\end{split}
	\end{equation} 
	for all ${\boldsymbol k}$, in place of $\langle x \vert v^n_{\boldsymbol k} \rangle$.
	Every matrix transformation is done on the bases $\{f^n_{\boldsymbol k}\}$, and the inner product
	must always be done with the links, as seen in Eq.~(\ref{eq:link-x}).
	\subsubsection{Peeling}\label{sec:g-peeling}
	In the graphene calculation, the initial condition in
	Eq.~(\ref{eq:provisional-initial-condition}) is imposed at
	$\boldsymbol k_0=(\pi,\pi)$ in the oblique reciprocal coordinates
	$(k_\xi,k_\eta)$. This choice fixes the initial local frame used for
	the adiabatic transport. It is different from the initial value of the
	Wannier-center parameters, which is set to
	$\boldsymbol\Lambda=\boldsymbol 0$ in the center-fixing update described
	below.
	
\begin{figure}[htbp]
		\centering
		\includegraphics[width=1.0\linewidth]{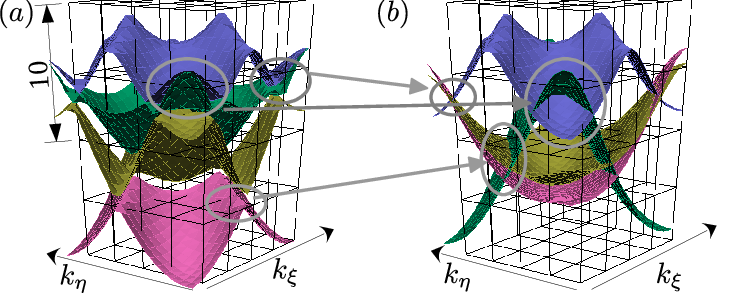}
		\caption{(Color online)
			Energy band \#0 to \#4 of graphene ($L$=22). The left half
			shows the original energy bands and the right half shows adiabatically
			peeled \textit{energy bands}. Strong bends are observed in the circled parts of
			the original energy band curvatures on the left side, while the bends are
			relaxed by the adiabatic transport via connecting multiple numbers of the
			original curvatures. $(k_\xi, k_\eta)$-axis corresponds to the oblique reciprocal coordinate spanning the primitive parallelogram cell $[-\pi, \pi)$, rather than a conventional high-symmetry path (such as $\Gamma$-M-K-$\Gamma$) in the hexagonal Brillouin zone.}
		\label{fig:gbands}
	\end{figure}

	The left half of Fig.~\ref{fig:gbands} shows the energy bands obtained from W90.
	The right half shows the peeled \textit{ energy bands} defined by Eq.~(\ref{eq:s-eband}).
	\begin{figure}[htbp]
		\centering
		\includegraphics[width=1\linewidth]{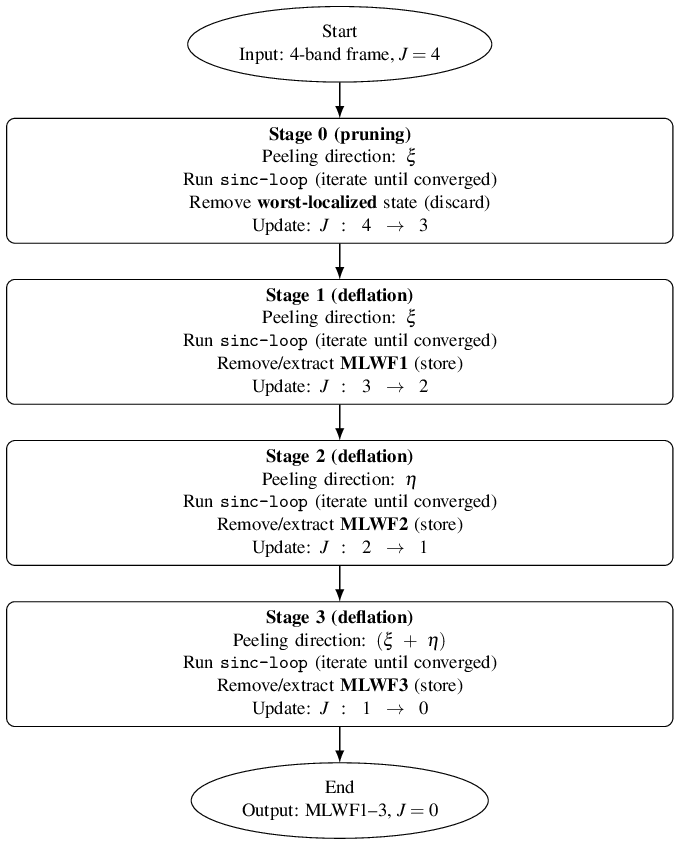}
		\caption{Flowchart of the 4-band peeling/deflation procedure used for graphene.
			Stage~0 (pruning): $\xi$-direction peeling and the sinc-loop, then discard the least localized state ($J:4\to3$).
			Stages~1--3 (deflation): repeat peeling and the sinc-loop, then extract/store MLWF1--3 sequentially
			using the $\xi$, $\eta$, and $(\xi+\eta)$ directions ($J:3\to2\to1\to0$).}
		\label{fig:flowchartdeflation4band}
	\end{figure}
	\begin{figure}[htbp]
		\centering
		\includegraphics[width=0.9\linewidth]{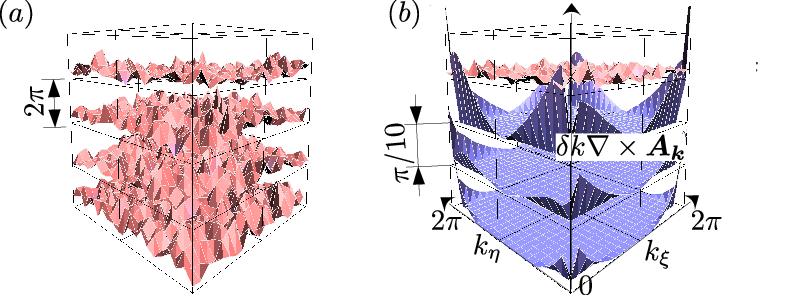}
		\caption{(Color online)
			Scaled curls of the diagonal connection (used here as frame-smoothness diagnostics) of graphene ($L$=22), defined as $\delta k \vert \nabla \times \boldsymbol A_{\boldsymbol k} \vert$, where $\boldsymbol A_{\boldsymbol k}$ denotes the corresponding diagonal connection of the frame being plotted, before and after the adiabatic transport. A factor of $40$ is applied to the smoother low-amplitude diagnostic surface in panel~(b) so that it can be shown together with the rough high-amplitude diagnostic surface in panel~(a); these two surfaces are distinguished by color in the online version. In panel~(a), the diagnostic is rough throughout the Brillouin zone because the raw energy-ordered frame is globally affected by band intertwinings across the composite subspace. In panel~(b), it becomes nearly zero in the interior for the lower three peeled series, while finite variations remain on the boundary seam. These seam-localized variations are the boundary gauge defects analyzed later in Figs.~\ref{fig:omega}--\ref{fig:sigmamin}.
		}\label{fig:grota}
	\end{figure}
	\begin{figure}[htbp]
		\centering
		\includegraphics[width=0.9\linewidth]{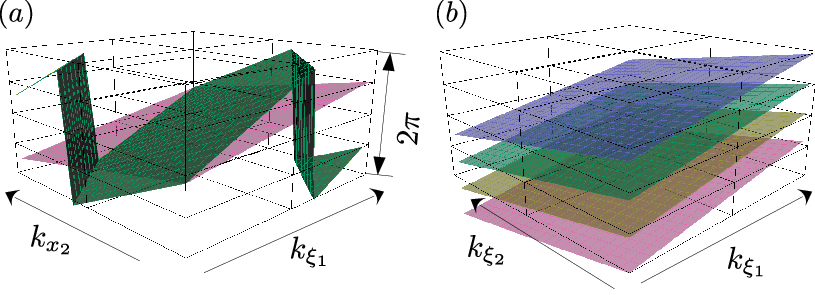}
		\caption{(Color online)
			Phase of the overlap of graphene ($L$=22), $\langle u^s_{\boldsymbol k_{1}} 
			\vert u^s_{\boldsymbol k_2} \rangle$, with fixed $k_\eta$. The figure on the left side shows the three overlap-phase surfaces of the three
			MLWFs with $k_\eta=0$. The slopes reflect $\xi_0$s of the corresponding MLWFs.
			The Wannier centers correspond to $\xi_0 = 1/3, 5/6, 5/6$. 
			The visible folds occur when the principal value crosses $\pm\pi$, because the plotted range is restricted
			to $[-\pi,+\pi)$. The four surfaces on the right correspond to one of the MLWFs at $k_\eta=-\pi, -\pi/2, 0, +\pi/2$, respectively. To show them
			distinctively, the planes are elevated and separated by $\pi/4$, although they are
			essentially the same plane.
		}
		\label{fig:gphase}
	\end{figure}
	\begin{figure}[htbp]
		\centering
		\includegraphics[width=0.9\linewidth]{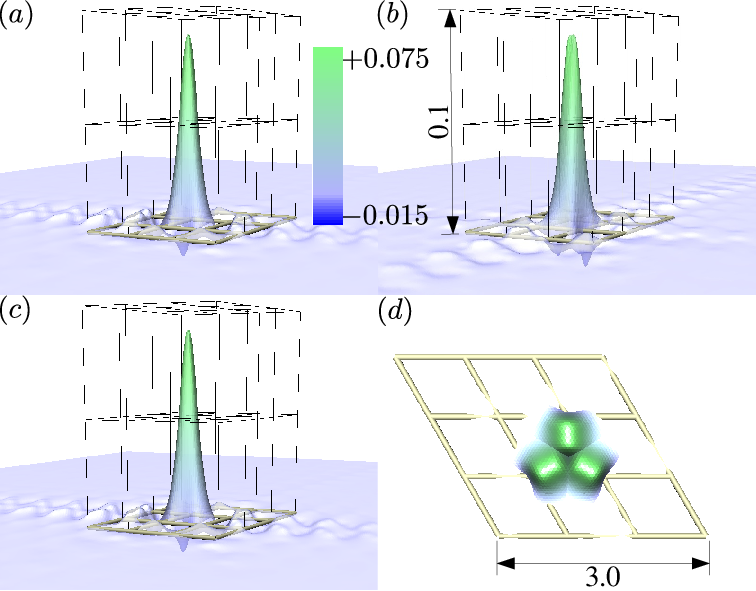}
		\caption{(Color online)
			MLWFs of graphene ($L$=22). The three MLWFs are shown in panels~(a), (b) and (c). To show the negative parts of 
			the MLWFs, they are drawn with transparent surfaces. Panel~(d) shows all three MLWFs together. The rectangular objects
			(gold-colored in the online version) are added to indicate $3\times3$ unit cells.
			The Wannier centers of the MLWFs seen in panels~(a),(b) and (c) are 
			(0.25, 0.15), (0.50, 0.58) and (0.75, 0.14), respectively. }
		\label{fig:wannierg}
	\end{figure}
	\begin{table}[htbp]
		\centering
		\caption{Centers aligned by matching (row permutation only). Units: \AA. Values rounded to two decimals.}
		\label{tab:centers_match_no_blank}
		\begin{ruledtabular}
			\begin{tabular}{c c | c c c | c c c}
				$L$ & WF\# &
				\multicolumn{3}{c|}{Current } &
				\multicolumn{3}{c}{W90} \\
				\cline{3-8}
				& & $\Omega$ (\AA$^2$) & $x_0$ (\AA) & $y_0$ (\AA)
				& $\Omega$ (\AA$^2$) & $x_0$ (\AA) & $y_0$ (\AA) \\
				\hline
				12 & 1 & 0.86 & -0.62 & -0.35 & 1.15 & -0.63 & -0.36 \\
				& 2 & 1.06 & 0.00 & 0.71 & 1.58 & 0.01 & 0.72 \\
				& 3 & 1.26 & 0.62 & -0.35 & 0.92 & 0.62 & -0.36 \\
				\hline
				22 & 1 & 1.16 & -0.62 & -0.35 & 0.91 & -0.63 & -0.36 \\
				& 2 & 1.58 & 0.00 & 0.72 & 0.99 & 0.00 & 0.72 \\
				& 3 & 1.91 & 0.62 & -0.37 & 1.03 & 0.63 & -0.36 \\
				\hline
				30 & 1 & 1.43 & -0.62 & -0.35 & 1.47 & -0.63 & -0.36 \\
				& 2 & 1.99 & 0.00 & 0.72 & 1.48 & 0.00 & 0.72 \\
				& 3 & 2.47 & 0.62 & -0.37 & 1.49 & 0.63 & -0.36 \\
			\end{tabular}
		\end{ruledtabular}
	\end{table}
	\begin{figure*}[t]
		\centering
		\includegraphics[width=\textwidth]{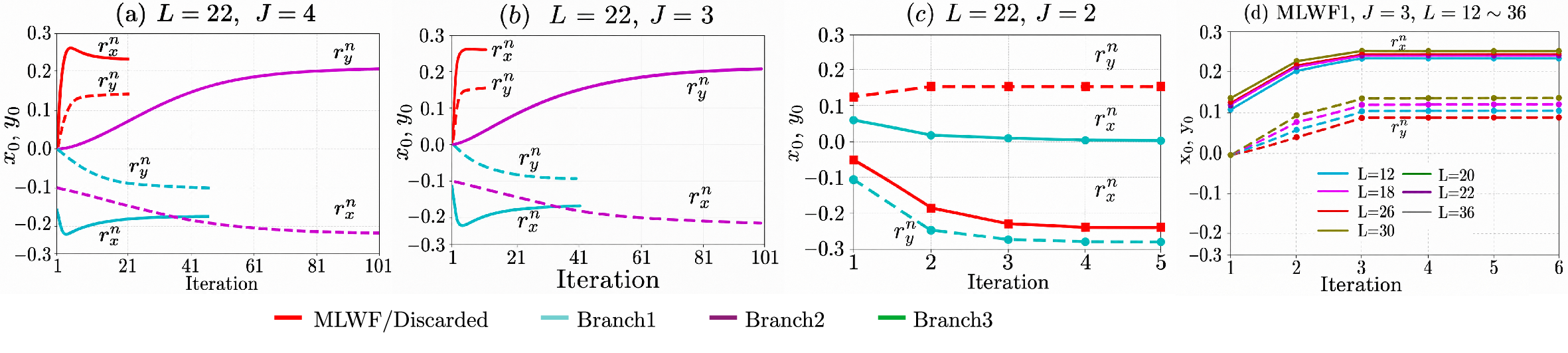}
		\caption{(Color online)
			Practical convergence of the center-fixing updates in graphene.
			The four panels, from left to right, show the iteration trajectories
			of $(x_0,y_0)$ for a representative $L=22$ run at the successive
			$J=4,3,2$ stages of the pruning/deflation sequence, followed by the
			mesh dependence of the selected $J=3$ branch corresponding to MLWF1.
			Here $J$ denotes the number of remaining branches in the composite
			subspace. Throughout the figure, solid and dashed curves denote $x_0$
			and $y_0$, respectively. In the first three panels, the red curves
			indicate, respectively, the precursor branch removed at the $J=4$
			stage, the MLWF1 branch selected at the $J=3$ stage, and the MLWF2
			branch selected at the $J=2$ stage. In panels (a) and (b), markers are
			omitted because the trajectories are too dense for marker symbols to be
			helpful. In the rightmost panel, the selected $J=3$ MLWF1 branch is
			shown for $L=12,18,20,22,26,30,$ and $36$; colors distinguish $L$,
			while the same solid/dashed convention distinguishes $x_0$ and $y_0$.
			The horizontal axis is the iteration count actually performed. No
			interpolation or artificial extension of the trajectories is used.
			The plotted coordinates are the dimensionless transformed coordinates
			$(x_0,y_0)=(\xi_0-\eta_0/2,\sqrt{3}\eta_0/2)$, not real-space
			Wannier-center coordinates in units of \AA. The selected MLWF1 center
			remains nearly unchanged over the full range of mesh sizes.
		}
		\label{fig:graphene-center-convergence}
	\end{figure*}
	As circled in Fig.~\ref{fig:gbands}, the original energy band curvatures have
	rather abrupt folds, for the band numbering is simply done according to the
	magnitude of the local energy in the k-space. After performing the procedure
	described by Eq.~(\ref{eq:simple-peeling}), the folds disappear by 
	\textit{naturally} connecting and bending the original energy band curvatures.
	
	The significance of Eq.~(\ref{eq:simple-peeling}) is seen in the discrete curl of the diagonal connection calculated with the original and processed periodic parts of the BFs, namely $\{\vert v^n_{\boldsymbol k} \rangle\}$ and $\{\vert u^s_{\boldsymbol k} \rangle\}$, as shown in Fig.~\ref{fig:grota}. As in the square-potential case, this diagnostic is not the gauge-invariant physical Berry curvature of a smoothly tracked Hamiltonian eigenband. In the present spinless $\mathcal{PT}$-symmetric setting, that physical curvature vanishes pointwise. Before transport, the raw energy-ordered frame is globally affected by band intertwinings across the composite subspace, so the corresponding diagnostic appears rough throughout the Brillouin zone. After transport, the diagnostic becomes nearly zero in the interior of the Brillouin zone, whereas finite $\mathcal{O}(L^0)$ variations remain concentrated on the boundary seam. Those seam-localized variations are precisely the geometric defects quantified later by $\omega_{\beta}^1(\boldsymbol{k})$ and $\sigma_{\min}(\boldsymbol{k};\delta\boldsymbol{k}_{\beta})$. 
	\subsubsection{Obtained MLWFs and their Wannier Centers}\label{sec:obtain-wannier-center}
	Following the adiabatic transport, the iteration based on
	Eqs.~(\ref{eq:s-XeqWxW}) and (\ref{eq:s-lambda-eigenvalue}) is applied.
	The same phase-plane interpretation applies to graphene, but now in the
	oblique coordinates used for the honeycomb lattice. Namely, the second
	line of Eq.~(\ref{eq:arg-plane}) is read with
	$(k_x,k_y;x_0,y_0)$ replaced by
	$(k_{\xi},k_{\eta};\xi_0,\eta_0)$. As seen in Fig.~\ref{fig:gphase},
	the processed overlap phase is well approximated by this planar surface.
	The visible folds are artifacts of principal-value phase wrapping. The
	same behavior is observed when the roles of $k_{\xi}$ and $k_{\eta}$ are
	interchanged. Hence the slopes of the phase plane give the individual
	components of $\boldsymbol{\xi}_0=(\xi_0,\eta_0)$.
	
	Figure~\ref{fig:wannierg} shows the MLWFs obtained from the periodic parts and the Wannier centers.
	The WFs in Fig.~\ref{fig:wannierg}(a)--(c) are solutions of the following equations,
	\begin{equation}
		\begin{split}
			\hat \eta \vert W^a_{\boldsymbol 0} \rangle &= \eta_0\vert W^a_{\boldsymbol 0} \rangle\\
			\hat \xi \vert W^b_{\boldsymbol 0} \rangle &= \xi_0\vert W^b_{\boldsymbol 0} \rangle\\
			(\hat \xi + \hat \eta) \vert W^c_{\boldsymbol 0} \rangle &= \zeta_0\vert W^c_{\boldsymbol 0} \rangle
		\end{split}
	\end{equation}
	The 4-band pruning/deflation sequence used for graphene is
	summarized in Fig.~\ref{fig:flowchartdeflation4band}.

	\subsubsection{Practical convergence of the center-fixing update}
	We briefly record the practical convergence behavior of the
	center-fixing updates used in the graphene calculation. The center
	update is not a minimization of the spread functional, but a
	fixed-point/self-consistent solution of the projected-position matrix
	elements and the associated Wannier-center parameters. Here $J$ denotes
	the number of remaining branches in the pruning/deflation sequence.
	In the calculations shown here, the center parameters are initialized by
	$\boldsymbol{\Lambda}=0$. For the branch-center update, convergence is
	monitored by the squared displacement in the underlying
	$(\xi_0,\eta_0)$ variables,
	\[
	\Delta_{\xi\eta}^2 =
	(\xi_0^{(n+1)}-\xi_0^{(n)})^2+
	(\eta_0^{(n+1)}-\eta_0^{(n)})^2 .
	\]
	In the implementation used for Fig.~\ref{fig:graphene-center-convergence},
	the branch update is iterated until
	$\Delta_{\xi\eta}^2<10^{-8}$, with at least five iterations
	imposed to avoid premature termination.
	
	Figure~\ref{fig:graphene-center-convergence} shows a representative
	$L=22$ graphene run for the $J=4,3,2$ stages, together with the
	mesh dependence of the selected $J=3$ branch corresponding to MLWF1.
	For the $L=22$ run, the highlighted precursor branch at the $J=4$
	stage, the MLWF1 branch at the $J=3$ stage, and the MLWF2 branch at
	the $J=2$ stage terminate after 14, 6, and 5 plotted iteration
	points, respectively. The final $J=1$ step yields MLWF3 and is omitted
	from the figure for compactness; it is a single-branch update and also
	converges within five iterations. For $L=12,18,20,22,26,30,$
	and $36$, the final values of the selected $J=3$ MLWF1 branch remain
	nearly unchanged, $x_0=0.2483$--$0.2512$ and
	$y_0=0.1403$--$0.1416$ in the plotted coordinate convention.
	Thus the center update is stable with respect to
	mesh refinement in this isolated graphene composite-band subspace.
	For stronger internal degeneracies, or when projected-position
	eigenvalues become nearly degenerate, branch assignment may require
	additional checks. The separate problem of selecting an optimal
	subspace from energetically entangled bands is outside the scope of the
	present work.

	\subsubsection{Spreads and reciprocal-space diagnostics}
\begin{figure}[thbp]
		\centering
		\includegraphics[width=\linewidth]{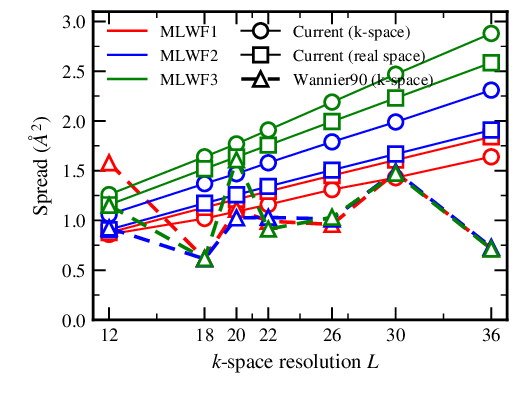}
		\caption{(Color online)
			Spreads of the three graphene MLWFs as a function of the $k$-mesh resolution $L$.
			Squares: results of the present method evaluated in real space via Eq.~(\ref{eq:spread-def}).
			Circles: results of the present method evaluated in reciprocal space from the link matrices.
			Triangles: \textsc{Wannier90} results evaluated in reciprocal space from the same link matrices.}
		\label{fig:spread}
	\end{figure}

	The obtained centers and spreads are compared with those from W90 in
	Table~\ref{tab:centers_match_no_blank}.
	Figure~\ref{fig:spread} compares the
	spreads of the graphene MLWFs as a function of $L$. The spreads obtained from
	the real-space integration (squares) and from the reciprocal-space
	finite-difference functional
	\cite{MarzariVanderbilt1997,Wannier90_2008,Wannier90_2020}
	(circles; see Appendix~\ref{ap:kspace-def}) agree well with
	those from W90, thereby confirming the validity of the adiabatic transport, projected position
	operator eigenvectors, sinc-loop and the pruning/deflation workflow.
	\begin{figure}[htbp]
		\centering
		\includegraphics[width=1\linewidth]{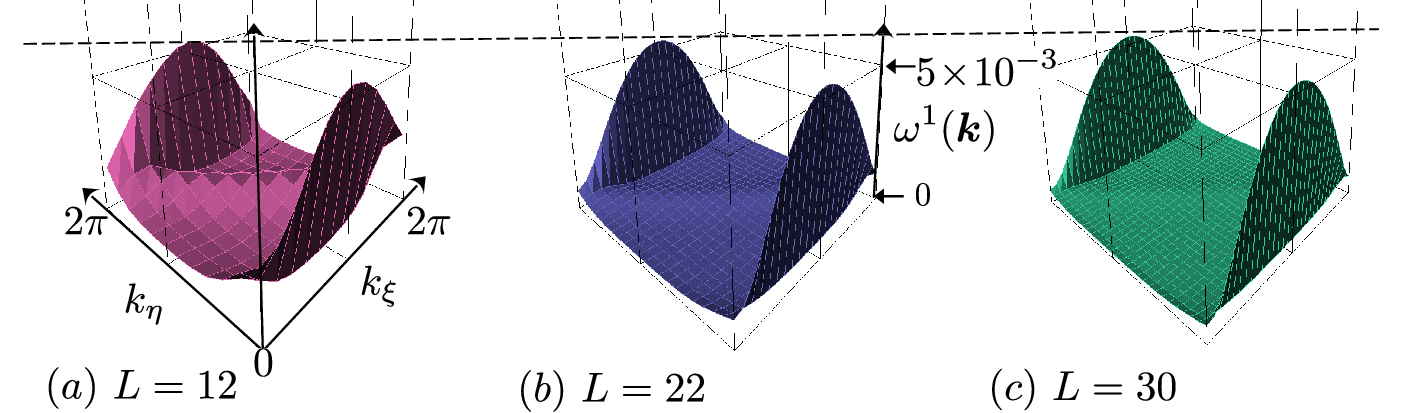}
		\caption{(Color online)
			Single-state geometric defect $\omega_{\beta}^1(\boldsymbol{k})$ carried by the extracted MLWF1 in graphene. 
			This quantity is defined as $\omega_{\beta}^1(\boldsymbol{k}) = 1-\vert \langle u_{\boldsymbol{k}}^1 \vert u_{\boldsymbol{k}+\delta\boldsymbol{k}_{\beta}}^1\rangle\vert^2$, isolating the geometric distortion of the one-dimensional subspace along the link direction $\beta$. 
			Panels (a)--(c) display its distribution in the oblique reciprocal coordinates $(k_{\xi},k_{\eta})$ for mesh resolutions $L=12$, $22$, and $30$. 
		}
		\label{fig:omega}
	\end{figure}
	
	\begin{figure}[htbp]
		\centering
		\includegraphics[width=1.0\linewidth]{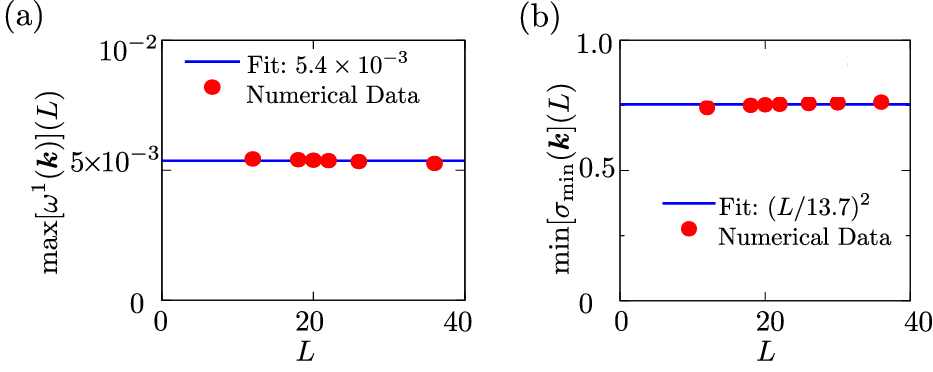}
		\caption{(Color online)
			Resolution dependence of the local geometric defects on the boundary seam in graphene.
			(a) The maximum value of the single-state spread kernel carried by MLWF1, $\max_{\boldsymbol{k}}\omega_{\beta}^1(\boldsymbol{k})$, plotted against the mesh resolution $L$.
			(b) The absolute minimum of the subspace-level diagnostic, $\min_{\boldsymbol{k}}\sigma_{\min}(\boldsymbol{k};\delta\boldsymbol{k}_{\beta})$, for the residual $J=2$ subspace.
			Both quantities appear $L$-independent, providing direct numerical evidence that the local gauge mismatch on the seam remains an $\mathcal{O}(L^0)$ geometric constant.
		}
		\label{fig:omega(k)}
	\end{figure}
	\begin{figure}[htpb]
		\centering
		\includegraphics[width=\linewidth]{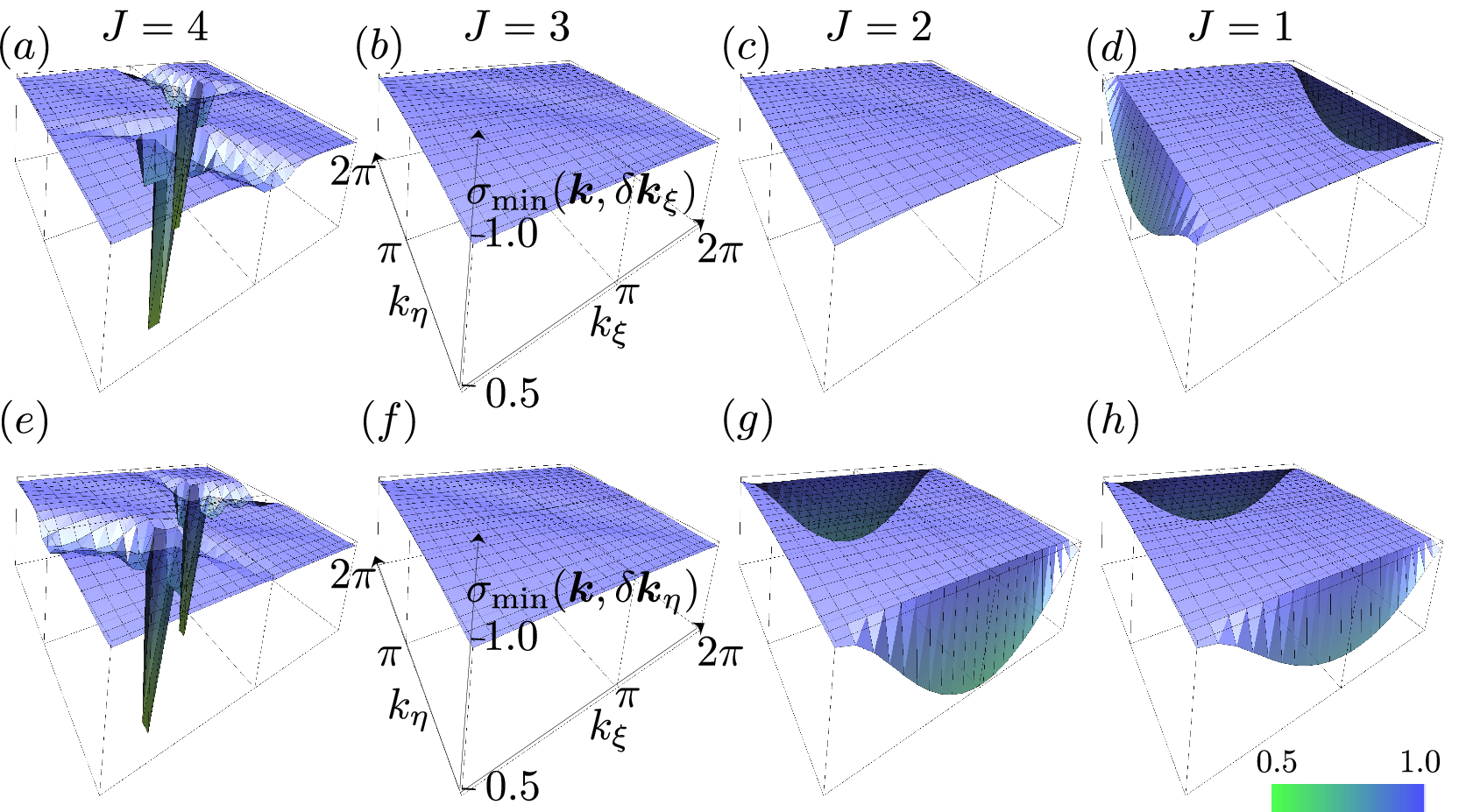}
		\caption{(Color online)
			Principal-angle diagnostic $\sigma_{\min}(\boldsymbol{k};\delta\boldsymbol{k}_{\alpha})$ of the transported subspaces in graphene ($L=22$). The upper panels (a)--(d) show the diagnostic evaluated along the $\xi$-direction ($\alpha=\xi$), while the lower panels (e)--(h) show it along the $\eta$-direction ($\alpha=\eta$). The initial transported $J=3$ subspace is globally flat near unity in both directions. After extracting MLWF1, the residual $J=2$ subspace remains flat along $\xi$ but exhibits broad dog-ear-like depressions near the boundaries along $\eta$.
		}
		\label{fig:sigmamin}
	\end{figure}
	Although the orbital shapes and Wannier centers agree well with those from W90,
	the spreads increase approximately linearly with $L$ as shown in Fig.~\ref{fig:spread}.
	This resolution dependence is understood by following the procedures shown in Fig.~\ref{fig:flowchartdeflation4band} one-by-one. 
	
	(i) $J=4\rightarrow 3$.
	Starting from the four-band subspace, we diagonalize $\hat{\xi}$ and remove the least localized state.
	At this stage, as seen in Figs.~\ref{fig:sigmamin}(b) and (f), the transported $J=3$ subspace becomes smooth in both link directions, as $\sigma_{\min}(\boldsymbol{k};\delta\boldsymbol{k}_{\beta})$, a measure of the closeness of adjacent local spaces spanned by $\{ \vert u^s_{\boldsymbol k} \rangle \}$ (see Appendix~\ref{ap:kspace-def}), is nearly flat and close to unity throughout the Brillouin zone. Thus, the neighboring local subspaces spanned by $\{\vert u^s_{\boldsymbol k}\rangle\}$ are almost identical. This $\sigma_{\min} \approx 1$ behavior serves as the direct visual confirmation of Eq.~(\ref{eq:Psi=0}).
	
	(ii) $J=3\rightarrow 2$.
	After MLWF1 is extracted, the seam is exposed and two complementary diagnostics become relevant.
	For the extracted series $s=1$, we define the diagonal link
	\begin{equation}
		m_{\beta}^1(\boldsymbol{k})
		=
		\langle u_{\boldsymbol{k}}^1\vert u_{\boldsymbol{k}+\delta\boldsymbol{k}_{\beta}}^1\rangle,
		\label{eq:result-m-single-clean}
	\end{equation}
	and its singular value $\sigma_{\beta}^1(\boldsymbol{k}) = \vert m_{\beta}^1(\boldsymbol{k})\vert$.
	Because this is a $1\times 1$ overlap matrix, $\sigma_{\beta}^1$ is simply the absolute value of the link itself.
	We then define the corresponding single-state kernel
	\begin{equation}
		\omega_{\beta}^1(\boldsymbol{k})
		=
		1-\left[\sigma_{\beta}^1(\boldsymbol{k})\right]^2
		=
		1-\left|m_{\beta}^1(\boldsymbol{k})\right|^2.
		\label{eq:result-omega-single-clean}
	\end{equation}
	Figure~\ref{fig:omega} shows that $\omega_{\beta}^1(\boldsymbol{k})$ is concentrated on the boundary seam,
	while Fig.~\ref{fig:omega(k)}(a) shows that its representative maximum value remains strictly constant as $L$ is increased, thus $\omega_{\beta}^1(\boldsymbol{k})\sim \mathcal{O}(L^0)$ on the boundary.
	
	The residual $J=2$ subspace left behind after the extraction shows the same seam geometry at the subspace level.
	As visualized in Fig.~\ref{fig:sigmamin}(c), it stays approximately level in the $\boldsymbol k_\xi$-direction, faithfully preserving the 1D uniformity mandated by Eq.~(\ref{eq:Psi=0}). However, it inevitably develops dog-ear-like cliffs in $\boldsymbol k_\eta$-direction along the boundary seam, as Fig.~\ref{fig:sigmamin}(g) clearly illustrates where the geometric frustration is concentrated.
	The magnitude of this geometric defect is measured by the floor of $\sigma_{\min}(\boldsymbol{k};\delta\boldsymbol{k}_{\beta})$,
	which is likewise $L$-independent and of the order of $\mathcal{O}(L^0)$, as explicitly shown in Fig.~\ref{fig:omega(k)}(b).
	
	The linear growth of the total spread $\Omega^1$ is therefore not caused by a local divergence of the spread kernel.
	Rather, what grows with $L$ is the macroscopic number of seam links that carry this finite defect.
	Using the discrete reciprocal-space measure $\nu_{\beta}=w_{\beta}/N_{\boldsymbol{k}}$ introduced in Appendix~\ref{ap:kspace-def}, the seam contribution to the spread is estimated as
	\begin{equation}
		\Delta\Omega_{\mathrm{seam}}^1
		\sim
		\sum_{\boldsymbol{k}\in\mathrm{seam}} \nu_{\beta}\,\omega_{\beta}^1(\boldsymbol{k}).
		\label{eq:result-seam-sum-clean}
	\end{equation}
	On a uniform $L\times L$ mesh, the discrete measure evaluates to $\nu_{\beta}=\mathcal{O}(L^0)$, and the number of grid points along the 1D boundary seam is $\#\{\boldsymbol{k}\in\mathrm{seam}\}=\mathcal{O}(L)$.
	As rigorously evaluated in Appendix~\ref{ap:border-term} and shown in Fig.~\ref{fig:omega(k)}, both the single-state and subspace-level diagnostics remain strictly finite ($\mathcal{O}(L^0)$) on this seam. Thus, the macroscopic integration strictly yields a linear divergence:
	\begin{equation}
		\Delta\Omega_{\mathrm{seam}}^1=\mathcal{O}(L).
		\label{eq:seam-linear-scaling-clean}
	\end{equation}
	The observed scaling in Fig.~\ref{fig:spread} is therefore a geometric consequence of a boundary-supported $\mathcal{O}(L^0)$ mismatch accumulated over an $\mathcal{O}(L)$ seam, not a numerical artifact.
	Ultimately, this $\mathcal{O}(L)$ spread scaling is an intrinsic physical signature of the 2D gauge frustration stemming from the non-commutativity of the projected position operators \cite{MarzariVanderbilt1997, Brouder2007}. As originally established by MV, iterative minimization schemes handle this non-commutativity variationally---distributing the geometric incompatibility globally across all reciprocal directions to achieve a minimum total spread \cite{MarzariVanderbilt1997}. In stark contrast, our deterministic sequential extraction mathematically forces the gauge to be flat along the chosen one-dimensional strings. This strict 1D flattening inevitably sweeps the entire geometric frustration into the 1D boundary seams, manifesting perfectly as the $\mathcal{O}(L)$ scaling. While subsequent multidimensional iterative optimization \cite{MarzariVanderbilt1997,Wannier90_2008} could be applied to redistribute this concentrated defect and reduce the total spread, isolating this exact geometric manifestation falls precisely within the scope of our constructive and analytical framework.
	\section{SUMMARY}\label{sec:summary}
	We proposed a non-variational constructive algorithm to build maximally localized Wannier
	functions (MLWFs) that unifies gauge smoothing and the eigenvalue problem of the
	projected position operator into a single deterministic framework. Rather than
	treating gauge alignment and center determination as separate steps or minimizing the spread functional over the gauge manifold, we demonstrated that discrete Kato parallel
	transport across band degeneracies emerges naturally as an integral part of the
	solution procedure for the position eigenvectors.
	
	By directly aligning the periodic parts of the Bloch functions, this transport
	suppresses band-index swaps and approximately linearizes the overlap phase in the interior of the Brillouin zone.
	Operating within this transport-aligned gauge, we introduced a sinc-kernel
	transformation that maps the Wannier center search to deterministic fixed-point iterations and self-consistent updates for explicit center equations and projected-position matrices, rather than spread-functional minimization over the gauge manifold.
	
	Benchmarks on 1D models, a 2D square potential, and graphene yield Wannier
	centers and orbital shapes in good agreement with standard gradient-based
	minimization (e.g., W90). Furthermore, because our sequential extraction works
	to flatten the interior gauge instead of globally distributing the geometric
	mismatch, it transparently isolates the physical origin of the $\mathcal{O}(L)$
	spread growth observed in graphene. The finite $\mathcal{O}(L^0)$ local defects
	accumulated on the one-dimensional boundary seam strictly integrate to an
	$\mathcal{O}(L)$ macroscopic spread growth. This confirms that the
	resolution-dependent spread divergence is an intrinsic geometric manifestation
	of non-commuting projected position operators. Remaining challenges include
	addressing the non-uniqueness of projection directions for co-centered WFs and
	managing directional incompatibilities during multi-step extractions.
	\section*{Acknowledgments}
	This work was supported by JSPS KAKENHI (Grants No. JP25K01609, No. JP22H05473,
	and No. JP21H01019), JST CREST (Grant No. JPMJCR19T1). K.W. acknowledges the
	financial support for Basic Science Research Projects (Grant No. 2401203) from
	the Sumitomo Foundation and a special individual research fund from Kwansei Gakuin University.
	\appendix
	\section{Supplemental Calculation for Adiabatic Expansion }\label{ap:adiabatic-supplement} 
	The eigenvalue equation of the system is defined as follows:
	\begin{equation}
		\begin{array}{c}
			\begin{aligned}\hat{H}(k)\vert v_{\boldsymbol{k}}^n\rangle & =\epsilon^n(t)\vert v_{\boldsymbol{k}}^n\rangle,\\
			\end{aligned}
		\end{array}
	\end{equation}
	where $\epsilon_{n}(t)$ is the instantaneous eigenvalue of the Bloch Hamiltonian.
	The transient solution of the system satisfies the following equation:
	\begin{equation}
		i\hbar\frac{d}{dt}\vert\Psi(t)\rangle=\hat{H}(k)\vert\Psi(t)\rangle,
	\end{equation}
	where the solution is decomposed into the instantaneous eigenstates of the system:
	
	\begin{equation}
		\begin{array}{c}
			\begin{aligned}\vert\Psi(t)\rangle & =\sum_{n}c_{\boldsymbol{k}}^n(t)e^{-i\theta^n(t)}\vert v_{\boldsymbol{k}}^n\rangle,
			\end{aligned}
		\end{array}
	\end{equation} 
	where, 
	\begin{equation}
		\begin{array}{c}
			\begin{aligned}
				\theta^n(t) & =\frac{1}{\hbar}\int^t\epsilon^n(\tau)d\tau.
			\end{aligned}
		\end{array}
	\end{equation} 
	
	The time derivative of $\vert \Psi(t)\rangle$ is calculated in two ways:
	\begin{equation}
		\begin{aligned}
			\frac{d}{dt}&\vert\Psi(t)\rangle\\
			&=\sum_{n}\left\{ \frac{dc_{\boldsymbol{k}}^n}{dt}(t)\vert v_{\boldsymbol{k}}^n\rangle-\frac{i}{\hbar}c_{\boldsymbol{k}}^n(t)\epsilon_{n}(t)\vert v_{\boldsymbol{k}}^n\rangle+c_{\boldsymbol{k}}^n\frac{d}{dt}\vert v_{\boldsymbol{k}}^n\rangle\right\} e^{-i\theta^n(t)}\\
			\frac{d}{dt}&\vert\Psi(t)\rangle =-\frac{i}{\hbar}\sum c_{\boldsymbol{k}}^n\epsilon_{n}\vert v_{\boldsymbol{k}}^n\rangle e^{-i\theta^n(t)}.
		\end{aligned}
	\end{equation}
	Hence, we obtain the equation for $c^n_{\boldsymbol k}$:
	\begin{equation}
		\begin{aligned}\sum_{n}\frac{dc_{\boldsymbol{k}}^n}{dt}(t)\vert v_{\boldsymbol{k}}^n\rangle e^{-i\theta^n(t)} & =-\sum_{n}c_{\boldsymbol{k}}^n(t)\vert\frac{dv_{\boldsymbol{k}}^n}{dt}(t)\rangle e^{-i\theta^n(t)}.
		\end{aligned}
	\end{equation}
	By multiplying $\langle v_{\boldsymbol{k}}^m\vert$ from left, we have: 
	\begin{equation}
		\begin{aligned}
			\frac{dc_{\boldsymbol{k}}^m}{dt}(t) & =-\sum_{n}c_{\boldsymbol{k}}^n(t)\langle v_{\boldsymbol{k}}^m\vert\frac{dv_{\boldsymbol{k}}^n}{dt}(t)\rangle e^{i( \theta^m(t)-\theta^n(t) )},
		\end{aligned}
	\end{equation}
	thus the first order forward differencing in time yields:
	\begin{equation}
		\begin{aligned}
			c_{\boldsymbol{k}}^m(t+\delta t)&=c_{\boldsymbol{k}}^m(t)\\
			&\quad-\delta t\sum_{n}c_{\boldsymbol{k}}^n(t)\langle v_{\boldsymbol{k}}^m\vert\frac{d v_{\boldsymbol{k}}^n}{dt}(t)\rangle e^{i(\theta^m(t)-\theta^n(t))}
			+\mathcal O(\delta t^2).
		\end{aligned}
	\end{equation}
	Hence, we obtain
	\begin{equation}
		\begin{aligned}
			c_{\boldsymbol{k}+\delta\boldsymbol{k}}^m & =c_{\boldsymbol{k}}^m-\sum_{n}c_{\boldsymbol{k}}^n\delta\boldsymbol{k} \cdot \langle v_{\boldsymbol{k}}^m\vert\nabla_{\boldsymbol{k}}\vert v_{\boldsymbol{k}}^n\rangle e^{i(\theta^m(t)-\theta^n(t))}\\
			& =c_{\boldsymbol{k}}^m+
			\sum_{n}c_{\boldsymbol{k}}^n\delta\boldsymbol{k}\cdot\langle\nabla_{\boldsymbol{k}} v_{\boldsymbol{k}}^m\vert v_{\boldsymbol{k}}^n\rangle e^{i(\theta^m(t)-\theta^n(t))}\\
			& =c_{\boldsymbol{k}}^m+\sum_{n}c_{\boldsymbol{k}}^n\left\{ \langle v_{\boldsymbol{k}+\delta\boldsymbol{k}}^m\vert v_{\boldsymbol{k}}^n\rangle-\langle v_{\boldsymbol{k}}^m\vert v_{\boldsymbol{k}}^n\rangle\right\} e^{i(\theta^m(t)-\theta^n(t))}\\
			& =c_{\boldsymbol{k}}^m+\sum_{n}c_{\boldsymbol{k}}^n\left\{ \langle v_{\boldsymbol{k}+\delta\boldsymbol{k}}^m\vert v_{\boldsymbol{k}}^n\rangle-\delta_{m,n}\right\} e^{i(\theta^m(t)-\theta^n(t))}\\
			& =\sum_{n}c_{\boldsymbol{k}}^n\langle v_{\boldsymbol{k}+\delta\boldsymbol{k}}^m\vert v_{\boldsymbol{k}}^n\rangle e^{i(\theta^m(t)-\theta^n(t))}+\mathcal O(\vert\delta\boldsymbol{k}\vert^2),\\
		\end{aligned}
	\end{equation}
	and hence:
	\begin{equation}
		\begin{aligned}
			c_{\boldsymbol{k}+\delta\boldsymbol{k}}^m & =\sum_{n}c_{\boldsymbol{k}}^n\langle v_{\boldsymbol{k}+\delta\boldsymbol{k}}^m\vert v_{\boldsymbol{k}}^n\rangle e^{i(\theta^m(t)-\theta^n(t))}+\mathcal O(\vert\delta\boldsymbol{k}\vert^2).
		\end{aligned}
	\end{equation}
	The general solution at $t=t+\delta t$ or $\boldsymbol k=\boldsymbol k+\delta \boldsymbol k$ is therefore
	\begin{equation}
		\vert \Psi(t+\delta t)\rangle
		=\sum_{m}c_{\boldsymbol{k}+\delta\boldsymbol{k}}^m\vert v_{\boldsymbol{k}+\delta\boldsymbol{k}}^m\rangle e^{-i\theta^m(t)}.
	\end{equation} 
	If the initial condition is,
	\begin{equation}
		c^n_{\boldsymbol k} = \delta_{n,n_0},
	\end{equation} 
	then, after removing the dynamical phase factor, we obtain
	\begin{equation}
		\begin{aligned}
			\vert u_{\boldsymbol{k}+\delta\boldsymbol{k}}^{n_{0}}\rangle
			&=
			\left\{ \sum_{m}\vert v_{\boldsymbol{k}+\delta\boldsymbol{k}}^m\rangle\langle v_{\boldsymbol{k}+\delta\boldsymbol{k}}^m\vert\right\}
			\vert u_{\boldsymbol{k}}^{n_{0}}\rangle
			+\mathcal O(\vert\delta\boldsymbol{k}\vert^2),
		\end{aligned}
	\end{equation} 
	which reproduces Eq.~(\ref{eq:composite-adiabatic-peeling}).
	\section{Equation for Berry Connection Derived from Eigenvalue Equation for Translation Operator}\label{ap:xeigen-supplement}
	When WFs defined by Eq.~(\ref{eq:wannier_series_def}) satisfy:
	\begin{equation}\label{eqap:single_wannier_x_eigen}
		\langle\psi_{\boldsymbol{k}}^m\vert e^{i\delta\boldsymbol{k}\cdot\hat{\boldsymbol{x}}}\vert W_{0}^s\rangle=e^{i\delta\boldsymbol{k}\cdot\boldsymbol{x}_{0}}\langle\psi_{\boldsymbol{k}}^m\vert W_{0}^s\rangle,\boldsymbol{x}_{0}\in[0,1)^d 
	\end{equation}
	for $\boldsymbol k \in \mathbb K_I$: 
	\begin{equation}\label{eq:wannier_shift}
		\begin{aligned}
			L\langle\psi_{\boldsymbol{k}}^m\vert e^{i\delta\boldsymbol{k}\cdot\hat{\boldsymbol{x}}}\vert W_{0}^s\rangle&=\sqrt{L}\langle\psi_{\boldsymbol{k}}^m\vert\sum_{\boldsymbol{p}}e^{i(\boldsymbol{p}+\delta\boldsymbol{k})\cdot\hat{\boldsymbol{x}}}\vert u_{\boldsymbol{p}}^s\rangle\\&=\langle v_{\boldsymbol{k}}^m\vert e^{-i\boldsymbol{k}\cdot\hat{\boldsymbol{x}}}\sum_{\boldsymbol{p}}e^{i\boldsymbol{p}\cdot\hat{\boldsymbol{x}}}\vert u_{\boldsymbol{p}-\delta\boldsymbol{k}}^s\rangle\\&=\langle v_{\boldsymbol{k}}^m\vert u_{\boldsymbol{k}-\delta\boldsymbol{k}}^s\rangle\\&=\langle v_{\boldsymbol{k}}^m\vert\sum_{n}f_{\boldsymbol{k}-\delta\boldsymbol{k}}^{s,n}\vert v_{\boldsymbol{k}-\delta\boldsymbol{k}}^n\rangle\\&=\sum_{n}f_{\boldsymbol{k}-\delta\boldsymbol{k}}^{s,n}\langle v_{\boldsymbol{k}}^m\vert v_{\boldsymbol{k}-\delta\boldsymbol{k}}^n\rangle,
		\end{aligned}
	\end{equation}
	and also, 
	\begin{equation}
		\begin{aligned}
			L\langle\psi_{\boldsymbol{k}}^m\vert e^{i\delta\boldsymbol{k}\cdot\hat{\boldsymbol{x}}}\vert W_{0}^s\rangle&=e^{i\delta\boldsymbol{k}\cdot\boldsymbol{x}_{0}}\langle v_{\boldsymbol{k}}^m\vert e^{-i\boldsymbol{k}\cdot\hat{\boldsymbol{x}}}\sum_{\boldsymbol{p}}e^{i\boldsymbol{p}\cdot\hat{\boldsymbol{x}}}\vert u_{\boldsymbol{p}}^s\rangle\\&=e^{i\delta\boldsymbol{k}\cdot\boldsymbol{x}_{0}}\langle v_{\boldsymbol{k}}^m\vert u_{\boldsymbol{k}}^s\rangle\\&=e^{i\delta\boldsymbol{k}\cdot\boldsymbol{x}_{0}}\langle v_{\boldsymbol{k}}^m\vert\sum_{n}f_{\boldsymbol{k}}^{s,n}\vert v_{\boldsymbol{k}}^n\rangle\\&=e^{i\delta\boldsymbol{k}\cdot\boldsymbol{x}_{0}}f_{\boldsymbol{k}}^{s,m},
		\end{aligned}
	\end{equation}
	
	and hence, it leads to the following equation composed of the periodic parts:
	\begin{equation}\tag{\ref{eq:composite-final-eigenvalue-equation}}
		f_{\boldsymbol{k}}^{s,m}=e^{-i\delta \boldsymbol k \cdot \boldsymbol x_{0}}\sum_{n}f_{\boldsymbol{k}-\delta\boldsymbol{k}}^{s,n}\langle v_{\boldsymbol{k}}^m\vert v_{\boldsymbol{k}-\delta\boldsymbol{k}}^n\rangle+\mathcal O(\vert\delta\boldsymbol{k}\vert^2),
	\end{equation}

	By multiplying $\vert v_{\boldsymbol{k}}^n\rangle$ to both sides and 
	summing over $n$, the energy bands are reconstructed in the following way: 
	\begin{equation}\label{eq:composite-peeling-equation}
		\begin{aligned}\vert u_{\boldsymbol{k}}^s\rangle & =\sum_{m}f_{\boldsymbol{k}}^{s,m}\vert v_{\boldsymbol{k}}^m\rangle\\
			& =\sum_{m}\left\{ \sum_{n}f_{\boldsymbol{k}-\delta\boldsymbol{k}}^{s,n}\langle v_{\boldsymbol{k}}^m\vert v_{\boldsymbol{k}-\delta\boldsymbol{k}}^n\rangle\right\} \vert v_{\boldsymbol{k}}^m\rangle e^{-i\delta \boldsymbol k \cdot \boldsymbol x_{0}}\\
			& =\left\{ \sum_{m}\vert v_{\boldsymbol{k}}^m\rangle\langle v_{\boldsymbol{k}}^m\vert\right\} \left\{ \sum_{n}\vert v_{\boldsymbol{k}-\delta\boldsymbol{k}}^n\rangle f_{\boldsymbol{k}-\delta\boldsymbol{k}}^{s,n} e^{-i\delta \boldsymbol k \cdot \boldsymbol x_{0}}\right\} \\
			& = e^{-i\delta \boldsymbol k \cdot \boldsymbol x_{0}}\left\{ \sum_{m}\vert v_{\boldsymbol{k}}^m\rangle\langle v_{\boldsymbol{k}}^m\vert\right\} \vert u_{\boldsymbol{k}-\delta\boldsymbol{k}}^s\rangle+\mathcal O(\vert\delta\boldsymbol{k}\vert^2),
		\end{aligned}
	\end{equation}
	which gives the following projection equation:
	\begin{equation}\tag{\ref{eq:composite-peeling-projection}}
		\begin{aligned}
			\vert u_{\boldsymbol{k}}^s\rangle & = e^{-i\delta \boldsymbol k \cdot \boldsymbol x_{0}}\left\{ \sum_{m}\vert v_{\boldsymbol{k}}^m\rangle\langle v_{\boldsymbol{k}}^m\vert\right\} \vert u_{\boldsymbol{k}-\delta\boldsymbol{k}}^s\rangle+\mathcal O(\vert\delta\boldsymbol{k}\vert^2),
		\end{aligned}
	\end{equation}
	which is also nearly identical to:
	\begin{equation}
		\begin{aligned}	
			\vert u_{\boldsymbol{k}+\delta\boldsymbol{k}}^{n_{0}}\rangle&=\vert\Psi(t+\delta t)\rangle\\
			&=\left\{ \sum_{m}\vert u_{\boldsymbol{k}+\delta\boldsymbol{k}}^m\rangle\langle u_{\boldsymbol{k}+\delta\boldsymbol{k}}^m\vert\right\} \vert u_{\boldsymbol{k}}^{n_{0}}\rangle+\mathcal O(\delta\boldsymbol{k}^2).
		\end{aligned}
		\tag{\ref{eq:composite-adiabatic-peeling}}
	\end{equation}
	\section{Supplemental Calculation Related to Kato's Formula }\label{ap:Kato}
	\subsection{Supplemental Calculation to Kato's Formula}\label{ap:kato} 
	Since 
	\begin{equation}
		\begin{aligned}-i\hat{P}(\boldsymbol{k}+\delta\boldsymbol{k})\hat{E}(\boldsymbol{k})\hat{P}(\boldsymbol{k})
			& =\hat{P}(t+\delta t)\left[\dot{\hat{P}}(t),\hat{P}(t)\right]\hat{P}(t)\\
			& =\hat{P}(t+\delta t)\dot{\hat{P}}(t)\hat{P}(t)\hat{P}(t)\\
			&\quad\quad-\hat{P}(t+\delta t)\hat{P}(t)\dot{\hat{P}}(t)\hat{P}(t)\\
			& =\left(\hat{P}(t)+\delta t\dot{\hat{P}}(t)+\mathcal O(\delta t^2)\right)\dot{\hat{P}}(t)\hat{P}(t)\\
			& =\delta t\dot{\hat{P}}(t)\dot{\hat{P}}(t)\hat{P}(t)+\mathcal O(\delta t^2)\\
			& =\delta t\dot{\boldsymbol{k}}\nabla \hat{P}(\boldsymbol{k})\dot{\boldsymbol{k}}\nabla \hat{P}(\boldsymbol{k})\hat{P}(t)+\mathcal O(\delta t^2)\\
			&=\mathcal O(\delta \boldsymbol k^2),
		\end{aligned}
	\end{equation}
	and,
	\begin{equation}
		\begin{aligned}
			\hat U(\boldsymbol{k}) 
			& =I-i\delta t\hat E(\boldsymbol k)+\mathcal O(\delta k^2),\\
		\end{aligned}
	\end{equation}
	the following holds,
	\begin{equation}
		\begin{aligned}
			\hat{P}(\boldsymbol{k}+\delta\boldsymbol{k})U(\boldsymbol{k})\hat{P}(\boldsymbol{k}) 
			& =\hat{P}(\boldsymbol{k}+\delta\boldsymbol{k})\hat{P}(\boldsymbol{k})+\left(\delta\boldsymbol{k}\nabla \hat{P}(\boldsymbol{k})\right)^2\hat{P}(\boldsymbol{k})\\
			& =\hat{P}(\boldsymbol{k}+\delta\boldsymbol{k})\hat{P}(\boldsymbol{k})+\mathcal O(\vert\delta\boldsymbol{k}\vert^2).
		\end{aligned}
	\end{equation}
	\subsection{Supplemental Calculation for (\ref{eq:k+dk-orthogonal})}\label{ap:k+dk-orthogonal}
	By definition,
	\begin{equation}
		\hat{P}(\boldsymbol{k})=\hat{P}(\boldsymbol{k}(t))=\hat{P}(t),
	\end{equation}
	where,
	\begin{equation}
		\hat{P}(\boldsymbol{k})=\sum_n \vert v_{\boldsymbol k}^n\rangle \langle v_{\boldsymbol k}^n\vert.
	\end{equation}
	Since,
	\begin{equation}
		\begin{aligned}
			&\hat{P}(t)\hat{P}(t) =\hat{P}(t), \\
			&\frac{d}{dt}\left(\hat{P}(t)\hat{P}(t)\right) =\dot{\hat{P}}(t)\hat{P}(t)+\hat{P}(t)\dot{\hat{P}}(t).
		\end{aligned}
	\end{equation}
	the following holds,
	\begin{equation}
		\dot{\hat{P}}(\boldsymbol{k})=\hat{P}(\boldsymbol{k})\dot{\hat{P}}(\boldsymbol{k})+\dot{\hat{P}}(\boldsymbol{k})\hat{P}(\boldsymbol{k}),
	\end{equation}
	where $\hat{P}(\boldsymbol{k}) \in C^1$ is assumed.
	By multiplying $\hat{P}(\boldsymbol k)$ on both sides of the above equation, we have,
	\begin{equation}
		\hat{P}(\boldsymbol{k})\dot{\hat{P}}(\boldsymbol{k})\hat{P}(\boldsymbol{k})=0.
	\end{equation}
	Thus,
	\begin{equation}
		\begin{aligned}
			\hat{P}(\boldsymbol{k}+\delta\boldsymbol{k})&=\hat{P}(\boldsymbol{k})+\delta t\dot{\hat{P}}(\boldsymbol{k})+\mathcal{O}(\delta \boldsymbol k^2)\\\hat{P}(\boldsymbol{k})\hat{P}(\boldsymbol{k}+\delta\boldsymbol{k})\hat{P}(\boldsymbol{k})&=\hat{P}(\boldsymbol{k})+\delta t\hat{P}(\boldsymbol{k})\dot{\hat{P}}(\boldsymbol{k})\hat{P}(\boldsymbol{k})+\mathcal{O}(\delta \boldsymbol k^2)\\&=\hat{P}(\boldsymbol{k})+\mathcal{O}(\delta \boldsymbol k^2).
		\end{aligned}
	\end{equation}
	And hence, 
	if the following holds,
	\begin{equation}
		\langle u_{\boldsymbol{k}}^{n_{0}}\vert u_{\boldsymbol{k}}^{n_{1}}\rangle=\delta_{n_{0},n_{1}},
	\end{equation}
	the equation below follows:
	\begin{equation}\label{eqap:delta_kk}
		\begin{split}
			\langle u_{\boldsymbol{k}+\delta\boldsymbol{k}}^{s_{0}}\vert u_{\boldsymbol{k}+\delta\boldsymbol{k}}^{s_{1}}\rangle
			&=\langle u_{\boldsymbol{k}}^{s_{0}}\vert\left\{ \sum_{m_{0}}\vert v_{\boldsymbol{k}+\delta\boldsymbol{k}}^{m_{0}}\rangle\langle v_{\boldsymbol{k}+\delta\boldsymbol{k}}^{m_{0}}\vert \right\}\\
			&\quad
			\left\{\sum_{m_{0},m_{1}} \vert v_{\boldsymbol{k}+\delta\boldsymbol{k}}^{m_{1}}\rangle\langle v_{\boldsymbol{k}+\delta\boldsymbol{k}}^{m_{1}}\vert\right\} \vert u_{\boldsymbol{k}}^{s_{1}}\rangle\\
			&=\langle u_{\boldsymbol{k}}^{s_{0}}\vert \hat{P}(\boldsymbol{k}+\delta\boldsymbol{k})\vert u_{\boldsymbol{k}}^{s_{1}}\rangle\\
			&=\langle u_{\boldsymbol{k}}^{s_{0}}\vert \hat{P}(\boldsymbol{k})\hat{P}(\boldsymbol{k}+\delta\boldsymbol{k})\hat{P}(\boldsymbol{k})\vert u_{\boldsymbol{k}}^{s_{1}}\rangle\\
			&=\langle u_{\boldsymbol{k}}^{s_{0}}\vert \hat{P}(\boldsymbol{k})\vert u_{\boldsymbol{k}}^{s_{1}}\rangle+\mathcal{O}(\delta \boldsymbol k^2)\\
			&=\delta_{s_{0},s_{1}}+\mathcal{O}(\delta \boldsymbol k^2).
		\end{split}
	\end{equation}
	\section{Sinc and Wannier Centers}\label{ap:sinc-wannier-centers}
	\subsection{Calculation related to sinc }\label{ap:sinc}
	In one dimension, the expansion of a WF by another set of WF
	becomes as follows:
	\begin{equation}\label{eqap:sinc-r-s}
		\begin{aligned}
			\vert W_{n}(s)\rangle &=\sum_{m}\vert W_{m}(r)\rangle\langle W_{m}(r)\vert W_{n}(s)\rangle\\
			&=\sum_{m}\left\{ \frac{1}{L}\sum_{k}e^{ik(m-n-s+r)}\right\} \vert W_{m}(r)\rangle\\
			&=\sum_{m}\mathrm{sinc}(m-n+r-s)\vert M_{m}(r)\rangle,
		\end{aligned}
	\end{equation}
	When the summation over the k-space is approximated by an integral, the following is used in the present paper:
	\begin{equation}\label{apeq:derv-sinc}
		\begin{aligned}
			\lim_{\delta k \rightarrow 0 }\frac{1}{L}\sum_k e^{ik(x -m) }
			&=\lim_{ L\rightarrow\infty }\frac{1}{2\pi} \sum_n e^{i\frac{2\pi n}{L}(x -m) } \left( \frac{2\pi}{L} \right) \\
			&=\frac{1}{2\pi} \int_{-\pi}^\pi e^{ik(x -m) }dk\\
			&=\mathrm{sinc} (x -m).
		\end{aligned}
	\end{equation}
	The following identities are used to calculate the matrix element transformation:
	\begin{equation}\label{eqap:sinc_sum}
		\begin{aligned}
			&\sum_{m}\mathrm{sinc}(s+m-n)\mathrm{sinc}(s+m)=\delta_{n,0}.
		\end{aligned}
	\end{equation}
	The above equation is easily checked by putting it back to the form of Eq.~(\ref{apeq:derv-sinc}).
	
	By utilizing the following identity~\cite{cotangent},
	\begin{equation}
		\sum_m \frac{1}{m+x}=\pi\cot\pi x,
	\end{equation}
	the following summation rule is obtained:
	\begin{equation}\label{eqap:sinc^2_sum}
		\begin{aligned}
			\sum_m(m+x)\mathrm{sinc}^2(m+x)
			&=\sum(m+x)\frac{\sin^2\pi x}{\pi^2(m+x)^2}\\
			&=\frac{\sin^2\pi x}{\pi^2}\sum\frac{1}{m+x}\\&=\frac{\sin^2\pi x}{\pi^2}\pi\cot\pi x\\
			&=\frac{1}{2\pi}\sin2\pi x,
		\end{aligned}
	\end{equation}
	\subsection{Convergence of $r^n$ to $x_0$}\label{ap:r-converges-to-x_0}
	The convergence of the iteration is examined by introducing the following function: 
	\begin{equation}
		F(r)=r+\frac{1}{2\pi}\sin\!\bigl(2\pi(x_0-r)\bigr)\,(\bmod\,1).
	\end{equation}
	Since 
	\begin{equation}\label{eq:dFdr}
		\bigl\vert \frac{dF(r)}{dr} \bigr\vert = 1-\cos\!\bigl(2\pi(x_0-r)\bigr) < 1,\quad ( 0 < x_0 - r < 1/4 ),
	\end{equation}
	$x_0$ is an attractor~\cite{Devaney2003}.
	Therefore, letting,
	\begin{equation}
		e_n=2\pi(x_0-r^n)\in(-\pi,\pi],
	\end{equation} 
	we have,
	\begin{equation}\label{en+1=en+}
		e_{n+1}=e_n-\sin e_n,\qquad
		|e_{n+1}| \le |e_n|\ \text{for } |e_n|<\pi,
	\end{equation}
	Since 
	\begin{equation}
		|e_n|<\pi,
	\end{equation}
	we find,
	\begin{equation}
		|e_{n+1}|<|e_n|,
	\end{equation}
	with equality only at \(e_n=\pm\pi\). This means the error becomes always
	smaller in any step of the iteration. Therefore, 
	\begin{equation}\tag{\ref{eq:x_0-convergence}}
		\lim_{n\rightarrow\infty} r^n = x_0.
	\end{equation}
	The order of convergence is cubic. When $e_n << 1$, from Eq.~(\ref{en+1=en+}),
	\begin{equation}
		\begin{split}
			e_{n+1}&=e_n-( e_n - \frac{1}{6}e_n^3 )\\
			&=\frac{1}{6}e_n^3
		\end{split}
	\end{equation}
	and hence, 
	\begin{equation}
		\lim_{e_n\to 0}\frac{|e_{n+1}|}{|e_n|^3}=\frac{1}{6}.
	\end{equation}
	\subsection{ Direct Derivation of Sinc-Loop }\label{ap:direct-derivation}
	\begin{figure*}[ht]
		\centering
		\includegraphics[width=\linewidth]{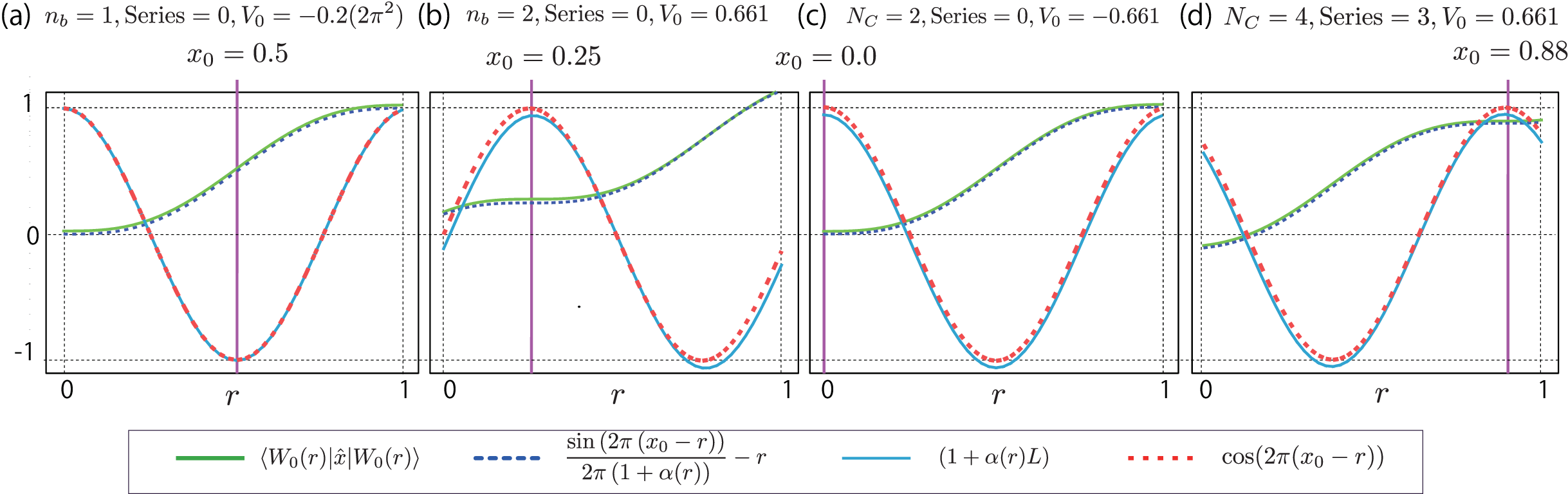}
		\caption{(Color online)
			Verification of Eq.~(\ref{eqap:gauge_check}) in single- and composite-band
			cases. $N_C$ denotes the number of the energy bands composing the composite band
			systems. $n_b$ indicates the energy band index of the single-band system. The
			series numbers for composite-band cases are labeled \#0, \#1, \#2, \#3 from
			the left side of a unit cell. Panels~(a) to (d) correspond to
			Figs.~\ref{fig:fraciona_sinc}(a), \ref{fig:fraciona_sinc}(b),
			\ref{fig:sel-sinc}(a) and \ref{fig:sel-sinc}(b). The MLWFs are found in Figs.~\ref{fig:fraciona_wannier} and \ref{fig:sel-wannier}.
		}\label{fig:gaugepictureinpaper}
	\end{figure*}
	Since the initial \textit{corrugated} Berry
	connection is swept flat, the swell developed over the border carries all the information.
	By segregating the \textit{cross-border} term out of the summation, the matrix element of $e^{i\delta k \hat x }$ is estimated by solving the following simplified equations: 
	\begin{equation}\label{eqap:ex-elem}
		\begin{split}
			&L\langle W_{0}(r)\vert e^{i\delta k\hat{x}}\vert W_{0}(r)\rangle
			=\!\!\!\!\!\!\sum_{k=-\pi+\delta k}^{\pi-\delta k}\!\!\!\!\!\!\langle u_{k}\vert u_{k-\delta k}\rangle e^{i\delta kr}\\
			&\quad\quad\quad\quad\quad\quad\quad\quad\quad	+\langle u_{-\pi}^0\vert e^{2i\pi \hat x}\vert u_{\pi-\delta k}^0\rangle e^{-2i\pi r}.
		\end{split}
	\end{equation}
	And we assume the phases of $\{\vert u_{k}\rangle\}$ are aligned and,
	\begin{equation}
		\langle u_{k}\vert u_{k-\delta k}\rangle = 1, ( k \in \mathbb K_I ),
	\end{equation}
	and, 
	\begin{equation}\label{eqap:ex-elem-2}
		L\langle W_{0}(r)\vert e^{i\delta k\hat{x}}\vert W_{0}(r)\rangle=\left\{ \begin{array}{c}
			L(1+\alpha(r))e^{i\delta kX_{0,0}(r)}\ (r\ne x_{0})\\
			L(1+\alpha(x_{0}))e^{i\delta kx_{0}}\ (r=x_{0})
		\end{array}\right.,
	\end{equation}
	where	$\alpha(r)$ is a factor expressing the deviation of $e^{i\delta k\hat{x}}$ from being unitary.
	\begin{equation}\label{eqap:def-alpha}
		\alpha(r) = \left| \langle W_{0}(r)\vert e^{i\delta k\hat{x}}\vert W_{0}(r)\rangle \right| - 1
		\sim \mathcal O(\delta k),
	\end{equation}	
	and by definition,
	\begin{equation}\label{eqap:alpha=0}
		\alpha(x_0) = 0.
	\end{equation}
	Except for the cross-border term, the right-hand-side of Eq.~(\ref{eqap:ex-elem}) is
	basically proportional to $e^{i\delta k r }$ regardless of the value of $r$.
	Thus, the task is to find $x_0$ which least disturbs the proportionality.
	By simultaneously solving Eqs.~(\ref{eqap:ex-elem}) and (\ref{eqap:ex-elem-2}) for $r=x_0$ and $r\ne x_0$, we have:
	\begin{equation}
		\begin{split}
			L(1+\alpha(r))e^{i\delta kX_{0,0}(r)}&=\left(L-1\right)e^{i\delta kr}+\langle u_{-\pi}^0\vert e^{2i\pi\hat{x}}\vert u_{\pi-\delta k}^0\rangle e^{-2i\pi r}\\L(1+\alpha(x_{0}))e^{i\delta kx_{0}}&=\left(L-1\right)e^{i\delta kx_{0}}+\langle u_{-\pi}^0\vert e^{2i\pi\hat{x}}\vert u_{\pi-\delta k}^0\rangle e^{-2i\pi x_{0}}.
		\end{split}
	\end{equation}
	By using $\delta k L=2\pi$, we have, 
	\begin{equation}\label{eqap:gauge_check}
		1+\alpha(r) L+i(1+\alpha(r))2\pi X_{00}(r)-2\pi ir=e^{2\pi ix_{0}}+O(\delta k).
	\end{equation}
	And hence, 
	\begin{equation}\label{eqap:gauge_sin_cos}
		\begin{split}
			&(1+\alpha(x_0))2\pi (X_{0,0}(r)-r)=\sin2\pi(x_{0}-r)\\
			&1+\alpha(r) L=\cos2\pi(x_{0}-r).
		\end{split}
	\end{equation}
	The first line of Eq.~(\ref{eqap:gauge_sin_cos}) is a reproduction of 
	Eq.~(\ref{eq:sinc-00}) with the deviation factor $\alpha(r)\sim \mathcal O(\delta k)$.
	And the second line shows
	the relationship between $\alpha(r)$ and $r$. The verification of 
	Eq.~(\ref{eqap:gauge_sin_cos}) is shown in Fig.~\ref{fig:gaugepictureinpaper}.
	\begin{widetext}
		\subsection{Supplemental Calculation for Eq.~(\ref{eq:sinc-00}) }\label{ap:multi-sinc} 
		For,
		\begin{equation}
			\begin{aligned}
				X_{\boldsymbol{M}_{1},\boldsymbol{M}_{2}}(\boldsymbol{x}_{0})\!=\delta_{M_{1,x},M_{2,x}}\delta_{M_{1,y},M_{2,y}}\delta_{M_{1,z},M_{2,z}}(M_{1,x}+x_{0})\end{aligned},
		\end{equation} 
		$X_{\boldsymbol 0\boldsymbol 0}(\boldsymbol r)$ is calculated as follows:
		\begin{equation}\label{eq:sinc-00-3d}
			\begin{aligned}
				X_{\boldsymbol 0\boldsymbol 0}(\boldsymbol r)
				&= \bigl\langle W_{\boldsymbol 0}(\boldsymbol r)\bigr|\hat x
				\bigl|W_{\boldsymbol 0}(\boldsymbol r)\bigr\rangle\\
				& =\left\{ \sum_{N_{1,x},N_{2,x}}\textrm{sinc}(N_{1,x}+r_{x}-x_{0})\textrm{sinc}(N_{2,x}+r_{x}-x_{0})\delta_{N_{1,x},N_{2,x}}(N_{1,x}+x_{0})\right\} \\
				& \quad\left\{ \sum_{N_{1,y},N_{2,y}}\textrm{sinc}(N_{1,y}+r_{y})\textrm{sinc}(N_{2,y}+r_{y})\delta_{N_{1,y},N_{2,y}}\right\} \left\{ \sum_{N_{1,z},N_{2,z}}\textrm{sinc}(N_{1,z}+r_{z})\textrm{sinc}(N_{2,z}+r_{z})\delta_{N_{1,z},N_{2,z}}\right\} \\
				& =\left\{ \sum_{N_{1,x}}\textrm{sinc}^2(N_{1,x}+r_{x}-x_{0})(N_{1,x}+x_{0})\right\} 
				\quad\left\{ \sum_{N_{1,y}}\textrm{sinc}^2(N_{1,y}+r_{y})\right\} \left\{ \sum_{N_{1,z}}\textrm{sinc}^2(N_{1,z}+r_{z})\right\} \\
				& =\left\{ \frac{\sin2\pi(x_{0}-r_{x})}{2\pi}+r_{x}\right\}.
			\end{aligned}
		\end{equation}
	\end{widetext}
	\section{Calculation Setting for Graphene }\label{ap:qe-conditions}
	A hexagonal cell ($\texttt{ibrav}=4$) with $a=4.602$~bohr ($2.435$~\AA) and
	$c=18.408$~bohr ($9.74$~\AA) was used, containing two carbon atoms at
	fractional coordinates $(0,0,0)$ and $(1/3,2/3,0)$.
	Plane-wave cutoffs of $\texttt{ecutwfc}=100$~Ry and $\texttt{ecutrho}=500$~Ry were adopted.
	Brillouin-zone sampling employed an $N\times N\times 1$ mesh (e.g., $26\times 26\times 1$),
	and Methfessel--Paxton smearing with $\texttt{degauss}=0.01$~Ry was used.
	For the Wannier analysis we used W90 with $\texttt{num\_bands}=5$ and
	$\texttt{num\_wann}=5$ on the same $k$-mesh (e.g., $\texttt{mp\_grid}=26~26~1$),
	starting from initial projections consisting of C-$p_z$ orbitals and three
	bond-centered $s$-like orbitals.
	
	\section{Reciprocal-space formulas and seam diagnostics for graphene}
	\label{ap:kspace-def}
	This appendix reformulates the standard reciprocal-space spread expressions~\cite{MarzariVanderbilt1997,Wannier90_2008,Wannier90_2020} to provide a rigorous theoretical basis for the $\mathcal{O}(L)$ scaling observed in Sec.~\ref{sec:graphene}. The purpose is to explicitly evaluate the geometric gauge defects using two complementary diagnostics, the single-state overlap $\sigma_{\alpha}^s$ and the subspace-level overlap $\sigma_{\min}$.
	\subsection{Nearest-neighbor overlaps and discrete measure}
	For each nearest-neighbor link $\delta\boldsymbol{k}_{\alpha}$ ($\alpha\in\mathcal N$),
	we define the overlap matrix
	\begin{equation}
		M_{sp}^{\alpha}(\boldsymbol{k})
		=
		\langle u_{\boldsymbol{k}}^s\vert u_{\boldsymbol{k}+\delta\boldsymbol{k}_{\alpha}}^p\rangle.
		\label{eq:link-overlap-new-clean}
	\end{equation}
	We also introduce the discrete measure
	\begin{equation}
		\nu_{\alpha}
		=
		\frac{w_{\alpha}}{N_{\boldsymbol{k}}},
		\qquad N_{\boldsymbol{k}}=L^2,
		\label{eq:measure-dS-clean}
	\end{equation}
	so that the weights do not appear explicitly in the local kernels discussed below.
	On a uniform mesh, $\nu_{\alpha}=\mathcal{O}(L^0)$.
	Using Eq.~(\ref{eq:measure-dS-clean}), the Marzari--Vanderbilt decomposition becomes
	\begin{equation}
		\Omega=\Omega_I+\Omega_{OD}+\Omega_D,
		\label{eq:MVspread-new-clean}
	\end{equation}
	with
	\begin{align}
		\Omega_I
		&=
		\sum_{\boldsymbol{k},\alpha} \nu_{\alpha}
		\left(
		J-\sum_{s,p}\left|M_{sp}^{\alpha}(\boldsymbol{k})\right|^2
		\right),\\
		\Omega_{OD}
		&=
		\sum_{\boldsymbol{k},\alpha} \nu_{\alpha}
		\sum_{s\neq p}\left|M_{sp}^{\alpha}(\boldsymbol{k})\right|^2,
		\\
		\Omega_D
		&=
		\sum_{\boldsymbol{k},\alpha} \nu_{\alpha}
		\sum_s
		\left[
		\Im\log M_{ss}^{\alpha}(\boldsymbol{k})
		+\delta\boldsymbol{k}_{\alpha}\cdot\boldsymbol{x}_0^s
		\right]^2.
		\label{eq:OmegaD-new-clean}
	\end{align}
	The seam analysis below uses only the amplitude part of the links.
	The phase contribution in $\Omega_D$ belongs to the full spread functional,
	but it is not needed for the curvature diagnostics introduced here.
	
	\subsection{Single-band(series) distortion carried by an extracted MLWF}
	For a fixed extracted series $s$, we consider the diagonal link
	\begin{equation}
		m_{\alpha}^s(\boldsymbol{k})
		=
		M_{ss}^{\alpha}(\boldsymbol{k})
		=
		\langle u_{\boldsymbol{k}}^s\vert u_{\boldsymbol{k}+\delta\boldsymbol{k}_{\alpha}}^s\rangle.
		\label{eq:diag-link-single-clean}
	\end{equation}
	Because this is a $1\times 1$ overlap matrix, its singular value is simply
	\begin{equation}
		\sigma_{\alpha}^s(\boldsymbol{k})
		=
		\left|m_{\alpha}^s(\boldsymbol{k})\right|.
	\end{equation}
	We then define the corresponding single-state kernel by
	\begin{equation}
		\omega_{\alpha}^s(\boldsymbol{k})
		=
		1-\left[\sigma_{\alpha}^s(\boldsymbol{k})\right]^2
		=
		1-\left|m_{\alpha}^s(\boldsymbol{k})\right|^2.
		\label{eq:omega-single-clean}
	\end{equation}
	Equation~(\ref{eq:omega-single-clean}) is the explicit relation between
	$\sigma_{\alpha}^s$ and $\omega_{\alpha}^s$.
	Thus $\omega_{\alpha}^s$ measures how far the one-dimensional subspace carried by the
	$s$-th Wannier series departs from perfect continuity across the link
	$\delta\boldsymbol{k}_{\alpha}$.
	No logarithmic phase term is included here, because the present diagnostic is meant to isolate the
	curvature of the carried one-dimensional subspace itself.
	\subsection{Subspace-level diagnostic for the residual transported space}
	The complementary diagnostic for the transported $J$-dimensional subspace is the minimum singular value of the full overlap matrix:
	\begin{equation}
		\sigma_{\min}(\boldsymbol{k};\delta\boldsymbol{k}_{\alpha})
		=
		\min_i\sigma_i\!\left(M^{\alpha}(\boldsymbol{k})\right)
		=
		\cos\theta_{\max}.
		\label{eq:sigmaMin-new-clean}
	\end{equation}
	The associated defect is
	\begin{equation}
		d_{\alpha}(\boldsymbol{k})
		=
		1-\sigma_{\min}^2(\boldsymbol{k};\delta\boldsymbol{k}_{\alpha}).
		\label{eq:subspace-defect-new-clean}
	\end{equation}
	Equation~(\ref{eq:sigmaMin-new-clean}) measures the mismatch of the neighboring $J$-dimensional transported subspaces, whereas Eq.~(\ref{eq:omega-single-clean}) measures the geometric curvature of the one-dimensional subspace carried by a particular extracted MLWF. Both quantities are evaluated strictly in reciprocal space using the unified discrete measure $\nu_{\alpha}$.
	
	\subsection{Order evaluation of the boundary seam defects}
	\label{ap:border-term}
	Appendix~\ref{ap:direct-derivation} shows that the interior links along each transported string are flattened,
	so the nontrivial geometric mismatch is pushed entirely to the boundary seam.
	Let $\beta$ denote the link direction in which the seam appears after the first extraction.
	The order evaluations of the single-state and subspace-level diagnostics on this boundary are strictly $\mathcal{O}(L^0)$:
	\begin{equation}
		\begin{cases}
			\omega_{\beta}^1(\boldsymbol{k})=\mathcal{O}(L^0)\\
			d_{\beta}(\boldsymbol{k})=1-\sigma_{\min}^2(\boldsymbol{k};\delta\boldsymbol{k}_{\beta})=\mathcal{O}(L^0)
		\end{cases}
		\quad
		(\boldsymbol{k}\in\mathrm{seam}).
		\label{eq:seam-orders-new-clean}
	\end{equation}
	Equivalently, the singular values themselves exhibit a constant $\mathcal{O}(L^0)$ floor:
	\begin{equation}
		\begin{cases}
			\sigma_{\beta}^1(\boldsymbol{k})=\mathcal{O}(L^0)\\
			\sigma_{\min}(\boldsymbol{k};\delta\boldsymbol{k}_{\beta})=\mathcal{O}(L^0)
		\end{cases}
		\quad
		(\boldsymbol{k}\in\mathrm{seam}).
		\label{eq:seam-sigma-orders-new-clean}
	\end{equation}
	These finite $\mathcal{O}(L^0)$ evaluations confirm that neither the single-state nor the subspace-level local kernels develop any anomalous $L$-dependent divergence. The macroscopic accumulation of these local defects over the seam, which yields the $\mathcal{O}(L)$ scaling of the total spread, is explicitly evaluated in Sec.~\ref{sec:graphene}.
	\bibliography{71886}
\end{document}